%% file: main/main.tex
\documentclass{article} 
\usepackage{main/iclr2026_conference,times}

\iclrfinalcopy 

\input{main/math_commands}

\input{main/macro}
\title{\system{}: Financial Fusion of Agentic Intelligence for Multimodal Trading}

\author{Changshuo Liu\textsuperscript{1$\dag$}, 
Yanzheng Jin\textsuperscript{1$\dag$}, 
Shangfeng Cai\textsuperscript{1$*$}, 
Peng Fang\textsuperscript{1,2}, 
Xiaokui Xiao\textsuperscript{1}, 
Beng Chin Ooi\textsuperscript{3}
\\
\textsuperscript{1}National University of Singapore\\
\textsuperscript{2}Huazhong University of Science and Technology\\
\textsuperscript{3}Zhejiang University\\
\texttt{\{changshuo, yanzheng, caisf\}@u.nus.edu}, \texttt{fangpeng@hust.edu.cn},\\ 
\texttt{xkxiao@nus.edu.sg}, \
\texttt{ooibc@zju.edu.cn} \\
}
\begin{document}

\maketitle
\let\thefootnote\relax\footnotetext{
$\dag$ Equal contribution. Alphabetical order.\\
}

\input{tex/0_abstract}
\input{tex/1_intro}
\input{tex/2_pre}
\input{tex/3_method}
\input{tex/4_system}
\input{tex/5_exp}
\input{tex/7_conclusion}

\bibliography{main/references} 
\bibliographystyle{main/iclr2026_conference}

\newpage
\appendix

\input{tex/8_appendix_catalog}
\newpage
\input{tex/8_appendix_full}

\end{document}

%% file: main/math_commands.tex
\usepackage{amsmath,amsfonts,bm}

\def\eqref#1{equation~\ref{#1}}

\def\1{\bm{1}}

\DeclareMathAlphabet{\mathsfit}{\encodingdefault}{\sfdefault}{m}{sl}
\SetMathAlphabet{\mathsfit}{bold}{\encodingdefault}{\sfdefault}{bx}{n}



%% file: main/macro.tex
\usepackage{xspace}
\usepackage[dvipsnames]{xcolor}
\newcommand{\system}{{F$^2$Agent}}

\newcommand{\highlight}[1]{\textbf{#1}}

\newcommand{\ignore}[1]{}

\newcommand{\best}[1]{\textbf{\underline{#1}}}
\newcommand{\second}[1]{\textbf{#1}}
\newcommand{\third}[1]{\underline{#1}}

\newcommand{\term}[1]{\textcolor{black}{#1}}

\usepackage[utf8]{inputenc} 
\usepackage[T1]{fontenc}    
\usepackage{hyperref}       
\usepackage{url}            
\usepackage{booktabs}       
\usepackage{amsfonts}       
\usepackage{nicefrac}       
\usepackage{microtype}      
\usepackage{xcolor}         
\usepackage{amsmath}
\usepackage{graphicx}
\usepackage{caption}
\usepackage{multirow}
\usepackage{amssymb}
\usepackage{rotating}
\usepackage{adjustbox}
\usepackage{subcaption}
\usepackage{float}
\usepackage{makecell}
\usepackage{amssymb}
\usepackage[table]{xcolor}
\usepackage{wrapfig}
\usepackage{booktabs}

\usepackage{arydshln}

\usepackage{pifont}
\newcommand{\cmark}{\checkmark}
\newcommand{\xmark}{\ensuremath{\times}}

\usepackage{tabularx}

\usepackage{array}

\newcolumntype{L}[1]{>{\raggedright\arraybackslash}p{#1}}
\newcolumntype{Y}{>{\raggedright\arraybackslash}X}

\definecolor{bestrow}{rgb}{0.9, 0.9, 0.9}       

%% file: tex/0_abstract.tex
\begin{abstract}
  With increasingly diverse and heterogeneous information sources, effectively leveraging multimodal data is becoming pivotal for high-quality financial trading. Although recent advancements in Large Language Model (LLM)-based agents have enabled the ingestion of multimodal inputs, existing methods \term{fail to capture nuanced cross-modal dependencies and remain vulnerable to market noise, due to limited multimodal modeling, ineffective fusion mechanisms, and inadequate robustness}. To address these challenges, we propose \system, a novel \term{\textit{multimodal agentic paradigm}} driven by the \underline{\bf F}inancial \underline{\bf F}usion of \underline{\bf Agent}ic Intelligence. \system{} first deploys a \term{hierarchy of specialized agents} to comprehensively extract modality-specific signals. It further introduces a \term{\textit{modality-aware adaptive fusion mechanism}} coupled with \term{\textit{noise-robust consistency regularization}} to dynamically capture fine-grained inter-modality dependencies and generate noise-resilient trading signals. Extensive experiments on six stocks and cryptocurrency assets demonstrate that \system{} consistently outperforms 16 competitive baselines across multiple trading metrics, with over 20\% relative improvement in annualized return on average. Notably, \system{} delivers returns of 120.48\% on GOOG and 148.41\% on TSLA, demonstrating its efficacy and robustness in varying market dynamics.
\end{abstract}

%% file: tex/1_intro.tex
\section{Introduction}
\label{intro}
Trading markets are driven by a complex interplay of heterogeneous information sources, ranging from quantitative historical prices to qualitative textual narratives~\cite{fama1970efficient}. 
While the efficient market hypothesis posits that prices incorporate available information, effectively digesting such multimodal signals remains challenging due to structural disparities across modalities and the high noise levels inherent in financial streams~\cite{feng2021hybrid}. 
Recent research has shifted toward AI-driven methods that fuse these disparate sources to model cross-modal dependencies and support more reliable trading decisions, beyond simple pattern extraction~\cite{feng2019temporal,koa2023diffusion,li2021modeling}.

Building on this trend, a broad spectrum of computational approaches has been explored for multimodal trading, including rule-based systems with expert-defined indicators~\cite{el2013sma}, ML/DL models that learn predictive patterns from statistical and neural architectures~\cite{yang2020qlib}, and reinforcement learning frameworks that optimize sequential trading policies~\cite{mnih2013playing,schulman2017proximal}. 
Despite progress, conventional methods are largely constrained by shallow semantics and limited capability to align unstructured textual signals with structured market data, making it difficult to capture nuanced cross-modal dependencies~\cite{koa2023diffusion,zhang2024multimodal}. 
Motivated by this gap, recent studies have begun to build LLM-based trading systems, such as FinGPT~\cite{liu2023fingpt}, FinAgent~\cite{zhang2024multimodal}, TradingAgents~\cite{xiao2024tradingagents}, and DeepFund~\cite{li2025time}, evolving from instruction tuning to agentic collaboration for decision making. 
However, existing LLM-based systems often emphasize role-playing and task coordination, yet lack a principled mechanism for fusing heterogeneous modalities. In practice, they commonly rely on prompt-level concatenation that reduces numerical time-series to textual tokens, which undermines reliable cross-modal dependency modeling and leaves decisions vulnerable to market noise in volatile markets~\cite{xiao2024tradingagents,li2025time}.

Achieving reliable multimodal trading for LLM-based agents still faces three key challenges:
{\bf i) Limited multimodal modeling.}
Financial markets are inherently multimodal, with signals derived from numerical market data, technical indicators, news articles, social media, and analyst reports~\cite{ding2015deep,xu2018stock,liu2023fingpt}. 
These heterogeneous sources jointly shape market dynamics~\cite{xiao2024tradingagents}. 
However, most existing methods~\cite{xiao2024tradingagents,yu2025finmem,liu2023fingpt} either focus primarily on price and technical indicators or rely on textual sentiment alone, underutilizing complementary information across modalities and overlooking their interactions. 
For example, in classic ``sell-the-news'' scenarios~\cite{tetlock2007giving}, positive news sentiment may coincide with bearish price movements. 
Processing modalities in isolation fails to detect cross-source misalignment, causing agents to chase sentiment traps that joint reasoning would avoid~\cite{zhang2024multimodal}.
{\bf ii) Ineffective fusion mechanisms.}
Even when multiple modalities are incorporated into LLM-based agents, aligning heterogeneous signals with different structures, semantics, and temporal granularities remains non-trivial. 
Most current solutions rely on prompt-level concatenation, treating numerical time-series as textual tokens in the context window~\cite{gruver2023large,zhang2024multimodal}. 
This ad-hoc fusion rarely resolves the modality gap: the LLM’s textual bias can dominate the joint representation and drown out subtle quantitative cues, leading to weak cross-modal interaction modeling~\cite{liang2022high,huang2023language}.
{\bf iii) Inadequate robustness.}
Financial streams are inherently stochastic, with low signal-to-noise ratios and frequent regime shifts~\cite{fama1970efficient,zhao2023doubleadapt}. 
However, existing LLM-based agents often lack explicit mechanisms to disentangle persistent market trends from transient perturbations (e.g., short-term fluctuations or microstructure noise). 
As a result, they may overreact to irrelevant news cues or minor price movements, producing spurious trading signals and unstable decisions under volatile market dynamics~\cite{shi2023large,zhu2025findeepresearch}.

To address these challenges, we propose \textbf{\system}
, a novel multimodal agentic paradigm driven by the \underline{\bf F}inancial \underline{\bf F}usion of \underline{\bf Agent}ic Intelligence. 
In contrast to prevailing approaches limited to shallow multimodal utilization, \system{} {\bf first} deploys a {\em Hierarchy of Specialized Agents} to extract complementary modality-specific signals, including a Market Analysis Agent, a Technical Analysis Agent, a News Analysis Agent, and a Sentiment Analysis Agent. 
{\bf Second}, \system{} introduces a \textit{Modality-aware Adaptive Fusion Mechanism} to dynamically capture fine-grained inter-modality dependencies across heterogeneous sources. 
{\bf Third}, \system{} leverages \textit{Noise-robust Consistency Regularization} to improve robustness against market noise and produce noise-resilient trading signals. 
{\bf Finally}, \system{} provides an end-to-end {\em Multimodal Agentic Trading System} that integrates a fine-tuned LLM with a news summarizer, a modality processing module, and a backtesting engine to produce trading decisions under practical trading constraints.

The main contributions of this paper are summarized as:
\textbf{1)} We introduce a \textbf{{\em Hierarchy of Specialized Agents}} to extract complementary modality-specific signals from heterogeneous sources. 
\textbf{2)} We propose a \textbf{\textit{Modality-aware Adaptive Fusion Mechanism}} that dynamically captures fine-grained inter-modality dependencies, mitigating the modality gap induced by prompt-level concatenation.
\textbf{3)} We introduce \textbf{\textit{Noise-robust Consistency Regularization}} to improve robustness against market noise and regime shifts, yielding stable trading signals.
\textbf{4)} We implement an end-to-end \textbf{{\em Multimodal Agentic Trading System}} and conduct extensive experiments on six assets, which demonstrate that \system{} consistently outperforms 16 competitive baselines (market/rule-based, ML/DL, RL, general and financial LLM agents) across multiple trading metrics, achieving over 20\% improvement in annualized return on average. Notably, it attains returns of 120.48\% on GOOG and 148.41\% on TSLA, and achieves robust performance across diverse market conditions.

%% file: tex/2_pre.tex
\section{Preliminaries}
\label{pre}
\noindent
\highlight{Multimodality.} Multimodal financial signals arise from heterogeneous data sources with distinct statistical properties and preprocessing requirements. Instead of enumerating categories, we treat each data source as an independent modality, allowing the modality set to evolve over time as new data sources become available. Formally, let \(\mathcal{M}\) denote the modality set, where each modality \(m\in\mathcal{M}\) is characterized by a temporal sequence of $T$ days, \(X^{m}=\{x^{m}_{1},x^{m}_{2},\ldots,x^{m}_{T}\}\). The multimodal input can be compactly represented as \(\mathcal{X}=\{X^{m}\}_{m\in\mathcal{M}}\). Each modality is processed by a modality-specific encoder $g^{m}(\cdot)$, followed by a projection layer $P^{m}(\cdot)$, which maps raw input into a shared latent space $h \in \mathbb{R}^{d}$ and serves as the basis for subsequent modality fusion. For each time step $t$, 
\vspace{-0.7mm}
\begin{equation}
\small
    h^{m}_{t}=P^{m}\!\big(g^{m}(x^{m}_{t})\big).
\end{equation}
\highlight{Problem Formulation.}
We study multimodal stock movement prediction over a stock universe $S$ and a time horizon $T$. 
At date $t$, the multimodal input for stock $s\in S$ is denoted as $\{x_t^{(i)}\}_{i=1}^{I}$
where $x_t^{(i)}$ is the feature of the $i$-th modality (market, technical, news, sentiment) available up to time $t$. 
Let $\{f_k\}_{k=1}^K$ be the $K$ agents, each producing a hidden representation $h_k(X_t^s)\in\mathbb{R}^{d_k}$. 
The multi-agent fusion module concatenates these features as $f(X_t^s) = h_1(X_t^s);\,h_2(X_t^s);\;\cdots;\,h_K(X_t^s).$

The prediction head $F_\theta$ maps the fused representation into class probabilities over the label set $L=\{\text{UP},\text{DOWN}\}$: $\Pr(y_t^s \in L \mid X_t^s) = F_\theta(f(X_t^s)).$
The ground-truth next-day movement label is defined from the adjusted closing price $p_t^c$ as
\vspace{-1mm}
\begin{equation}
\small
Y_t^s =
\begin{cases}
0, & \text{if } p_t^c < p_{t-1}^c\\
1, & \text{if } p_t^c \geq p_{t-1}^c
\end{cases},
\label{eq:yt}
\end{equation}
where $Y_t^s=0$ indicates a price decrease and $Y_t^s=1$ indicates a price increase.

%% file: tex/3_method.tex
\section{Methodology}
\label{method}

\subsection{Overview}
\begin{figure*}[t]
    \centering
    \includegraphics[width=\textwidth]{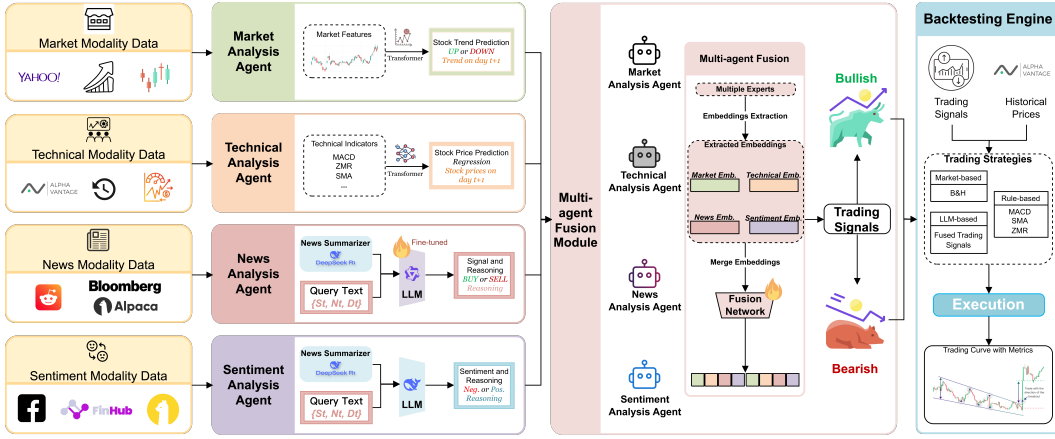}
    \captionof{figure}{The overall architecture of \system{}, which orchestrates a hierarchy of specialized agents to extract latent features from heterogeneous multimodal inputs and synthesizes these representations via a Multi-agent Fusion module to generate robust trading signals.}
    \label{fig:fagents_framework}
    \vspace{-4mm}
\end{figure*}

As illustrated in Fig.~\ref{fig:fagents_framework}, \system{} is a multimodal agentic trading system that integrates a hierarchy of specialized agents with a modality-aware adaptive fusion mechanism to produce noise-resilient trading signals, which are then evaluated by an end-to-end backtesting engine.
Specifically, the market and technical analysis agents encode OHLCV and technical indicator sequences using trained forecasting models to capture temporal patterns and market dynamics.
For textual modalities, the sentiment analysis agent leverages DeepSeek-R1 (Distill-Llama-8B)~\cite{huang2025explainable} to infer market sentiment from ticker-conditioned daily news summaries, while the news analysis agent uses a Qwen-based LLM~\cite{team2024qwen2,yang2025qwen3} fine-tuned in two stages and prompted with CoT-style instructions~\cite{wei2022chain} to generate faithful reasoning explanations.
The multi-agent fusion module then aggregates agent embeddings through adaptive, modality-aware attention and enforces robustness via consistency regularization, yielding unified representations for downstream backtesting evaluation.

\subsection{Hierarchy of Specialized Agents}
\highlight{Market Analysis Agent.}
The market analysis agent encodes raw OHLCV (Open, High, Low, Close, Volume) sequences into high-level latent representations.
We treat OHLCV as a multivariate time series with non-stationary volatility, where predictive cues can arise from non-linear price-volume coupling~\cite{cont2001empirical}.
To capture long-range temporal dependencies and lead-lag effects within the lookback window, we adopt a causal Transformer encoder~\cite{zerveas2021transformer}.
For a stock $s$ at prediction date $t$, we construct a $T$-day window
$\mathbf{X}^{\mathrm{MA}}_{s,t}=[x^{\mathrm{MA}}_{s,t-T},\dots,x^{\mathrm{MA}}_{s,t-1}]\in\mathbb{R}^{T\times F_{\mathrm{MA}}}$,
where $x^{\mathrm{MA}}_{s,\tau}$ contains OHLCV and $F_{\mathrm{MA}}$ is the feature dimension.
We embed the input with positional encoding as
$\mathbf{H}^{(0)}_{\mathrm{MA}}=\mathrm{PE}(\mathbf{X}^{\mathrm{MA}}_{s,t})\,\mathbf{W}^{E}_{\mathrm{MA}}\in\mathbb{R}^{T\times d_{\mathrm{MA}}}$,
and apply $L_{\mathrm{MA}}$ causal Transformer blocks (mask $\mathbf{M}_{\mathrm{causal}}$) to obtain
$\mathbf{H}^{\mathrm{MA}}_{s,t}\in\mathbb{R}^{T\times d_{\mathrm{MA}}}$.
We use the \texttt{[CLS]} representation as a fixed-length summary $h^{\mathrm{MA}}_{s,t}\in\mathbb{R}^{d_{\mathrm{MA}}}$ when needed, with projection to the shared space performed by the fusion module (Eq.~\ref{eq:embed_extract_align}).

\highlight{Technical Analysis Agent.}
The technical analysis agent focuses on expression-based alpha factors~\cite{kakushadze2016101}, which provide structured quantitative descriptions of trend, momentum, and volatility and complement textual modalities in multimodal trading.
For a stock $s$ at prediction date $t$, we assemble a $T$-day window of technical indicators
$\mathbf{X}^{\mathrm{TA}}_{s,t}=[x^{\mathrm{TA}}_{s,t-T},\dots,x^{\mathrm{TA}}_{s,t-1}]\in\mathbb{R}^{T\times F_{\mathrm{TA}}}$,
where $x^{\mathrm{TA}}_{s,\tau}\in\mathbb{R}^{F_{\mathrm{TA}}}$ includes MACD, RSI, and other technical indicators.
We adopt a causal Transformer encoder to model temporal dependencies and non-linear interactions among indicators.
Analogous to the market encoder, we compute
$\mathbf{H}^{(0)}_{\mathrm{TA}}=\mathrm{PE}(\mathbf{X}^{\mathrm{TA}}_{s,t})\,\mathbf{W}^{E}_{\mathrm{TA}}\in\mathbb{R}^{T\times d_{\mathrm{TA}}}$
and obtain the technical representation
$\mathbf{H}^{\mathrm{TA}}_{s,t}\in\mathbb{R}^{T\times d_{\mathrm{TA}}}$ after $L_{\mathrm{TA}}$ masked Transformer blocks.
We follow the same fusion interface as the market agent: we expose the hidden-state sequence $\mathbf{H}_{s,t}^{\mathrm{TA}}$ and, when needed, a fixed-length summary $h^{\mathrm{TA}}{s,t}$. The fusion module then performs alignment via Eq.~\ref{eq:embed_extract_align}.

\highlight{Sentiment Analysis Agent.}
The sentiment analysis agent captures sentiment signals from financial news and social media.
We use DeepSeek-R1~\cite{huang2025explainable} and enable tool-augmented retrieval: given $(s,t)$, the agent queries external news APIs (e.g., Alpaca and Alpha Vantage), and the retrieved articles are deduplicated and summarized by the News Summarizer (Fig.~\ref{fig:news_summarizer}) to form
$\mathbf{X}^{\mathrm{SA}}_{s,t}=\{n_{s,t,1},\dots,n_{s,t,K_{s,t}}\}$.

\underline{Per-item Sentiment Inference.}
For each summary $n_{s,t,j}$, the agent predicts a binary label $\hat c_{s,t,j}\in\{0,1\}$ and a brief rationale $\hat q_{s,t,j}$:
$(\hat c_{s,t,j},\hat q_{s,t,j}) = g_{\mathrm{SA}}(n_{s,t,j}),\ j=1,\dots,K_{s,t}$,
where $g_{\mathrm{SA}}(\cdot)$ is instantiated by DeepSeek-R1.

\noindent
\underline{Hidden Representation for Fusion.}
We extract last-layer hidden states $\mathbf{H}^{\mathrm{SA}}_{s,t,j}\in\mathbb{R}^{L_{s,t,j}\times d_{\mathrm{SA}}}$, apply last-token pooling $u_{s,t,j}=\mathbf{H}^{\mathrm{SA}}_{s,t,j}(L_{s,t,j})$, and aggregate across items to obtain
\vspace{-2mm}
\begin{equation}
\small
\label{eq:sa_agg}
h^{\mathrm{SA}}_{s,t}=\frac{1}{K_{s,t}}\sum_{j=1}^{K_{s,t}} u_{s,t,j}\in\mathbb{R}^{d_{\mathrm{SA}}}.
\end{equation}
The fusion module maps $h^{\mathrm{SA}}_{s,t}$ to the shared space via Eq.~\ref{eq:embed_extract_align}.

\label{sec:news_analysis}
\highlight{News Analysis Agent.}
As illustrated in Fig.~\ref{fig:fagents_framework}, the news analysis agent (i) retrieves and consolidates news evidence for a queried pair $(s,t)$ and (ii) generates transparent explanations to support downstream prediction.
Given $(s,t)$, the agent queries external news APIs, including Alpaca News and the New York Times API.
Retrieved articles are deduplicated and summarized by the News Summarizer (Fig.~\ref{fig:news_summarizer}), yielding top-$K_{s,t}$ summaries
$\mathbf{X}^{\mathrm{NA}}_{s,t}=\{n_{s,t,1},\dots,n_{s,t,K_{s,t}}\}$.

\noindent
\underline{Explanation Prompting.}
Conditioned on $(s,t)$ and $\mathbf{X}^{\mathrm{NA}}_{s,t}$, the news reasoning model $g_{\mathrm{NA}}(\cdot)$ produces an auxiliary directional signal $\hat c^{\mathrm{NA}}_{s,t}\in\{\textsc{UP},\textsc{DOWN}\}$, a confidence score $\hat p^{\mathrm{NA}}_{s,t}\in[0,1]$, and a natural-language explanation $\hat e^{\mathrm{NA}}_{s,t}$, where the backbone LLM in \system{} instantiates
$g_{\mathrm{NA}}(\cdot)$ to produce
$(\hat c^{\mathrm{NA}}_{s,t}, \hat p^{\mathrm{NA}}_{s,t}, \hat e^{\mathrm{NA}}_{s,t})
= g_{\mathrm{NA}}(s,t,\mathbf{X}^{\mathrm{NA}}_{s,t})$.

\noindent
\underline{Model Fine-tuning.}
Following Chain-of-Thought prompting~\cite{wei2022chain}, we fine-tune $g_{\mathrm{NA}}$ using a two-stage instruction dataset (Fig.~\ref{fig:system_training_pipeline}).
We first collect topic-focused QA pairs (e.g., trading decisions under different cases) generated with GPT-4o-mini~\cite{hurst2024gpt}, and then collect simulated application trajectories designed to mimic real usage of \system.
We optimize a combined objective:
\vspace{-2mm}
\begin{equation}
\label{eq:na_ft_loss}
\small
\mathcal{L}_{\mathrm{NA}}(\theta)
=-\mathbb{E}_{(\mathcal{P},\mathcal{Y},\mathbf{y})\sim\mathcal{D}_{\mathrm{NA}}}
\Big[
\log P_{\theta}(\mathcal{Y}\mid \mathcal{P})
+
\lambda \sum_{k=1}^{C} y_k \log P_{\theta}(k\mid \mathcal{P})
\Big],
\end{equation}
where $\mathcal{P}$ is the instruction prompt, $\mathcal{Y}$ is the target response, and $\mathbf{y}\in\{0,1\}^{C}$ is the one-hot label for the auxiliary classification (e.g., \textsc{UP}/\textsc{DOWN}).
The first term corresponds to the language modeling loss, while the second term is an auxiliary classification cross-entropy weighted by $\lambda$.
To mitigate majority-class bias, prompts explicitly require selecting from $\{\textsc{UP},\textsc{DOWN}\}$~\cite{zhao2021calibrate}.
Our prompt format and interaction design draw inspiration from FinMEM~\cite{yu2025finmem}, while inference is executed in a single summarization-reasoning step (see case studies in Fig.~\ref{fig:case_study} in Appendix~\ref{app:sentiment_module}).

\underline{Hidden Representation for Fusion.}
Following the sentiment analysis agent, we apply last-token pooling to the LLM's last-layer hidden states to obtain $h^{\mathrm{NA}}_{s,t}\in\mathbb{R}^{d_{\mathrm{NA}}}$, which is projected into the shared fusion space via Eq.~\ref{eq:embed_extract_align}. Detailed implementations are provided in Appendix~\ref{app:agents_details}.

\subsection{Modality-aware Adaptive Fusion Mechanism}
Inspired by multimodal LLMs and MoE~\cite{zhang2024multimodal, zhang2024magiclens, mu2025comprehensive}, we design a modality-aware fusion layer to coordinate specialized agents and robustly model cross-modal dependencies.
Although recent LLM-based agents can ingest diverse inputs, existing systems often rely on shallow, ad hoc fusion, making them brittle to cross-modal mismatch and corrupted signals.
To this end, \system{} introduces a modality-aware adaptive fusion mechanism as the coordination layer.
As shown in Fig.~\ref{fig:fagents_framework}, it enables structured information exchange across agents, captures inter-modality dependencies, and forms a unified representation for trading decisions under noise-robust regularization.

\noindent
\highlight{Modality Embedding Extraction and Alignment.}
Given the multimodal input $x=\{X^{(m)}\}_{m=1}^{M}$ 
\begin{wrapfigure}[18]{r}{0.4\linewidth}
    \vspace{-4mm}
    \centering
    \includegraphics[
        width=\linewidth,
        keepaspectratio
    ]{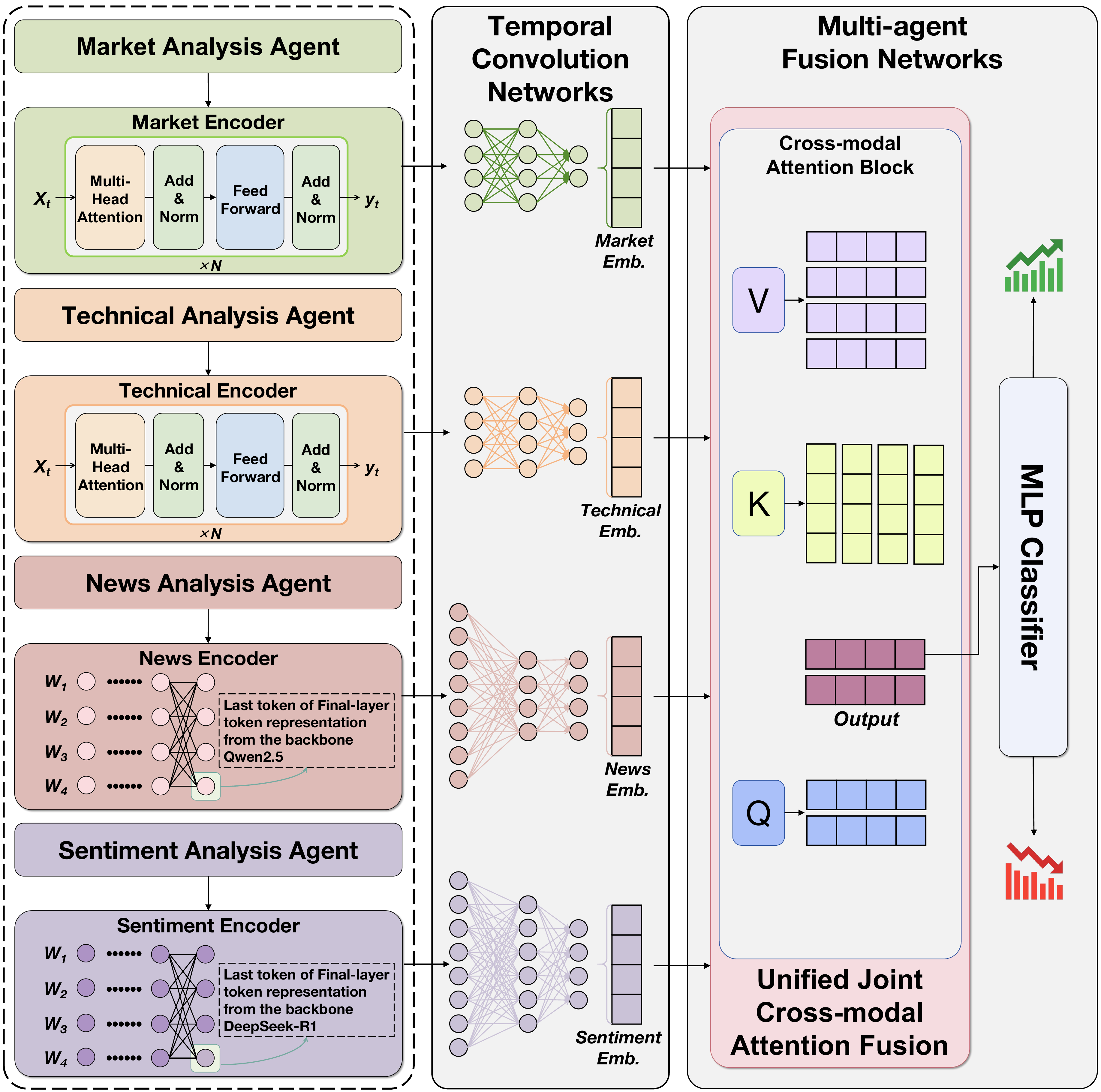}
    \caption{Modality-aware adaptive fusion mechanism.}
    \label{fig:rjcma_framework}
    \vspace{-2mm}
\end{wrapfigure}
over $M$ modalities, each modality sequence $X^{(m)}$ is processed by a modality-specific agent $f_m$ to produce a sequence representation $H_m(x)=f_m(X^{(m)})\in\mathbb{R}^{T_m\times \tilde d_m}$, where $T_m$ denotes the sequence length and $\tilde d_m$ is the hidden dimension of agent $m$. We then extract a fixed-length summary $h_m(x)\in\mathbb{R}^{\tilde d_m}$. For LLM-based agents, we use last-token pooling~\cite{behnamghader2024llm2vec} to summarize the output (e.g., $h_m(x)=H_m(x)_{T_m}\in\mathbb{R}^{\tilde d_m}$), where $H_m(x)_{T_m}$ is the last-layer hidden state of the last token. For non-LLM agents, we use a standard sequence summarization to obtain $h_m(x)$ (use the \texttt{[CLS]} token for Transformers encoders). Since different agents may output representations of different dimension $\tilde d_m$, we project each $h_m(x)$ into a shared $d$-dimensional space via a learnable projection $P_m\in\mathbb{R}^{d\times \tilde d_m}$, yielding the aligned modality embedding $e_m = P_m h_m(x)\in\mathbb{R}^{d}$.
For notational convenience, we summarize the above extraction and alignment as:
\vspace{-1mm}
\begin{equation}
\small
e_m \;=\; P_m\,\phi_m\!\big(f_m(X^{(m)})\big)\in\mathbb{R}^{d},
\quad P_m\in\mathbb{R}^{d\times \tilde d_m}.
\label{eq:embed_extract_align}
\end{equation}
Here $\phi_m(\cdot)$ denotes the pooling operator: for LLM-based agents we use last-token pooling, while for Transformer encoders we take the \texttt{[CLS]} representation.

\noindent
\highlight{Adaptive Modality Attention.} Given aligned embeddings $\{e_m\}_{m=1}^{M}$, we compute modality-specific queries, keys, and values as $Q_m = e_m W_Q^{(m)}$, $K_m = e_m W_K^{(m)}$, and $V_m = e_m W_V^{(m)}$. We form a joint query by concatenating modality queries, $Q_{\mathrm{all}}=[Q_1;\dots;Q_M]$. The modality attention weight is then computed by $\alpha_m=\mathrm{softmax}_m\!\big(\frac{Q_{\mathrm{all}}K_m^\top}{\sqrt{d}}\big)$ (softmax over $m$), and obtain the fused representation: $\small R_{\mathrm{final}}=\sum_{m=1}^{M}\alpha_m V_m.$
To add modality-level priors beyond instance attention, we learn a modality vector $r_m$ and inject it into each modality's key and value, with $Q_{\mathrm{all}}$ fixed:
\vspace{-1mm}
\begin{equation}
\small
\label{eq:kv_injection}
r_m = u_m A,\;
\tilde K_m = e_m W_K^{(m)} + r_m U_K^{(m)},\;
\tilde V_m = e_m W_V^{(m)} + r_m U_V^{(m)},
\end{equation}
where $A$ denotes a shared modality space and $u_m$ is the projection direction for modality $m$. By injecting $r_m$ into $(\tilde K_m,\tilde V_m)$, the fusion preserves instance-independent evidence encoded in $e_m$ while introducing an instance-independent, modality-specific bias via $r_m$. We further regularize $\{r_m\}$ to encourage modality-wise diversity and avoid trivial scaling solutions:
\vspace{-2mm}
\begin{equation}
\small
\label{eq:l_modality}
L_{\mathrm{modality}}
=
\sum_{m\neq n}\Big(\frac{r_m^\top r_n}{\|r_m\|\,\|r_n\|}\Big)^2
+
\sum_{m=1}^{M}\|r_m\|^2,
\end{equation}
where $m,n$ index modalities, $m\neq n$ denotes distinct pairs, and $\|\cdot\|$ is the $\ell_2$ norm.

\noindent
\highlight{Noise-robust Consistency Regularization.} Financial signals are noisy and certain modalities may be missing or corrupted. To improve robustness, we randomly select a modality $m^\star\in\{1,\dots,M\}$ and perturb its embedding (e.g by adding gaussian noise) to obtain a perturbed input $x'$ with embeddings $\{e'_m\}_{m=1}^{M}$. Let $s(x)\in\mathbb{R}^{C}$ and $s(x')\in\mathbb{R}^{C}$ denote the predicted logits for $x$ and $x'$, where $C$ is the number of classes. We then let $\alpha_m$ and $\alpha'_m$ be the modality attention weights computed from $x$ and $x'$, respectively, with $\sum_{m=1}^{M}\alpha_m=\sum_{m=1}^{M}\alpha'_m=1$. We enforce logit-level consistency and discourage the model from increasing reliance on the perturbed modality via:
\vspace{-1mm}
\begin{equation}
\small
\label{eq:l_robust}
L_{\mathrm{robust}}
=
\|s(x)-s(x')\|
+
\gamma \max\!\big(0,\,\alpha'_{m^\star}-\alpha_{m^\star}\big),
\end{equation}
where $\|\cdot\|$ denotes the $\ell_2$ norm and $\gamma>0$ controls the strength of the attention suppression term.

\noindent
\highlight{Overall  Objective.}
Given $R_{\mathrm{final}}$, the prediction head outputs logits $s(x)\in\mathbb{R}^{C}$ and probabilities $\hat p(x)=\mathrm{softmax}(s(x))$, and is trained by:
\vspace{-2mm}
\begin{equation}
\small
\label{eq:ce_loss}
L_{\mathrm{CE}}
=
-\sum_{c=1}^{C} y^{(c)} \log \hat{p}^{(c)}(x),
\end{equation}
where $y\in\{0,1\}^{C}$ is the one-hot label and $C$ is the number of classes. We then augment the supervised objective with two proposed regularizers and optimize the final loss function:
\vspace{-1mm}
\begin{equation}
\small
\label{eq:overall_loss}
L
=
L_{\mathrm{CE}}
+
\lambda_{\mathrm{mod}}\,L_{\mathrm{modality}}
+
\lambda_{\mathrm{rob}}\,L_{\mathrm{robust}},
\end{equation}
where $\lambda_{\mathrm{mod}}$ and $\lambda_{\mathrm{rob}}$ control the modality and robustness regularization strengths.

%% file: tex/4_system.tex
\section{Multimodal Agentic Trading System}
\label{sec:system}

\system{} provides an end-to-end {\em Multimodal Agentic Trading System} that integrates a fine-tuned LLM with a news summarizer, a modality processing module, and a backtesting engine to produce trading decisions under practical trading constraints. Detailed implementations are provided in Appendix~\ref{app:system_workflow}.

\begin{wrapfigure}[9]{r}{0.52\linewidth}
    \vspace{-6mm}
    \centering
    \includegraphics[
        width=\linewidth,
        keepaspectratio
    ]{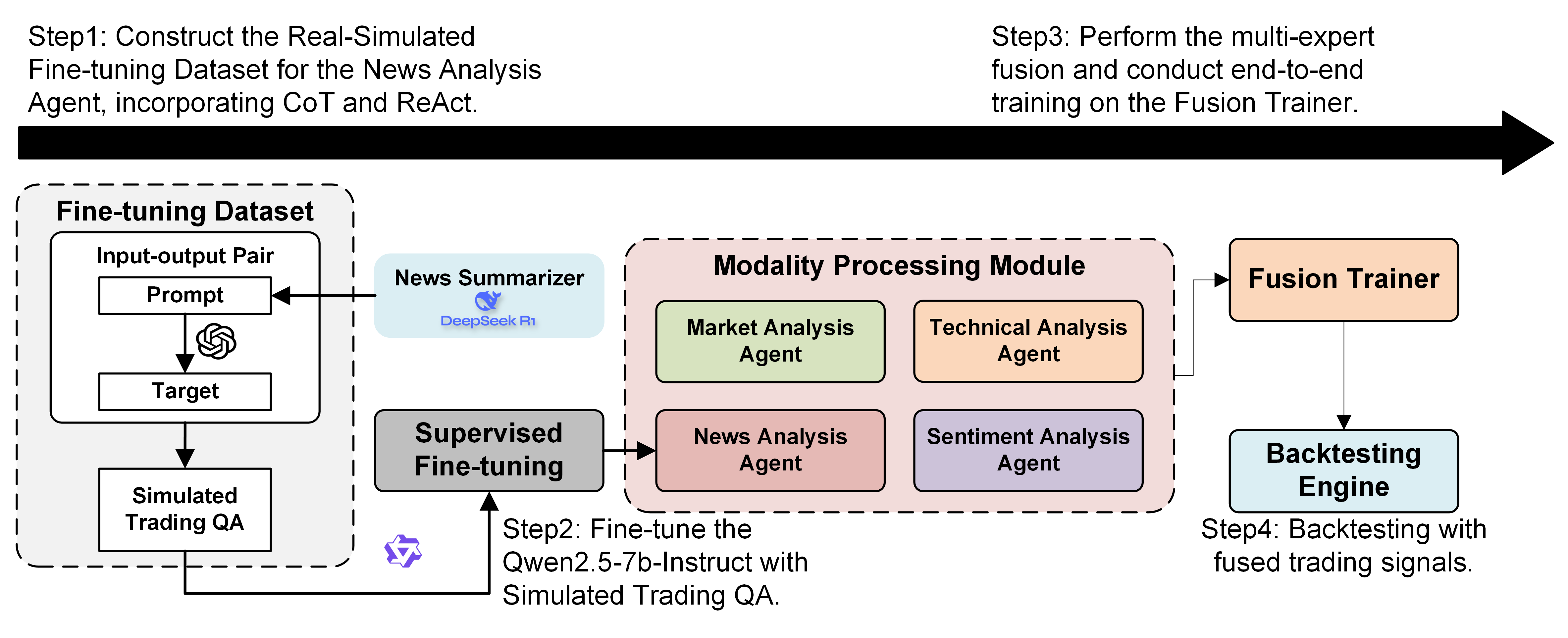}
    \vspace{-4mm}
    \caption{System workflow of \system{}.}
    \label{fig:system_training_pipeline}
    \vspace{-2mm}
\end{wrapfigure}

\highlight{News Summarizer.}
To reduce noise and control token budgets, the news summarizer serves as a preprocessing filter.
It retrieves ticker-relevant articles via semantic vector search, ranks them by estimated market impact using DeepSeek-R1, and compresses the most relevant items into concise daily summaries for downstream reasoning.

\highlight{Modality Processing Module.}
\system{} encodes four modalities with specialized agents: causal Transformers for OHLCV and technical indicators, a fine-tuned Qwen2.5-7B-Instruct model for news reasoning and explanations, and DeepSeek-R1 for sentiment inference. Each agent outputs a fixed-length latent vector for fusion.

\highlight{Backtesting Engine.}
The backtesting engine converts predicted signals into executable actions under a trading policy (detailed policy is provided in Appendix~\ref{app:backtest}) and evaluates performance on historical data using standard profitability and risk metrics. Overall, \system{} forms a pipeline that encodes each modality with specialized agents, fuses the resulting representations, and outputs trading signals with explanations and evaluation metrics. 
Formally, for a queried pair $(s,t)$ with inputs $\mathcal{X}_{s,t}=\{\mathbf{X}^{\mathrm{MA}}_{s,t},\mathbf{X}^{\mathrm{TA}}_{s,t},\mathbf{X}^{\mathrm{NA}}_{s,t},\mathbf{X}^{\mathrm{SA}}_{s,t}\}$:
\begin{equation}
\small
(y,e,\mathcal{M})
=
\Psi\!\Big(
\Phi\big(\{g_k(\mathbf{X}^{k}_{s,t},s,t)\}_{k\in\{\mathrm{MA},\mathrm{TA},\mathrm{NA},\mathrm{SA}\}}\big),
\,s,t
\Big),
\label{eq:system_workflow}
\end{equation}
where $g_k$ denotes the modality-specific agent, $\Phi(\cdot)$ is the fusion module, and $\Psi(\cdot)$ maps the fused representation to the trading output $y$, explanation $e$, and metrics $\mathcal{M}$.

%% file: tex/5_exp.tex
\section{Experiments}
\label{exp}
\subsection{Experimental Setup}
\label{exp_setup}
\highlight{Implementation Details.}
We use six NVIDIA A40 GPUs for all experiments. To ensure fair comparison, we standardize hyperparameters and optimization settings within each model category (gradient-based, RL, and LLM) under a unified environment. Additional details on the benchmark and experimental setup are provided in Appendix~\ref{app:imple_details}.

\highlight{Metrics.}
We evaluate \system{} using four standard trading metrics following~\cite{sun2023trademaster,zong2024macrohft}: one profitability metric, annualized rate of return (ARR); two performance metrics, cumulative return (CR) and Sharpe ratio (SR); and one risk metric, maximum drawdown (MDD). 
Definitions are given in Appendix~\ref{app:formula_metrics}.

\highlight{Baselines.}
We comprehensively evaluate \system{} against 16 diverse baselines categorized into traditional quantitative methods and LLM-based agents.
The traditional category includes widely used rule-based strategies (\textbf{B\&H}, \textbf{MACD}, \textbf{ZMR}, \textbf{SMA}), deep learning methods (\textbf{LSTM}, \textbf{Transformer}), and reinforcement learning algorithms (\textbf{DQN}, \textbf{PPO}).
For LLM-based comparisons, we select four state-of-the-art general-purpose models (\textbf{Qwen3-8B}, \textbf{DeepSeek-R1-0528}, \textbf{Llama4-Scout-17B}, \textbf{GPT-5-mini}) and four specialized financial agents (\textbf{FinGPT}, \textbf{FinAgent}, \textbf{TradingAgents}, \textbf{DeepFund}).
Detailed baseline settings are provided in Appendix~\ref{app:baselines}.

\input{tables/main_exp}
\begin{figure}[!ht]
\vspace{-4mm}
    \centering
    \includegraphics[width=\textwidth]{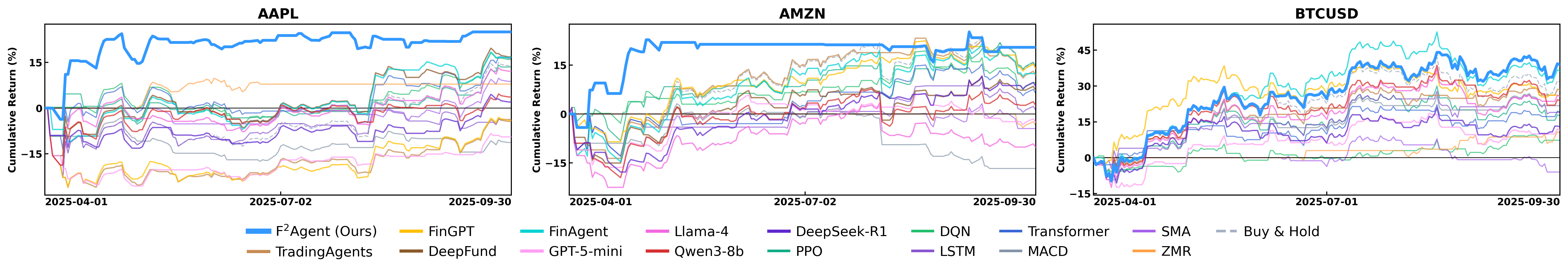}
    \caption{Cumulative return comparison of \system{} and competing benchmarks on AAPL, AMZN, and BTCUSD. Results for the remaining three assets are provided in Fig.~\ref{fig:cr_full_plot} of Appendix~\ref{app:full_results}.}
    \label{fig:cr_plot}
    \vspace{-4mm}
\end{figure}

\subsection{Overall Results}
We evaluate \system{} against 16 baselines across six diverse financial assets, including five stocks and BTCUSD, using CR, ARR, SR, and MDD. Table~\ref{tab:main_results} shows that \system{} consistently outperforms existing methods across most evaluation dimensions. In terms of ARR, \system{} achieves an average rank of 1.00 across all tested assets, indicating consistent return advantages over the nearest competitor, FinAgent (rank 3.17), and the market benchmark Buy \& Hold (rank 3.50). Specifically, \system{} obtains the best ARR on all six assets and the highest SR on five assets. 
The gains are particularly evident on TSLA (148.41\% ARR, 1.94 SR) and GOOG (120.48\% ARR, 2.55 SR), where \system{} surpasses the strongest baselines by clear margins. As shown in the improvement row, \system{} delivers substantial gains over the best-performing alternative, with improvements of up to 48.03\% in ARR and 18.60\% in SR.

\input{tables/ablation_study}

Figure~\ref{fig:cr_plot} further illustrates the cumulative return trajectories on representative assets. 
On AAPL and AMZN, \system{} establishes an early advantage and maintains stable return accumulation over most of the test horizon. On BTCUSD, where all methods exhibit larger fluctuations, \system{} remains among the strongest performers while avoiding the severe reversals observed in several baselines. This suggests that the proposed multimodal agentic design improves not only final profitability but also temporal robustness under different market conditions.

While conservative rule-based methods occasionally achieve lower MDDs at the cost of returns, \system{} strikes a stronger balance between profitability and drawdown control. In contrast, several baselines exhibit instability or asset-specific behavior.
Traditional ML/DL and RL methods show higher cross-asset variance, while FinAgent trails \system{} in profitability and generalizability. These findings suggest multimodal fusion improves decision robustness across diverse market dynamics.

\subsection{Ablation Study}
Table~\ref{tab:fusion_ablation_aapl_btc} shows that the proposed fusion design plays an important role in robust performance. The full \system{} model clearly outperforms the concatenation-based fusion baseline on both AAPL and BTCUSD, suggesting that simple feature concatenation is less effective for integrating heterogeneous financial signals. The improvement is especially pronounced on the higher-volatility BTCUSD asset in terms of ARR and SR, indicating that the proposed design can better handle noisy or conflicting modality cues. The ablation variants further support the contribution of each component. Removing $M_P$ or $M_R$ weakens return and risk-adjusted performance, while removing both leads to the largest degradation. This suggests that the modality prior guides the fusion process, whereas the robustness regularizer improves stability under noisy cross-modal interactions. The full model performs best overall, which indicates component complementarity.

\subsection{Robustness Study}
To assess robustness under heterogeneous information sources, we compare \system{} with the \input{tables/robustness_analysis}strongest general LLM baseline (Llama4-Scout-17B) and financial LLM baseline (FinAgent) across different modality combinations. Results on AAPL and the highly volatile BTCUSD asset (Table~\ref{tab:ablation_combined}) underscore the value of our modality-aware adaptive fusion and noise-robust consistency regularization. The baselines often degrade as more modalities are added, suggesting difficulty in resolving conflicting cross-modal cues: on AAPL, Llama4-Scout-17B worsens when technical indicators are appended to news and market inputs, and on BTCUSD, FinAgent likewise shows reduced risk-adjusted performance under multimodal integration.
In contrast, \system{} remains stable as more modalities are incorporated and benefits from richer multimodal inputs. Across both assets, the full multimodal setting achieves the strongest or near-strongest risk-adjusted performance, showing that hierarchical agent specialization, adaptive fusion, and consistency regularization improve inter-modality modeling and robustness.

\subsection{Token Consumption Analysis}

\noindent We evaluate practical deployment costs by quantifying the average daily token consumption required 
\begin{wrapfigure}{r}{0.40\linewidth}
    \vspace{-2mm}
    \centering
    \includegraphics[
        width=\linewidth,
        keepaspectratio
    ]{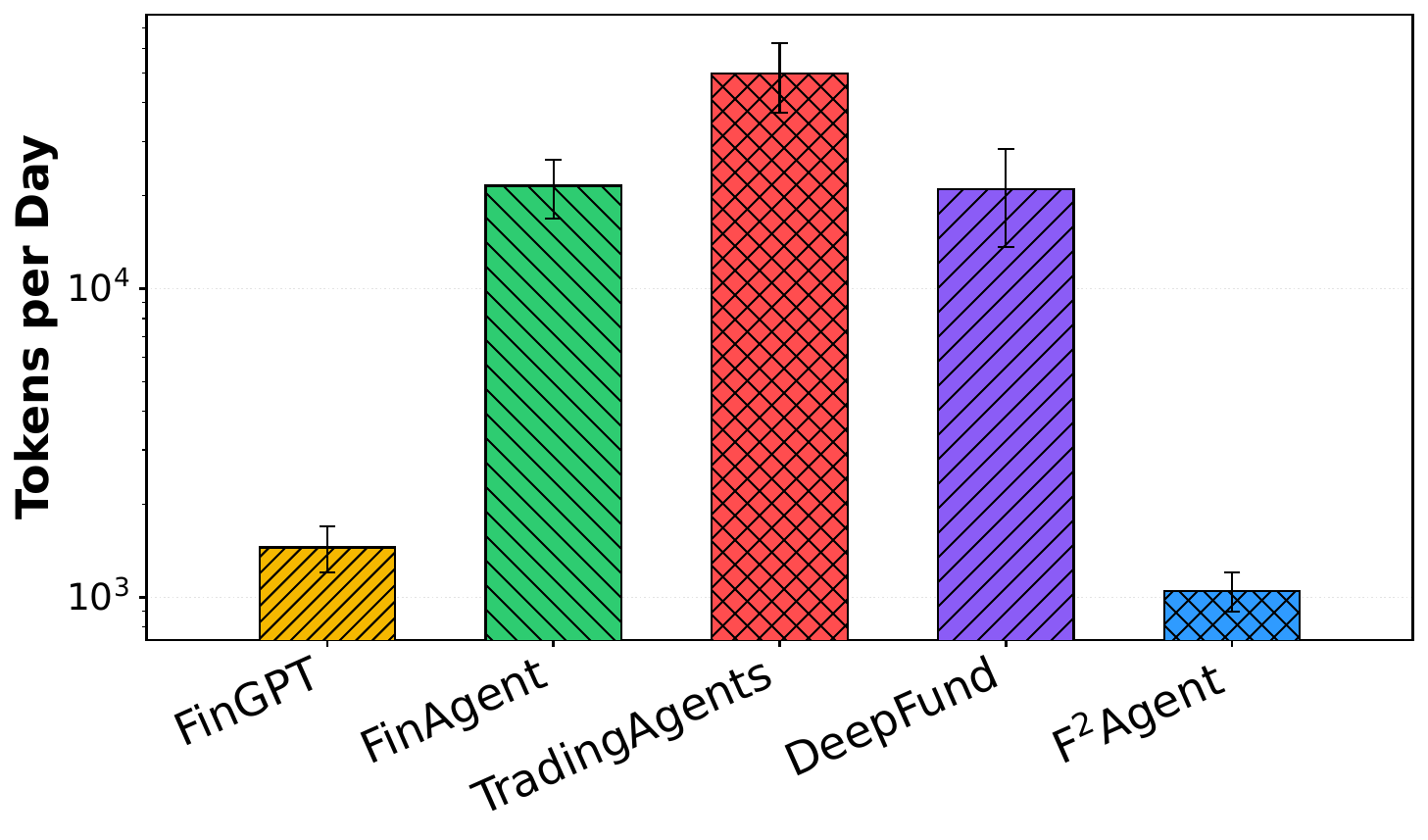}
    \vspace{-3mm}
    \caption{Token consumption analysis.}
    \label{fig:token_analysis}
    \vspace{-2mm}
\end{wrapfigure}
for signal generation across varying assets. As shown in Figure~\ref{fig:token_analysis}, empirical results demonstrate that \system{} achieves superior token efficiency compared to competing frameworks like TradingAgents and FinAgent, which shows that complex architectures need not incur excessive overhead. This efficiency stems from the strategic architectural design of \system{}, which aims to minimize unnecessary reliance on general-purpose LLMs. Unlike baselines that process heterogeneous data streams through monolithic LLM calls, our framework allocates computation across specialized agents. Token-intensive reasoning is confined mainly to the news and sentiment agents, where textual understanding is essential, while other components rely on structured processing to reduce overhead without degrading performance.

\subsection{Stock Movement Prediction Enhancement}
\begin{wrapfigure}{r}{0.40\linewidth}
    \centering
    \includegraphics[
        width=\linewidth,
        keepaspectratio
    ]{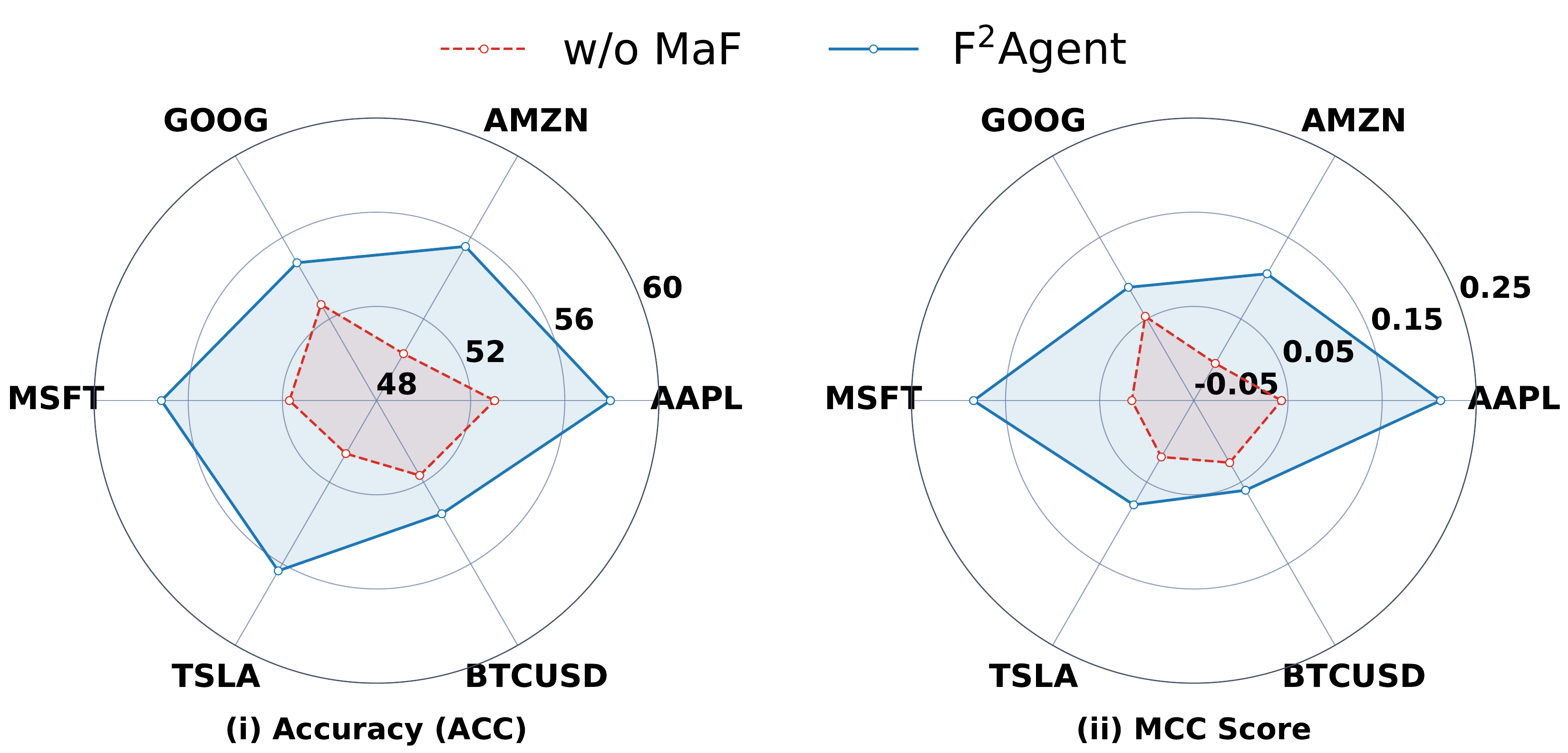}
    \caption{MaF impact on performance.}
    \label{fig:smp_radar}
    \vspace{-4mm}
\end{wrapfigure}

To intuitively assess the contribution of the Multi-agent Fusion (MaF) module, we present a radar chart comparison in Figure~\ref{fig:smp_radar}. The visualization provides empirical support for the importance of the module. Geometrically, the \system{} polygon extends beyond the ablated variant on all axes, which indicates consistent performance gains across assets. This consistency suggests that MaF improves robustness beyond isolated asset-specific performance effects. This gap is even more pronounced for the Matthews Correlation Coefficient (MCC), where the ablated model lies much closer to the origin, with MCC values near zero, suggesting limited ability to capture directional trends. In contrast, the full model shows a substantially larger coverage area. Overall, this visual evidence suggests that the MaF module plays an important role in fusing heterogeneous inputs and capturing fine-grained inter-modality dependencies.

\subsection{Qualitative Analysis: Case Study on AAPL}
\begin{figure}[H]
    \centering
    \includegraphics[width=1\textwidth]{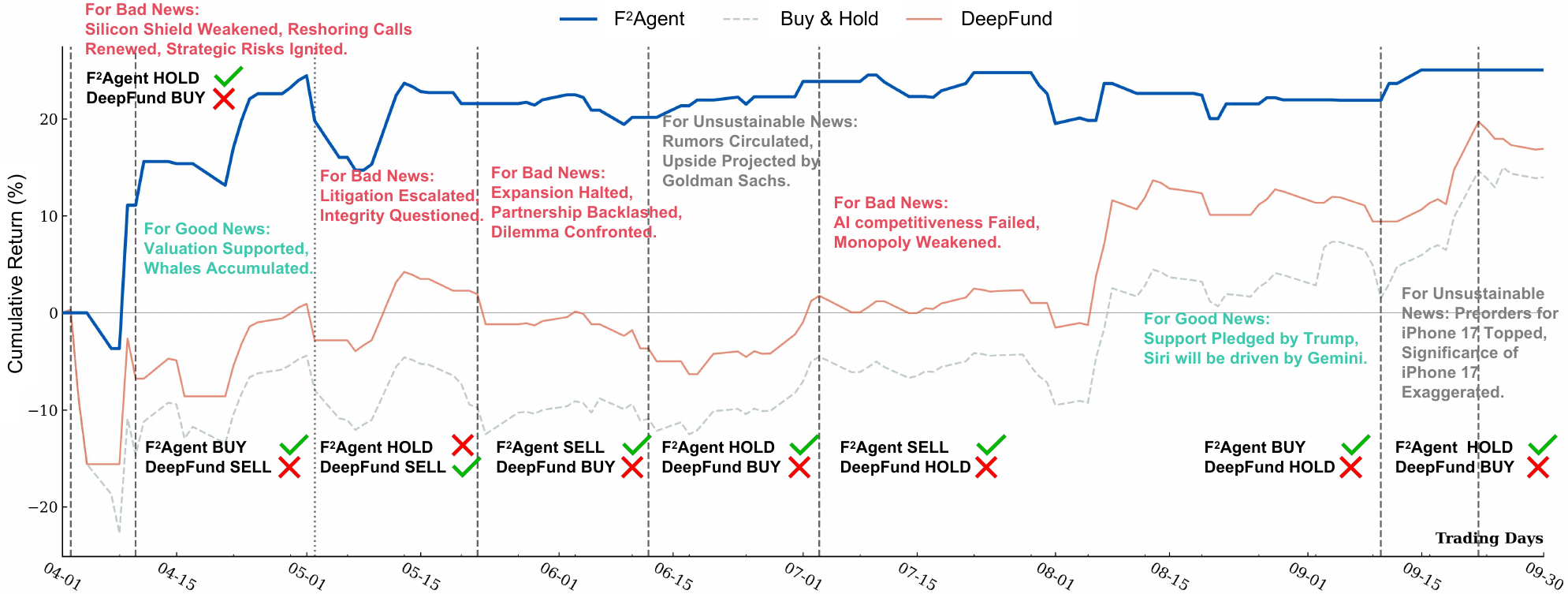}
    \caption{Case study of \system{} vs. DeepFund when performing single asset trading. DeepFund is the strongest baseline on AAPL among the compared alternatives.}
    \label{fig:case_analysis}
    \vspace{-4mm}
\end{figure}

To examine \system{}'s robustness to noise and cross-modal reasoning, we study AAPL from April to October 2025 (Fig.~\ref{fig:case_analysis}). Compared with DeepFund, \system{} better handles noisy narratives, structural risks, and signal confirmation. On Sep.~22, DeepFund issues BUY on an overstated iPhone 17 preorder narrative, while \system{} finds weak cross-modal support and chooses HOLD. On May~22, halted expansion and partnership backlash indicate deterioration; DeepFund still issues BUY from oversold technical cues, whereas \system{} chooses SELL based on news-driven risk evidence. On Apr.~10, aligned valuation and accumulation signals prompt \system{} to BUY, while DeepFund issues SELL. Overall, this case study (see Appendix~\ref{app:case_study}) suggests that \system{} relies on cross-modal evidence rather than isolated modality cues.

%% file: tables/main_exp.tex
\begin{table}[H]
\centering
\caption{Performance comparison across six assets. Best, second, and third results are marked by \best{bold-underlined}, \second{bold}, and \third{underlined} text, respectively.
The \textit{Improvement} row reports the relative performance gain of \system{} over the best-performing baseline.
Average Rank is based on annualized return (ARR). Additional tests on a broader stock pool spanning six sectors and an extended test horizon are reported in Appendix~\ref{app:add_stocks}.
}
\label{tab:main_results}
\resizebox{\textwidth}{!}{
\begin{tabular}{llcccccccccccccccccc c}
\toprule
\multirow{2}{*}{Category} & \multirow{2}{*}{Model}
& \multicolumn{3}{c}{AAPL}
& \multicolumn{3}{c}{AMZN}
& \multicolumn{3}{c}{GOOG}
& \multicolumn{3}{c}{MSFT}
& \multicolumn{3}{c}{TSLA}
& \multicolumn{3}{c}{BTCUSD} 
& \multirow{2}{*}{Avg Rank} \\ 
\cmidrule(lr){3-5} \cmidrule(lr){6-8} \cmidrule(lr){9-11}
\cmidrule(lr){12-14} \cmidrule(lr){15-17} \cmidrule(lr){18-20}
&
& ARR\%$\uparrow$ & SR$\uparrow$ & MDD\%$\downarrow$
& ARR\%$\uparrow$ & SR$\uparrow$ & MDD\%$\downarrow$
& ARR\%$\uparrow$ & SR$\uparrow$ & MDD\%$\downarrow$
& ARR\%$\uparrow$ & SR$\uparrow$ & MDD\%$\downarrow$
& ARR\%$\uparrow$ & SR$\uparrow$ & MDD\%$\downarrow$
& ARR\%$\uparrow$ & SR$\uparrow$ & MDD\%$\downarrow$ 
& (ARR) \\ 
\midrule

Market & B\&H
& 27.95 & 0.61 & 22.99
& \third{28.51} & 0.63 & 14.64
& 106.38 & 1.98 & 7.98
& \third{70.52} & 1.76 & 8.03
& \second{131.28} & 1.32 & 21.54
& \third{43.76} & 1.27 & 11.70 
& 3.50 \\ 
\midrule

\multirow{3}{*}{Rule-based} & MACD
& -22.48 & -0.73 & 17.46
& -33.43 & -0.94 & 19.25
& 30.80 & 0.94 & 12.06
& 13.61 & 0.67 & 6.79
& 51.48 & 0.94 & 17.29
& 29.92 & \third{1.40} & 8.29 
& 13.67 \\ 
 & ZMR
& 15.79 & \third{0.96} & \best{4.06}
& -7.08 & -0.97 & \best{4.16}
& 7.99 & 1.26 & \best{1.03}
& 3.76 & 0.50 & \best{2.10}
& 20.84 & 0.60 & \third{14.26}
& 14.46 & 1.10 & \best{3.54} 
& 14.33 \\ 
 & SMA
& 17.70 & 0.86 & \second{5.42}
& -8.93 & -0.49 & \third{7.36}
& 75.42 & \second{2.15} & 8.94
& 3.50 & 0.67 & \best{2.10}
& 25.95 & 0.70 & 18.68
& -8.26 & -0.51 & 13.91 
& 13.17 \\ 
\midrule

\multirow{2}{*}{ML/DL} & LSTM
& -7.46 & -0.02 & 22.99
& 2.26 & 0.16 & 14.64
& 60.88 & 1.52 & \third{7.74}
& 58.26 & 1.55 & 8.03
& 19.13 & 0.42 & 21.53
& 26.09 & 1.00 & 9.07 
& 11.17 \\ 
 & Transformer
& 32.90 & 0.72 & 13.77
& 12.60 & 0.38 & 14.64
& \third{107.85} & \third{2.12} & 8.32
& 45.64 & 1.35 & 8.03
& 77.88 & 1.09 & 16.99
& 26.35 & 1.18 & \third{8.26} 
& 7.00 \\ 
\midrule

\multirow{2}{*}{RL} & DQN
& 32.07 & \second{1.06} & 12.18
& 19.14 & \third{0.72} & 9.84
& -11.54 & -0.47 & 13.77
& 13.91 & 0.81 & 9.70
& 40.02 & 0.82 & 19.70
& 10.05 & 0.57 & \second{6.97} 
& 11.83 \\ 
 & PPO
& 27.20 & 0.68 & 9.48
& \second{29.03} & 0.70 & 12.93
& 43.73 & 1.14 & 7.98
& 38.54 & 1.51 & 7.55
& 34.68 & 0.80 & 20.10
& 24.24 & 0.89 & 11.59 
& 9.33 \\ 
\midrule

\multirow{4}{*}{General LLMs} & Qwen3-8B
& 7.34 & 0.27 & 18.95
& 5.12 & 0.23 & 19.52
& 49.70 & 1.43 & 10.97
& 21.19 & 0.80 & 9.28
& 87.48 & 1.20 & 23.66
& 32.93 & 1.10 & 12.03 
& 10.00 \\ 
 & DeepSeek-R1-0528
& 4.02 & 0.20 & 13.42
& 16.80 & 0.51 & 19.41
& 68.40 & 1.67 & 9.90
& 20.81 & 0.97 & 7.52
& 78.83 & 1.15 & 24.13
& 17.77 & 0.71 & 10.58 
& 10.00 \\ 
 & Llama4-Scout-17B
& 24.23 & 0.58 & 18.95
& -20.46 & -0.39 & 22.50
& 79.69 & 1.87 & 9.96
& 31.90 & 1.19 & 9.73
& 103.09 & 1.34 & 21.24
& 35.42 & 1.15 & 13.98 
& 8.33 \\ 
 & GPT5-mini
& -18.88 & -0.42 & 25.83
& -5.27 & -0.06 & 18.01
& 23.85 & 0.74 & 9.59
& 39.21 & 1.63 & 7.55
& 105.17 & \third{1.55} & 19.44
& 16.08 & 0.71 & 12.39 
& 11.50\\ 
\midrule

\multirow{4}{*}{Financial LLMs} & FinGPT
& -7.85 & -0.06 & 26.25
& 27.78 & \second{0.86} & 8.53
& 65.99 & 1.40 & 9.26
& 26.52 & 0.98 & 10.03
& 94.35 & 1.20 & 21.54
& 35.46 & 1.20 & 11.28 
& 8.17 \\ 
 & FinAgent
& \third{33.77} & 0.90 & 13.16
& 24.76 & 0.57 & 14.63
& \second{108.16} & 2.00 & 7.98
& 69.87 & \best{2.19} & 8.02
& \third{116.87} & 1.24 & 24.43
& \second{50.07} & \second{1.47} & 14.82 
& 3.17 \\ 
 & TradingAgents
& -8.58 & -0.09 & 26.25
& 23.72 & 0.66 & 15.31
& 66.85 & 1.57 & 9.76
& 48.29 & 1.59 & \second{5.31}
& 51.80 & 0.93 & 22.64
& 38.96 & 1.37 & 8.74 
& 8.17 \\ 
 & DeepFund
& \second{33.83} & 0.74 & 15.86
& 14.41 & 0.43 & 16.70
& 54.13 & 1.40 & 12.86
& \second{80.60} & \third{2.07} & \third{5.94}
& 51.18 & \second{1.72} & \best{6.52}
& 8.05 & 0.38 & 16.11 
& 8.67 \\ 
\midrule

\textbf{Ours} & \textbf{\system}
& \best{50.08} & \best{1.22} & \third{7.83}
& \best{40.87} & \best{1.18} & \second{5.57}
& \best{120.48} & \best{2.55} & \second{7.01}
& \best{84.14} & \second{2.08} & 8.03
& \best{148.41} & \best{1.94} & \second{12.60}
& \best{53.57} & \best{1.52} & 9.79 
& \textbf{1.00} \\ 

\midrule
\multicolumn{2}{c}{Improvement(\%)}
& 48.03 & 15.09 & --
& 40.79 & 37.21 & --
& 11.39 & 18.60 & --
& 4.39 & -- & --
& 13.05 & 12.79 & --
& 6.99 & 3.40 & -- 
& -- \\ 

\bottomrule
\end{tabular}
}
\vspace{-2mm}
\end{table}

%% file: tables/ablation_study.tex
\begin{wraptable}{r}{0.50\linewidth}
    \centering
    \scriptsize
    \setlength{\tabcolsep}{2pt}
    \renewcommand{\arraystretch}{1.02}

    \caption{Ablation study: $M_F$, $M_P$, $M_R$ denote the adaptive fusion module, modality prior, robustness regularizer, respectively. \textbf{Bold} denotes best results.}
    \label{tab:fusion_ablation_aapl_btc}

    \resizebox{\linewidth}{!}{
    \begin{tabular}{lccc ccc}
        \toprule
        \multirow{2}{*}{\textbf{Method}}
        & \multicolumn{3}{c}{\textbf{Modules}}
        & \multicolumn{3}{c}{\textbf{Metrics}} \\
        \cmidrule(lr){2-4}
        \cmidrule(lr){5-7}
        & \textbf{$M_F$}
        & \textbf{$M_P$}
        & \textbf{$M_R$}
        & ARR(\%) $\uparrow$
        & SR $\uparrow$
        & MDD(\%) $\downarrow$ \\
        \midrule

        \multicolumn{7}{c}{\textbf{AAPL}} \\
        \midrule
        Concat. Fusion
            & \xmark & \xmark & \xmark
            & 19.21 & 0.33 & 22.99 \\
        \hdashline

        \rowcolor{bestrow}
        \textbf{\system}
            & \cmark & \cmark & \cmark
            & \textbf{50.08} & \textbf{1.22} & \textbf{7.83} \\

        \rowcolor{bestrow}
        \quad w/o $M_P,M_R$
            & \cmark & \xmark & \xmark
            & 27.28 & 0.57 & 22.99 \\

        \rowcolor{bestrow}
        \quad w/o $M_R$
            & \cmark & \cmark & \xmark
            & 34.77 & 0.86 & 8.46 \\

        \rowcolor{bestrow}
        \quad w/o $M_P$
            & \cmark & \xmark & \cmark
            & 38.04 & 0.80 & 10.69 \\

        \midrule

        \multicolumn{7}{c}{\textbf{BTCUSD}} \\
        \midrule
        Concat. Fusion
            & \xmark & \xmark & \xmark
            & 7.18 & 0.40 & 10.76 \\
        \hdashline

        \rowcolor{bestrow}
        \textbf{\system}
            & \cmark & \cmark & \cmark
            & \textbf{53.57} & \textbf{1.52} & \textbf{9.79} \\

        \rowcolor{bestrow}
        \quad w/o $M_P,M_R$
            & \cmark & \xmark & \xmark
            & 15.32 & 0.72 & 16.28 \\

        \rowcolor{bestrow}
        \quad w/o $M_R$
            & \cmark & \cmark & \xmark
            & 32.45 & 1.20 & \textbf{9.79} \\

        \rowcolor{bestrow}
        \quad w/o $M_P$
            & \cmark & \xmark & \cmark
            & 46.22 & 1.16 & 11.70 \\

        \bottomrule
    \end{tabular}
    }
    \vspace{-5mm}
\end{wraptable}

%% file: tables/robustness_analysis.tex
\begin{wraptable}{r}{0.60\linewidth}
\vspace{-2mm}
\centering
\scriptsize
\setlength{\tabcolsep}{0.5pt}
\renewcommand{\arraystretch}{0.92}

\caption{Robustness analysis of \system{} under different modality configurations. Additional random seed variance and significance analyses are provided in Appendix~\ref{app:random_seed}.}
\label{tab:ablation_combined}

\resizebox{\linewidth}{!}{
\begin{tabular}{@{}cccc@{\hspace{2pt}}ccc@{\hspace{2pt}}ccc@{\hspace{2pt}}ccc@{}}
\toprule
\multicolumn{4}{c}{\textbf{Modalities}}
& \multicolumn{3}{c}{\textbf{Llama4-Scout}}
& \multicolumn{3}{c}{\textbf{FinAgent}}
& \multicolumn{3}{c}{\textbf{\system{}}} \\
\cmidrule(lr){1-4}
\cmidrule(lr){5-7}
\cmidrule(lr){8-10}
\cmidrule(lr){11-13}
N & M & T & S
& ARR(\%)$\uparrow$ & SR$\uparrow$ & MDD(\%)$\downarrow$
& ARR(\%)$\uparrow$ & SR$\uparrow$ & MDD(\%)$\downarrow$
& ARR(\%)$\uparrow$ & SR$\uparrow$ & MDD(\%)$\downarrow$ \\
\midrule

\multicolumn{13}{c}{\textbf{AAPL}} \\
\midrule

\cmark & \xmark & \xmark & \xmark
& 18.78 & 0.55 & 15.73
& 22.36 & 0.59 & 19.43
& 26.61 & 0.60 & 22.99 \\

\cmark & \cmark & \xmark & \xmark
& 36.31 & 1.00 & 13.10
& --    & --   & --
& 47.29 & 1.01 & 12.58 \\

\cmark & \xmark & \cmark & \xmark
& 38.34 & 0.82 & 18.95
& --    & --   & --
& 47.18 & \textbf{1.22} & \textbf{5.09} \\

\cmark & \xmark & \xmark & \cmark
& --    & --   & --
& 19.12 & 0.50 & 22.25
& 39.42 & 0.98 & 8.00 \\

\cmark & \cmark & \cmark & \xmark
& 24.23 & 0.58 & 18.95
& 26.23 & 0.58 & 22.99
& 33.97 & 0.70 & 22.98 \\

\cmark & \cmark & \xmark & \cmark
& --    & --   & --
& --    & --   & --
& 24.03 & 0.59 & 18.95 \\

\cmark & \xmark & \cmark & \cmark
& --    & --   & --
& --    & --   & --
& 28.36 & 0.61 & 22.99 \\

\cmark & \cmark & \cmark & \cmark
& --    & --   & --
& 33.77 & 0.90 & 13.16
& \textbf{50.08} & \textbf{1.22} & 7.83 \\

\midrule

\multicolumn{13}{c}{\textbf{BTCUSD}} \\
\midrule

\cmark & \xmark & \xmark & \xmark
& 40.20 & 1.20 & 10.23
& 27.85 & 0.82 & 21.39
& 41.68 & 1.27 & 9.79 \\

\cmark & \cmark & \xmark & \xmark
& 20.63 & 0.76 & 11.83
& --    & --   & --
& 28.63 & 1.08 & 11.88 \\

\cmark & \xmark & \cmark & \xmark
& 18.00 & 0.68 & 17.65
& --    & --   & --
& 22.00 & 1.04 & \textbf{8.48} \\

\cmark & \xmark & \xmark & \cmark
& --    & --   & --
& 51.98 & 1.50 & 14.25
& 43.57 & \textbf{1.71} & 8.67 \\

\cmark & \cmark & \cmark & \xmark
& 35.42 & 1.15 & 13.98
& 19.43 & 0.72 & 13.40
& 47.10 & 1.58 & 11.27 \\

\cmark & \cmark & \xmark & \cmark
& --    & --   & --
& --    & --   & --
& 40.92 & 1.20 & 11.91 \\

\cmark & \xmark & \cmark & \cmark
& --    & --   & --
& --    & --   & --
& 48.56 & 1.57 & 10.52 \\

\cmark & \cmark & \cmark & \cmark
& --    & --   & --
& 50.07 & 1.47 & 14.82
& \textbf{53.57} & 1.52 & 9.79 \\

\bottomrule
\end{tabular}
}
\vspace{-4mm}
\end{wraptable}

%% file: tex/7_conclusion.tex
\section{Conclusion}
\label{conclu}
\system{} is a multimodal agentic paradigm for financial trading that improves cross-modal understanding and robustness under noisy markets. \system{} organizes a hierarchy of specialized agents to extract modality-specific signals from heterogeneous sources, which provides a holistic view of market dynamics. It further introduces a modality-aware adaptive fusion mechanism to capture fine-grained inter-modality dependencies and produce coherent trading representations. In addition, noise-robust consistency regularization improves stability under corrupted or shifting modalities, enhancing reliability in information-rich environments. Future work will extend \system{} to portfolio management for multimodal, risk-aware allocation under realistic trading frictions.

\section*{Acknowledgements}
\label{ack}
We would like to express our gratitude to  Yumo Yao and Yiqiu Liu for their contributions to the initial design and implementation of the trading system.
We also thank the anonymous reviewers for their valuable feedback and constructive comments that helped improve this manuscript.

%% file: tex/8_appendix_catalog.tex
{\bf \large Appendix}
\begin{itemize}
  \item \textbf{\ref{app:related_work}. Related Work} \dotfill \pageref{app:related_work}

  \item \textbf{\ref{app:notations}. Notations} \dotfill \pageref{app:notations}

  \item \textbf{\ref{app:details_problem_ormulation}. Details of Problem Formulation} \dotfill \pageref{app:details_problem_ormulation}

  \item \textbf{\ref{app:agents_details}. Details of Multiple Agents in \system{}} \dotfill \pageref{app:agents_details}

  \item \textbf{\ref{app:system_workflow}. Details of \system{} Design} \dotfill \pageref{app:system_workflow}
  \begin{itemize}
      \item[$\cdot$] \textit{\ref{app:news}. News Summarizer} \dotfill \pageref{app:system_workflow}
      \item[$\cdot$] \textit{\ref{app:modality_processing}. Modality Processing Module} \dotfill \pageref{app:system_workflow}
      \item[$\cdot$] \textit{\ref{app:backtest}. Backtesting Engine} \dotfill \pageref{app:system_workflow}
  \end{itemize}

  \item 
  \textbf{\ref{app:experiments_baselines}. Details of Comparison with Baselines} \dotfill \pageref{app:imple_details}
  \begin{itemize}
      \item[$\cdot$] \textit{\ref{app:imple_details}. Implementation Details} \dotfill \pageref{app:imple_details}
      \item[$\cdot$] \textit{\ref{app:modality_def}. Modality Definitions} \dotfill \pageref{app:modality_def}
      \item[$\cdot$] \textit{\ref{app:formula_metrics}. Formulations of Used Metrics} \dotfill \pageref{app:formula_metrics}
      \item[$\cdot$] \textit{\ref{app:baselines}. Details of Baselines} \dotfill \pageref{app:baselines}
      \item[$\cdot$] \textit{\ref{app:full_results}. Full Experimental Results of Stocks Mentioned in Paper} \dotfill \pageref{app:baselines}
      \item[$\cdot$] \textit{\ref{app:add_stocks}. Experimental Results of Broader Asset Pool and Extended Test Horizons} \dotfill \pageref{app:baselines}
  \end{itemize}

  \item \textbf{\ref{app:random_seed}. Details of Random-seed Robustness and Statistical Significance}

  \item \textbf{\ref{app:case_study}. Details of Case Study} \dotfill \pageref{app:case_study}

  \item \textbf{\ref{app:details_Stock_Movement}. Details of Stock Movement Predictions Enhancement} \dotfill \pageref{app:details_Stock_Movement}



\end{itemize}

%% file: tex/8_appendix_full.tex
\input{tex/6_related_work}

\section{Notations}
\label{app:notations}
To improve clarity and maintain consistent notation throughout the paper, we provide a notation table as a quick reference (see Table~\ref{tab:notation}) for the definitions of all symbols used.
\begin{table}[htbp]
\centering
\caption{Notation table for the multimodal fusion module.}
\label{tab:notation}
\begin{tabular}{p{0.28\linewidth} p{0.67\linewidth}}
\toprule
\textbf{Symbol} & \textbf{Description} \\
\midrule
$M$ & Number of modalities (and modality-specific agents). \\[2pt]
$m,n$ & Modality indices, $m,n\in\{1,\dots,M\}$. \\[2pt]
$x=\{X^{(m)}\}_{m=1}^{M}$ & Multimodal input; $X^{(m)}$ is the input sequence of modality $m$. \\[2pt]
$X^{(m)}$ & Modality-$m$ input sequence (e.g., over a window of trading days or tokens). \\[2pt]
$f_m(\cdot)$ & Modality-specific agent/encoder for modality $m$. \\[2pt]
$T_m$ & Sequence length for agent $m$ (time steps or tokens). \\[2pt]
$\tilde d_m$ & Hidden dimension of agent $m$ before projection. \\[2pt]
$H_m(x)\in\mathbb{R}^{T_m\times \tilde d_m}$ & Sequence representation produced by agent $m$. \\[2pt]
$\phi_m(\cdot)$ & Pooling operator that maps $H_m(x)$ to a fixed-length summary (last-token for LLMs; \texttt{[CLS]} for Transformer encoders). \\[2pt]
$h_m(x)\in\mathbb{R}^{\tilde d_m}$ & Fixed-length summary of modality $m$ after pooling. \\[2pt]
$d$ & Shared embedding dimension after projection/alignment. \\[2pt]
$P_m\in\mathbb{R}^{d\times \tilde d_m}$ & Learnable projection mapping $h_m(x)$ to the shared $d$-dimensional space. \\[2pt]
$e_m\in\mathbb{R}^{d}$ & Aligned modality embedding used for fusion. \\[2pt]
$Q_m,K_m,V_m$ & Modality-specific query, key, and value representations derived from $e_m$. \\[2pt]
$\tilde K_m,\tilde V_m$ & Key/value after modality bias injection. \\[2pt]
$\alpha_m,\alpha'_m$ & Modality attention weights for $x$ and perturbed input $x'$, respectively. \\[2pt]
$R_{\mathrm{final}}$ & Fused representation produced by modality attention. \\[2pt]
$r_m$ & Learnable modality-specific bias vector injected into key/value. \\[2pt]
$A$ & Shared modality space used to parameterize $\{r_m\}$. \\[2pt]
$u_m$ & Projection direction (modality-specific) used to generate $r_m=u_mA$. \\[2pt]
$s(x)\in\mathbb{R}^{C}$ & Predicted logits from the classification head for input $x$. \\[2pt]
$C$ & Number of classes (logit dimension). \\[2pt]
$\hat p(x)=\mathrm{softmax}(s(x))$ & Predicted class probability vector. \\[2pt]
$y\in\{0,1\}^{C}$ & One-hot ground-truth label. \\[2pt]
$L_{\mathrm{CE}}$ & Cross-entropy loss for supervised learning. \\[2pt]
$L_{\mathrm{modality}}$ & Modality regularizer encouraging diversity of $\{r_m\}$ and preventing trivial scaling. \\[2pt]
$m^\star$ & Index of the perturbed modality in robust training. \\[2pt]
$x'$ & Perturbed input obtained by corrupting/modifying modality $m^\star$. \\[2pt]
$L_{\mathrm{robust}}$ & Robust training loss (logit consistency + attention suppression). \\[2pt]
$\gamma$ & Weight for the attention suppression term in $L_{\mathrm{robust}}$. \\[2pt]
$\lambda_{\mathrm{mod}},\lambda_{\mathrm{rob}}$ & Loss weights for $L_{\mathrm{modality}}$ and $L_{\mathrm{robust}}$ in the overall objective. \\[2pt]
$\|\cdot\|_2$ & $\ell_2$ norm; $\|v\|_2^2=\sum_i v_i^2$. \\
\bottomrule
\end{tabular}
\end{table}

\section{Details of Problem Formulation}
\label{app:details_problem_ormulation}
In addition to numerical modalities, we incorporate textual news. 
For stock $s$ at date $t$, the summarized news set is 
$N_t^s = \{n_{t,1}^s,\ldots,n_{t,|N_t^s|}^s\}$, 
with $|N_t^s|$ the number of daily news items. 
The model outputs a binary prediction $\hat{y}_t^s \in \{0,1\}$ together with a human-readable explanation $\hat{e}_t^s$. 
The parameters $\theta$ are trained by empirical risk minimization with a task-appropriate loss such as cross-entropy:
$\min_\theta \; \mathbb{E}_{(X_t^s,Y_t^s)} \big[ \ell(F_\theta(f(X_t^s)),\, Y_t^s)\big]$.

\section{Details of Multiple Agents in \system{}}
\label{app:agents_details}
\label{app:encoders}
\highlight{Market Analysis Agent.} The Market Analysis Agent is tasked with encoding raw OHLCV (Open, High, Low, Close, Volume) sequences into high-level latent representations. We treat this data as a multivariate time series characterized by non-stationary volatility, where predictive cues often emerge from the non-linear coupling between price movements and trading volume, known as volume-price divergence~\cite{cont2001empirical}. The core advantage lies in its self-attention mechanism, which provides a global receptive field to capture long-range temporal dependencies and effectively model the intricate lead-lag relationships within the lookback window~\cite{zerveas2021transformer}. This allows the agent to dynamically prioritize critical market events regardless of their temporal distance, generating robust embeddings optimized for downstream multimodal fusion. For a stock $s$ at predictions date $t$, we construct a $T$-day window $\mathbf{X}^{\mathrm{MA}}_{s,t}=[x^{\mathrm{MA}}_{s,t-T},\dots,x^{\mathrm{MA}}_{s,t-1}]\in\mathbb{R}^{T\times F_{\mathrm{MA}}}$, where contains $\{open, high, low, adj\_close,volume\}$ fields. Here $T$ is the window length and $F_{\mathrm{MA}}$ is the market feature dimension.

\noindent
\underline{Transformer-based Market Encoder.} Let $d_{\mathrm{MA}}$ denote the model dimension. We first project the input features into a hidden sequence with positional encoding, $\mathbf{H}^{(0)}_{\mathrm{MA}}=\mathrm{PE}(\mathbf{X}^{\mathrm{MA}}_{s,t})\,\mathbf{W}^{E}_{\mathrm{MA}}\in\mathbb{R}^{T\times d_{\mathrm{MA}}}$, where $\mathbf{W}^{E}_{\mathrm{MA}}\in\mathbb{R}^{F_{\mathrm{MA}}\times d_{\mathrm{MA}}}$ is learnable. We then apply $L_{\mathrm{MA}}$ causal Transformer blocks with a causal mask $\mathbf{M}_{\mathrm{causal}}$ to prevent information leakage:
\vspace{-2mm}
\begin{equation}
\small
\label{eq:ma_block}
\mathbf{H}_{\mathrm{MA}}^{(\ell)}
=
\mathrm{FFN}\!\Big(
\mathrm{MHA}\big(\mathbf{H}_{\mathrm{MA}}^{(\ell-1)};\mathbf{M}_{\mathrm{causal}}\big)
\Big).
\end{equation}
We expose the final hidden state sequence
$\mathbf{H}^{\mathrm{MA}}_{s,t}=\mathbf{H}^{(L_{\mathrm{MA}})}_{\mathrm{MA}}\in\mathbb{R}^{T\times d_{\mathrm{MA}}}$
as the market modality representation, $\ell=1,\dots,L_{\mathrm{MA}}$.
Following the implementation, we use the \texttt{[CLS]} representation as a fixed-length summary $h^{\mathrm{MA}}_{s,t}\in\mathbb{R}^{d_{\mathrm{MA}}}$ when a vector representation is required.
The subsequent projection into the shared fusion space is performed by the fusion module (Eq.~\ref{eq:embed_extract_align}).


\highlight{Technical Analysis Agent.}
The technical analysis agent focuses on expression-based alpha factors~\cite{kakushadze2016101}. Compared with purely statistical features, such indicators provide structured quantitative descriptions of trend, momentum, and volatility, which are complementary to news and sentiment signals in multimodal trading. For a stock $s$ at prediction date $t$, we assemble a window of technical features from the past $T$ trading days, $\mathbf{X}^{\mathrm{TA}}_{s,t}=[x^{\mathrm{TA}}_{s,t-T},\dots,x^{\mathrm{TA}}_{s,t-1}]\in\mathbb{R}^{T\times F_{\mathrm{TA}}}$.
Each $x^{\mathrm{TA}}_{s,\tau}\in\mathbb{R}^{F_{\mathrm{TA}}}$ contains indicators such as MACD, RSI, and other engineered alpha factors.

\noindent
\underline{Transformer-based Technical Encoder.} To capture non-linear interactions among technical indicators and their temporal dependencies within the window, we adopt a Transformer encoder with model dimension $d_{\mathrm{TA}}$.
We first embed the input as $\mathbf{H}^{(0)}=\mathrm{PE}(\mathbf{X}^{\mathrm{TA}}_{s,t})\,\mathbf{W}^{E}_{\mathrm{TA}}\in\mathbb{R}^{T\times d_{\mathrm{TA}}}$, where $\mathbf{W}^{E}_{\mathrm{TA}}\in\mathbb{R}^{F_{\mathrm{TA}}\times d_{\mathrm{TA}}}$ is learnable and $\mathrm{PE}(\cdot)$ denotes positional encoding. The encoder stacks $L$ blocks of causal multi-head self-attention followed by a position-wise feed-forward network:
\vspace{-2mm}
\begin{equation}
\small
\label{eq:ta_block}
\mathbf{H}_{\mathrm{TA}}^{(\ell)}
=
\mathrm{FFN}\!\Big(
\mathrm{MHA}\big(\mathbf{H}_{\mathrm{TA}}^{(\ell-1)};\mathbf{M}_{\mathrm{causal}}\big)
\Big),
\end{equation}
where $\mathbf{M}_{\mathrm{causal}}$ masks future positions to information leakage, $\ell=1,\dots,L_{\mathrm{TA}}$. For attention head $h\in\{1,\dots,H\}$ in layer $\ell$, we compute $\mathbf{Q}^{(\ell,h)}=\mathbf{H}^{(\ell-1)}\mathbf{W}^{Q}_{\ell,h}$, $\mathbf{K}^{(\ell,h)}=\mathbf{H}^{(\ell-1)}\mathbf{W}^{K}_{\ell,h}$, and $\mathbf{V}^{(\ell,h)}=\mathbf{H}^{(\ell-1)}\mathbf{W}^{V}_{\ell,h}$, and apply masked attention:
\vspace{-2mm}
\begin{equation}
\small
\label{eq:ta_attn}
\mathrm{Attn}^{(\ell,h)}
=\mathrm{softmax}\!\Big(\frac{\mathbf{Q}^{(\ell,h)}(\mathbf{K}^{(\ell,h)})^\top}{\sqrt{d_k}}+\mathbf{M}_{\mathrm{causal}}\Big)\mathbf{V}^{(\ell,h)},
\end{equation}
where $d_k$ is the per-head key dimension. The multi-head outputs are concatenated and linearly projected following the standard Transformer formulation. We follow the same fusion interface as market analysis agent, exposing $\mathbf{H}^{\mathrm{TA}}_{s,t}$ and the optional \texttt{[CLS]} summary $h^{\mathrm{TA}}_{s,t}$, then alignment is performed by Eq.~\ref{eq:embed_extract_align}.

\label{app:sentiment_module}
\highlight{Sentiment Analysis Agent.} The sentiment analysis agent processes financial news and social media signals to capture sentiment information that may influence market dynamics. We adopt DeepSeek-R1~\cite{huang2025explainable} as the backbone model.

\noindent
\underline{Tool-augmented News Retrieval.}
Given a stock ticker $s$ and date $t$, the agent can autonomously retrieve raw news articles on demand by calling external news download APIs (e.g., Alpaca News API and Alpha Vantage News API), conditioned on $(s,t)$.
The retrieved articles are deduplicated and summarized by the News Summarizer (integrated in sentiment analysis agent as a tool) (Fig.~\ref{fig:news_summarizer}), yielding $K_{s,t}$ summarized items.
We denote the resulting sentiment input as
$\mathbf{X}^{\mathrm{SA}}_{s,t}=\{n_{s,t,1},\dots,n_{s,t,K_{s,t}}\}$.

\noindent
\underline{Per-item Sentiment Inference.}
For each summary $n_{s,t,j}$, the agent predicts a binary sentiment label and a brief rationale.
We denote the label by $\hat c_{s,t,j}\in\{0,1\}$ (1 for positive and 0 for negative) and the rationale by $\hat q_{s,t,j}$:
\vspace{-2mm}
\begin{equation}
\label{eq:sa_pred}
(\hat c_{s,t,j},\hat q_{s,t,j}) = g_{\mathrm{SA}}(n_{s,t,j}),
\end{equation}
where $g_{\mathrm{SA}}(\cdot)$ is instantiated by DeepSeek-R1, $j=1,\dots,K_{s,t}$.

\underline{Hidden Representation for Fusion.}
Beyond discrete labels, we extract a continuous representation from the last-layer hidden states of the LLM.
Let $\mathbf{H}^{\mathrm{SA}}_{s,t,j}\in\mathbb{R}^{L_{s,t,j}\times d_{\mathrm{SA}}}$ denote the last-layer hidden states produced when encoding $n_{s,t,j}$, where $L_{s,t,j}$ is the token length and $d_{\mathrm{SA}}$ is the hidden dimension.
Using last-token pooling, we obtain $u_{s,t,j}=\mathbf{H}^{\mathrm{SA}}_{s,t,j}(L_{s,t,j})\in\mathbb{R}^{d_{\mathrm{SA}}}$ and aggregate across items to form a fixed-length summary:
\vspace{-2.7mm}
\begin{equation}
\small
\label{eq:sa_agg}
h^{\mathrm{SA}}_{s,t}=\frac{1}{K_{s,t}}\sum_{j=1}^{K_{s,t}} u_{s,t,j}\in\mathbb{R}^{d_{\mathrm{SA}}}.
\end{equation}
The fusion module then applies its projection step (Eq.~\ref{eq:embed_extract_align}) to map $h^{\mathrm{SA}}_{s,t}$ into the shared fusion space.

\label{sec:news_analysis}
\label{app:news_analysis}
\highlight{News Analysis Agent.} As illustrated in Fig.~\ref{fig:fagents_framework}, the news analysis agent serves two objectives. First, it retrieves and consolidates news evidence relevant to a queried stock-date pair $(s,t)$. Second, it produces transparent, human-readable explanations to support downstream prediction. Given a ticker $s$ and date $t$, the agent retrieves raw articles from external news APIs, including the Alpaca News API (financial news with ticker annotations) and the New York Times API (general news).
The retrieved articles are deduplicated and condensed by the News Summarizer (Fig.~\ref{fig:news_summarizer}), yielding a set of top-$K_{s,t}$ summarized items,
$\mathbf{X}^{\mathrm{NA}}_{s,t}=\{n_{s,t,1},\dots,n_{s,t,K_{s,t}}\}$,
which serves as the textual input to the news analysis agent.
\begin{figure}[H]
    \centering
    \includegraphics[width=\textwidth]{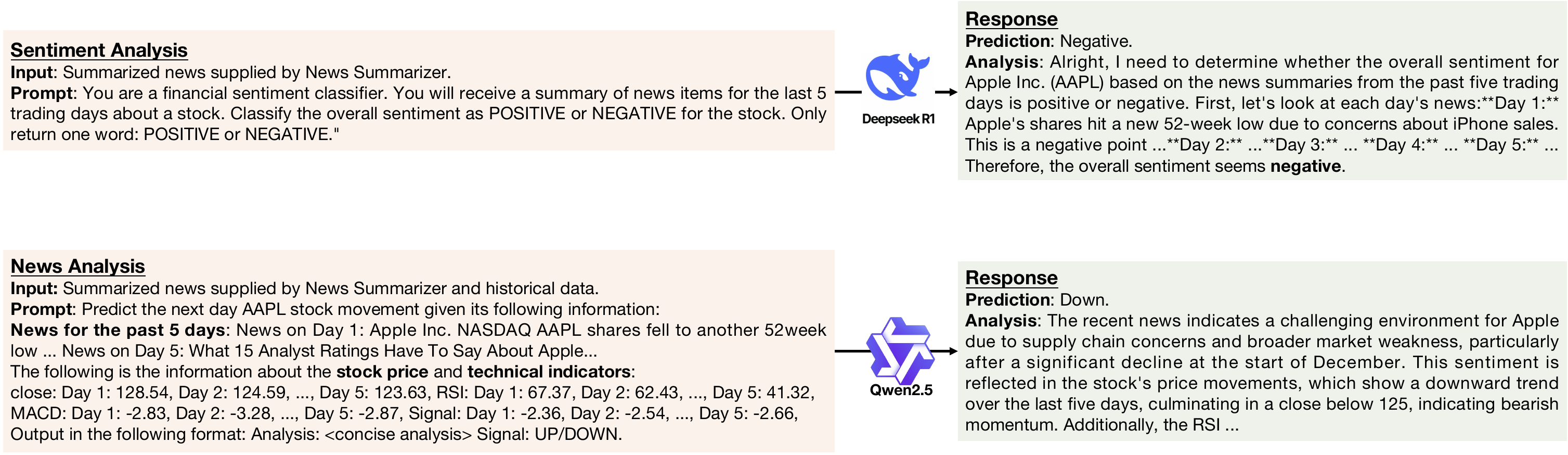}
    \caption{Case studies of our stock prediction system, we only display partial prompt for brevity.}
    \label{fig:case_study}
\end{figure}

\noindent
\underline{Explanation Prompting.}
Conditioned on $(s,t)$ and the summarized news set $\mathbf{X}^{\mathrm{NA}}_{s,t}$, the news reasoning model $g_{\mathrm{NA}}(\cdot)$ generates (i) an auxiliary directional signal $\hat c^{\mathrm{NA}}_{s,t}\in\{\textsc{UP},\textsc{DOWN}\}$, (ii) a confidence score $\hat p^{\mathrm{NA}}_{s,t}\in[0,1]$, and (iii) a natural-language explanation $\hat e^{\mathrm{NA}}_{s,t}$:
\vspace{-1mm}
\begin{equation}
\small
\label{eq:na_reason}
(\hat c^{\mathrm{NA}}_{s,t},\hat p^{\mathrm{NA}}_{s,t},\hat e^{\mathrm{NA}}_{s,t})
=
g_{\mathrm{NA}}\!\big(s,t,\mathbf{X}^{\mathrm{NA}}_{s,t}\big),
\end{equation}
where $g_{\mathrm{NA}}(\cdot)$ is instantiated by the backbone LLM used in \system.

\noindent
\underline{Model Fine-tuning.}
Following previous works that utilizes Chain of Thought(CoT)~\cite{wei2022chain}, we use a similar two-step process to fine-tune a LLM (see Fig.~\ref{fig:system_training_pipeline}). During the training, we use the binary evaluations to choose the better response (see Fig.~\ref{fig:system_training_pipeline}). In the first step, we collect the QA pairs related to selected topics (e.g. “trading decisions under different cases”, “momentum strategy and mean-reverse strategy”, etc.), which are generated with GPT-4o-mini~\cite{hurst2024gpt}. In the second step, we collect the real application simulated QA pairs with GPT-4o-mini, which are designed to simulate real application of the stock prediction system. Specifically, we fine-tune the model by minimizing following combined loss:
\vspace{-2mm}
\begin{equation}
\label{eq:na_ft_loss}
\scriptsize
\mathcal{L}_{\mathrm{NA}}(\theta)
=-\mathbb{E}_{(\mathcal{P},\mathcal{Y},\mathbf{y})\sim\mathcal{D}_{\mathrm{NA}}}
\Big[
\log P_{\theta}(\mathcal{Y}\mid \mathcal{P})
+
\lambda \sum_{k=1}^{C} y_k \log P_{\theta}(k\mid \mathcal{P})
\Big],
\end{equation}
where $\mathcal{P}$ denotes the instruction prompt and $\mathcal{Y}$ is the target response.
The first term corresponds to the language modeling loss (${lm\_loss}$), i.e., the negative log-likelihood of generating $\mathcal{Y}$ conditioned on $\mathcal{P}$.
The second term corresponds to the auxiliary classification cross-entropy loss (${cls\_loss}$), where $\mathbf{y}\in\{0,1\}^{C}$ is the one-hot label (e.g., \textsc{UP}/\textsc{DOWN}) and $\lambda$ weights the classification objective. To mitigate majority-class bias, prompts require choosing a signal from $\{\textsc{UP},\textsc{DOWN}\}$~\cite{zhao2021calibrate}.
Our prompt format (see Table~\ref{tab:news_agent_prompt_example_part1},~\ref{tab:news_agent_prompt_example_part2},~\ref{tab:sentiment_agent_prompt_example}) and interactive trajectory (see Fig.~\ref{fig:case_study}) draw inspiration from FinMEM~\cite{yu2025finmem}, while the execution is performed in a single summarization-reasoning step.

Beyond the textual explanation, we extract a fixed-length representation from the last-layer hidden states of the LLM and expose it to the fusion module.
Let $\mathbf{H}^{\mathrm{NA}}_{s,t}\in\mathbb{R}^{L_{s,t}\times d_{\mathrm{NA}}}$ denote the last-layer hidden states when encoding the prompt constructed from $(s,t,\mathbf{X}^{\mathrm{NA}}_{s,t})$, where $L_{s,t}$ is the token length and $d_{\mathrm{NA}}$ is the hidden dimension.
We apply last-token pooling to obtain $h^{\mathrm{NA}}_{s,t}\in\mathbb{R}^{d_{\mathrm{NA}}}$, which corresponds to the agent summary $h_m(x)$ used in Eq.~\ref{eq:embed_extract_align} (with $m=\mathrm{NA}$).
The fusion module then applies its projection step in Eq.~\ref{eq:embed_extract_align} to map $h^{\mathrm{NA}}_{s,t}$ into the shared fusion space.

\input{tables/prompt_template}

\section{Details of \system{} Design}
\label{app:system_workflow}
\subsection{News Summarizer}
\label{app:news}

\begin{figure}[htbp]
    \centering
    \includegraphics[width=0.5\linewidth]{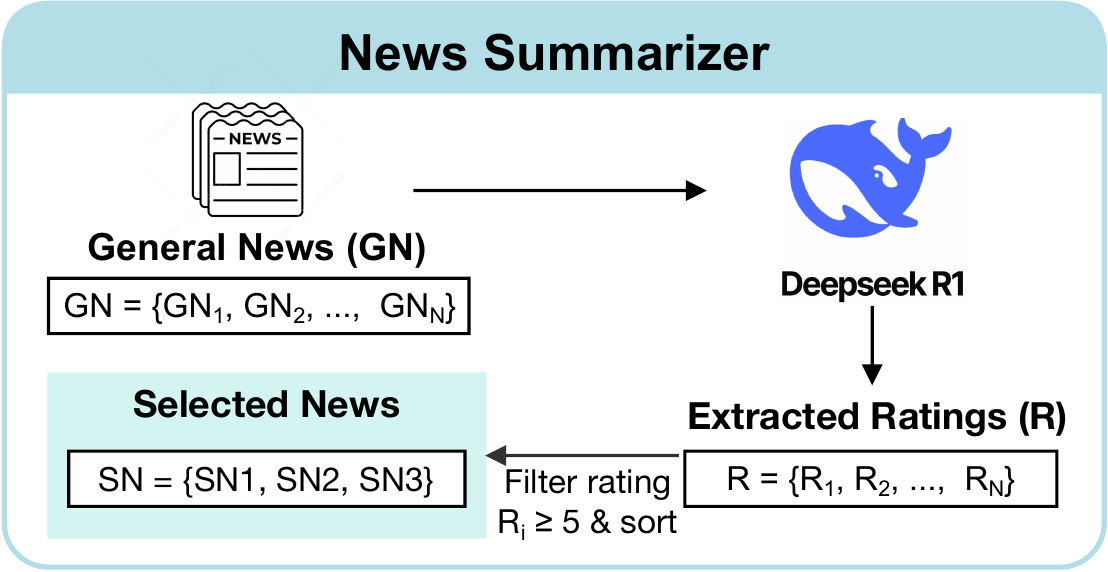}
    \caption{Illustration of News Summarizer.}
    \label{fig:news_summarizer}
\end{figure}

The News Summarizer, depicted in Fig.~\ref{fig:news_summarizer}, comprises three main components: a news extractor, a news evaluator, and a SQLite cache database. For news extraction, we utilize two distinct sources: the Alpaca News API, which provides financial news explicitly labeled by stock tickers, and the New York Times API, which offers broader news content without explicit ticker annotations. We directly query Alpaca News using stock symbols due to its inherent labeling. Conversely, for news retrieval from the New York Times API, we employ semantic vector search, matching embeddings of stock-related keywords against embeddings of news articles.
To enhance efficiency and minimize redundant API requests, the retrieved and summarized news data are stored in an SQLite cache database. As shown in Fig.~\ref{fig:fagents_framework}, when the system receives a query for a specific date and stock ticker, it first checks the cache database. If the corresponding news summary exists, it is retrieved directly from the cache; otherwise, relevant news articles are fetched through the APIs, summarized by the summarization model, and then saved into the cache database.

Regarding the summarization process, we perform evaluations on multiple advanced summarization models, including LLaMA-3~\cite{grattafiori2024llama} and DeepSeek-R1 \cite{rehman2025green, huang2025explainable}, to determine the most effective summarization approach. After retrieving all news items for a given day, we further prompt DeepSeek-R1 to rate each article according to its potential impact on the target stock. Based on these ratings, only the top three articles above a defined threshold are retained. The chosen summarization model then condenses these selected articles into concise, high-correlation summaries, ensuring that the final outputs capture the most influential news signals.
\subsection{Modality Processing Module}
\label{app:modality_processing}
Our proposed framework consists of four specialized agents designed to extract latent features from distinct data modalities: market data, technical indicators, news text, and sentiment signals. Each agent employs a specific architecture tailored to the nature of its input data.

\noindent
\highlight{Market Analysis Agent.} 
The market analysis agent encodes historical OHLCV observations to capture temporal dependencies in market dynamics.
For stock $s$ at date $t$, its input is a lookback window $\mathbf{X}^{\mathrm{MA}}_{s,t}\in\mathbb{R}^{T\times F_{\mathrm{MA}}}$ constructed from standard fields (Open, High, Low, Close, Volume) collected from data providers such as Yahoo Finance and Alpha Vantage APIs.
We employ a Transformer-based encoder to produce latent market representations, exposing the sequence hidden states $\mathbf{H}^{\mathrm{MA}}_{s,t}$ and a fixed-length summary $h^{\mathrm{MA}}_{s,t}$ for downstream fusion (cf. Eq.~\ref{eq:embed_extract_align}).

\noindent
\highlight{Technical Analysis Agent.}
The technical analysis agent encodes expression-based alpha factors derived from technical indicators.
For stock $s$ at date $t$, we construct a technical feature window $\mathbf{X}^{\mathrm{TA}}_{s,t}\in\mathbb{R}^{T\times F_{\mathrm{TA}}}$, which includes indicators such as MACD, SMA, and Z-score mean reversion (ZMR)~\cite{nti2020systematic}.
We employ a Transformer-based encoder to model the temporal structure of these indicators and their non-linear interactions, producing latent technical representations.
Following the same fusion interface, we expose the hidden states $\mathbf{H}^{\mathrm{TA}}_{s,t}$ and a fixed-length summary $h^{\mathrm{TA}}_{s,t}$ for downstream fusion (cf. Eq.~\ref{eq:embed_extract_align}).

\noindent
\highlight{News Analysis Agent.}
The news analysis agent extracts semantic evidence and trading-relevant implications from unstructured financial text retrieved from news providers (e.g., Bloomberg and Alpaca news APIs).
We leverage Qwen2.5-7B-Instruct~\cite{team2024qwen2,yang2025qwen3}, which is fine-tuned on financial corpora, to perform news reasoning conditioned on the queried stock-date context $(s,t)$.
Given a set of summarized news items $\mathbf{X}^{\mathrm{NA}}_{s,t}$, the agent generates an auxiliary trading signal (e.g., \textsc{UP}/\textsc{DOWN}) together with a natural-language reasoning chain.
In addition, we extract a fixed-length latent representation $h^{\mathrm{NA}}_{s,t}$ from the LLM hidden states, which is subsequently projected by the fusion module (Eq.~\ref{eq:embed_extract_align}) and used as the news modality input for downstream fusion.

\noindent
\highlight{Sentiment Analysis Agent.}
The Sentiment Analysis Agent assesses market sentiment by analyzing news and social media streams (e.g., FinHub).
Given a stock $s$ and date $t$, it constructs a set of summarized textual items $\mathbf{X}^{\mathrm{SA}}_{s,t}=\{n_{s,t,1},\dots,n_{s,t,K_{s,t}}\}$ and uses DeepSeek-R1~\cite{guo2025deepseek} to perform binary sentiment classification (positive/negative) for each item.
Beyond discrete polarity, the agent extracts a fixed-length latent representation $h^{\mathrm{SA}}_{s,t}$ from the LLM hidden states, which is subsequently projected by the fusion module (Eq.~\ref{eq:embed_extract_align}) and serves as a complementary signal to the news analysis component.

\subsection{Backtesting Engine}
\label{app:backtest}
The Backtesting Tool takes the predictions produced by \system{} and transforms them into bullish or bearish trading decisions. First, the predicted signals and actual market prices from earlier modules are loaded. Based on these inputs, customized trading strategies are applied to generate buy, hold, or sell actions. The resulting trading signals are then backtested on historical market data with specified stock tickers, simulating how the strategy would have performed in practice. Finally, the module outputs comprehensive evaluation results, including equity curves, risk–return plots, and standard performance metrics such as cumulative return, annualized return, Sharpe ratio, and maximum drawdown.

\highlight{HOLD signals in binary price prediction.}
The key distinction is between the binary price-direction label used for supervised learning and the downstream trading action in the backtesting. In our formulation, $\{\mathrm{UP}, \mathrm{DOWN}\}$ is the supervised prediction target defined directly from the adjusted closing price: following Eq.~\ref{eq:yt}, $Y_t=1$ if $p_t^c \geq p_{t-1}^c$, and $Y_t=0$ otherwise. 
\begin{wraptable}[9]{r}{0.5\linewidth}
\vspace{-2mm}
\centering
\footnotesize
\caption{Mapping from consecutive predicted signals to execution actions.}
\label{tab:signal_action}
\begin{tabularx}{\linewidth}{@{}ccX@{}}
\toprule
\makecell{\textbf{Signal} \textbf{at $t$}} &
\makecell{\textbf{Signal} \textbf{at $t+1$}} &
\textbf{Execution action} \\
\midrule
BUY  & BUY  & Hold the position \\
BUY  & SELL & Sell / liquidate \\
SELL & BUY  & Buy / enter position \\
SELL & SELL & Stay in cash \\
\bottomrule
\end{tabularx}
\vspace{2mm}
\end{wraptable}
Therefore, a \emph{no price change} case is grouped into the UP class by definition, and we do not introduce a separate HOLD class at the label level. The notion of HOLD appears only in the backtesting/execution stage, where consecutive predicted signals are translated into buy/sell/hold actions, as shown in Table~\ref{tab:signal_action}. Thus, HOLD is an execution-state outcome induced by consecutive predicted signals, rather than a third class in the binary price-direction prediction task. 

\highlight{Assumptions used in the backtesting setup.}
\noindent Details of the backtesting protocol are specified below. Our experiments use a daily, low-frequency, long-or-flat backtesting setting. A signal generated on day $t$ is executed on day $t+1$, which avoids look-ahead bias. Positions are implemented using integer shares, with
$\text{shares}=\left\lfloor \text{cash}/\text{price} \right\rfloor$,
and any remaining capital is kept as cash. When the signal changes from positive to non-positive, the position is fully liquidated. Thus, the setup is not long-short and does not involve short borrowing constraints. We apply a uniform transaction cost of $0.003$ to all methods. We do not impose an explicit turnover cap, nor explicitly model slippage or market impact, following common practice in prior work~\cite{zhang2024multimodal,yu2024fincon}. These factors are not the primary focus of this study, which targets prediction performance under a shared and consistent backtesting engine. To mitigate their potential impact, we evaluate at daily frequency on highly liquid assets, where such effects are typically reduced.

\section{Details of Comparison with Baselines}
\label{app:experiments_baselines}
\subsection{Implementation Details}
\label{app:imple_details}
To ensure reproducibility, all experiments are conducted on NVIDIA A40 GPUs using PyTorch and HuggingFace Transformers. We adopt a unified optimization strategy for gradient-based models (LSTM, Transformer, and \system{}'s fusion module) and RL agents, utilizing the Adam optimizer with a learning rate of $2 \times 10^{-3}$, a batch size of 64, and a 30-day look-back window, with training capped at 100 epochs under early stopping (patience=10). Specific hyperparameters include a discount factor of $\gamma=0.99$ for RL agents (with a 10,000-step buffer for DQN and $\epsilon=0.2$ for PPO) and a generation temperature of 0.5 with a repetition penalty of 1.1 for all LLM components. To strictly prevent look-ahead bias, data is chronologically split into training (2023-10-01--2024-09-30), validation (2024-10-01--2025-03-31), and testing (2025-04-01--2025-09-30) periods.

\subsection{Modality Definitions}
\label{app:modality_def}
Specifically, these modalities include: i) \textbf{Asset Price} at the day-level, covering open, high, low, adj close, and volume. ii) \textbf{Refined Alphas} refer to a compact collection of high-signal technical indicators identified through preliminary feature-selection analysis. iii) \textbf{Asset news} coverage with daily updates from various esteemed sources such as New York Times API and Alpaca API, ensuring a diverse and thorough perspective on the financial markets. iv) \textbf{Summarized Daily News} updated daily via the Alpaca News API by news summarizer, ensuring broad and in-depth coverage of financial market information \cite{yu2025finmem}. 

\subsection{Formulations of Used Metrics}
\label{app:formula_metrics}
\noindent
\highlight{Cumulative Return (CR)} measures the total value change of an investment over time by summing daily logarithmic returns, shown in Equation~\ref{eq:cum_return}. Higher values indicate better strategy effectiveness:
\vspace{-2mm}
\begin{align}
\small
       \label{eq:cum_return}
       \textbf{CR} &= \sum_{t=1}^{n} r_i = \sum_{t=1}^{n} \left[ \ln\left(\frac{p_{t+1}}{p_t}\right) \cdot \text{action}_t \right].
\end{align}

\noindent
\highlight{Annual Rate of Return (ARR)} normalizes the cumulative return over one year:
\vspace{-2mm}
\begin{equation}
\small
    \textbf{ARR} = \left( \left( \frac{V_{\text{end}}}{V_{\text{start}}} \right)^{\frac{C}{T}} - 1 \right) \times 100\%.
\end{equation}

\noindent
\highlight{Sharpe Ratio (SR)} assesses risk-adjusted returns by dividing the average excess return ($R_p$) over the risk-free rate ($R_f$) by its volatility ($\sigma_p$), detailed in Equation~\ref{eq:sharpe}. Higher ratios signify better performance.
\vspace{-1mm}
 \begin{equation}
 \small
        \textbf{SR} = \frac{R_p - R_f}{\sigma_p}.
        \label{eq:sharpe}
\end{equation}

\noindent
\highlight{Maximum Drawdown (MDD)} calculates the largest portfolio value drop from peak to trough, as given in Equation~\ref{eq:maxdrawdown}. Lower values indicate lesser risk and higher strategy robustness. 
\vspace{-2mm}
\begin{align}
\small
    \label{eq:maxdrawdown}
    \textbf{MDD} = \text{max}(\frac{P_{\text{peak}} - P_{\text{trough}}}{P_{\text{peak}}}).
\end{align}

\subsection{Details of Baselines}
\label{app:baselines}
We compare the trading performance of \system{} against four widely used rule-based strategies (\textbf{B\&H}, \textbf{MACD}, \textbf{ZMR}, \textbf{SMA}), two ML\&DL-based price prediction models, two canonical RL-based agents, four general-purpose LLMs, and four financial-domain LLM agents. 
The ML\&DL models comprise \textbf{LSTM} \cite{yang2020qlib} and \textbf{Transformer} \cite{yang2020qlib}. 
The RL-based agents include \textbf{DQN} \cite{mnih2013playing} and \textbf{PPO} \cite{schulman2017proximal}. 
The general LLM baselines are \textbf{Qwen3-8B}, \textbf{DeepSeek-R1-0528}, \textbf{Llama4-Scout-17B}, and \textbf{GPT-5-mini}. 
The financial LLM baselines include \textbf{FinGPT} \cite{liu2023fingpt}, \textbf{FinAgent}, \textbf{TradingAgents} \cite{xiao2024tradingagents}, and \textbf{DeepFund}. 
Below we briefly describe each baseline:

\noindent\textbf{Rule-based Methods.} We select representative traditional strategies:
\textbf{Buy-and-Hold (B\&H)} holds the asset throughout the evaluation horizon, ignoring short-term fluctuations.
\textbf{MACD} generates signals from MACD--signal line crossovers to capture trend momentum.
\textbf{Z-score Mean Reversion (ZMR)} assumes prices revert to a statistical mean, with entries/exits determined by Z-score thresholds.
\textbf{SMA} trades based on moving-average levels and crossovers to reflect smoothed trend direction.

\noindent\textbf{ML \& DL-based Models.} We employ sequential modeling baselines:
\textbf{LSTM}~\cite{yang2020qlib} models sequential dependencies with gated memory to improve next-step price forecasting.
\textbf{Transformer}~\cite{yang2020qlib} leverages self-attention to capture long-range temporal interactions for price prediction.

\noindent\textbf{RL-based Methods.} We utilize standard reinforcement learning algorithms:
\textbf{DQN}~\cite{mnih2013playing} approximates the action-value function with deep networks to select trades from market states.
\textbf{PPO}~\cite{schulman2017proximal} optimizes a clipped surrogate objective to update policies stably and sample-efficiently.

\noindent\textbf{General LLMs.} We include strong general LLMs as generic text-driven trading agents:
\textbf{Qwen3-8B}, \textbf{DeepSeek-R1-0528}, \textbf{Llama4-Scout-17B}, and \textbf{GPT-5-mini} are prompted to infer directional signals from the same textual inputs (e.g., news summaries) and to produce trading decisions under a unified prompting and execution protocol.

\noindent\textbf{Financial LLMs.} We compare against LLMs and agent systems specialized for financial tasks (introduced in chronological order):
\textbf{FinGPT}~\cite{liu2023fingpt} is an open-source framework converting financial news and price information into trading decisions.
\textbf{FinAgent}~\cite{zhang2024multimodal} is a finance-oriented LLM agent designed to perform market reasoning and decision making with domain-specific instructions and tools.
\textbf{TradingAgents}~\cite{xiao2024tradingagents} is a multi-agent LLM framework designed for automated trading decision making.
\textbf{DeepFund}~\cite{li2025time} is a recent finance-focused LLM agent that integrates market information and textual evidence for fund-style trading decisions.

\subsection{Full Experimental Results of Stocks Mentioned in Paper}

We summarize the quantitative results in Table~\ref{tab:main_results_split}, evaluating profitability and risk across four key metrics: CR, ARR, SR, and MDD. \system{} establishes a new state-of-the-art by achieving the highest cumulative returns across all six evaluated assets. On high-volatility assets, the system demonstrates exceptional profitability, securing a 74.21\% return on TSLA and 38.90\% on BTCUSD, significantly outperforming the Buy-and-Hold benchmark (65.64\% and 31.78\%, respectively). Furthermore, on stable tech giants, \system{} generates substantial alpha where other methods struggle. For instance, on AAPL, while the market yielded only 13.97\% and the closest competitor (DeepFund) reached 16.91\%, our system delivered a remarkable 25.04\%, translating to a relative improvement of 48.08\%. This consistent dominance validates that the multi-agent fusion mechanism successfully captures long-term appreciation trends that single-modality agents often miss.

\begin{figure}[!h]
\vspace{-4mm}
    \centering
    \includegraphics[width=\textwidth]{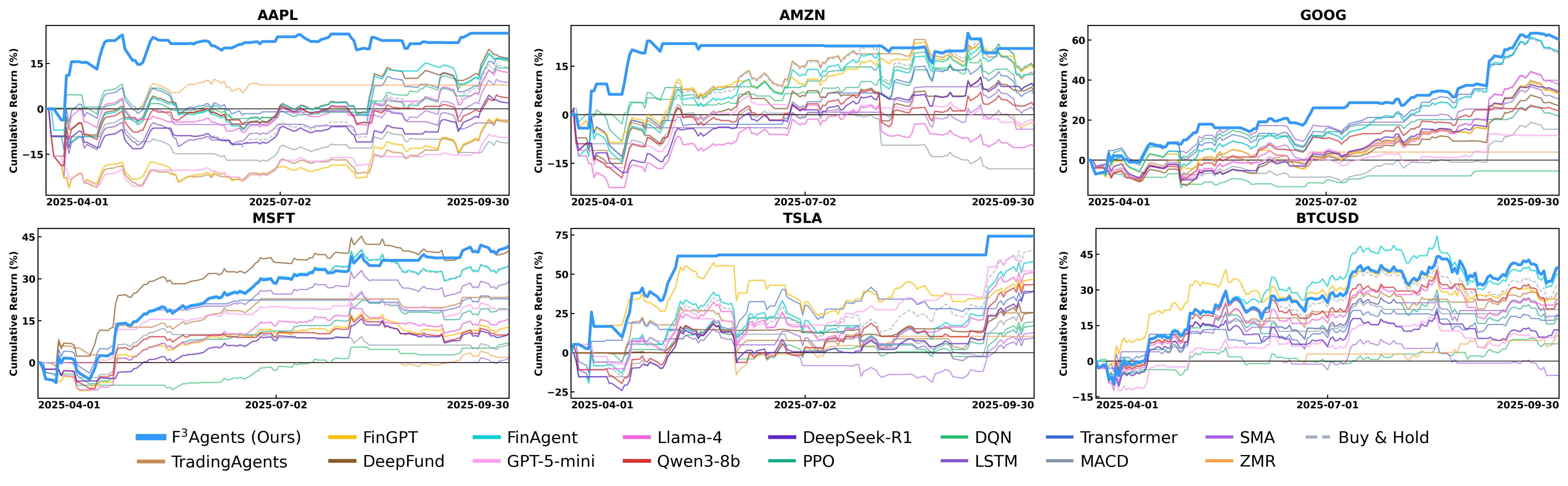}
    \caption{Performance comparison on cumulative return  over time between \system{} and other benchmarks across all assets.}
    \label{fig:cr_full_plot}
    \vspace{-4mm}
\end{figure}

Figure~\ref{fig:cr_full_plot} further illustrates the cumulative return trajectories on representative assets. On AAPL and AMZN, \system{} establishes an early advantage and maintains stable return accumulation over most of the test horizon. On BTCUSD, where all methods exhibit larger fluctuations, \system{} remains among the strongest performers while avoiding the severe reversals observed in several baselines. This suggests that the proposed multimodal agentic design improves not only final profitability but also temporal robustness under different market conditions.

\label{app:full_results}
\begin{table}[t]
\centering
\caption{Performance comparison across four metrics on six assets. Top block reports results for AAPL, AMZN, and GOOG; Bottom block reports results for MSFT, TSLA, and BTCUSD. Best, second, and third results are marked by \best{bold-underlined}, \second{bold}, and \third{underlined} text, respectively. CR: Cumulative Return; ARR: Annualized Return; SR: Sharpe Ratio; MDD: Maximum Drawdown. The \textit{Improvement} row reports the relative gain of \system{} over the best-performing baseline.}
\label{tab:main_results_split}

\resizebox{\textwidth}{!}{
\begin{tabular}{ll cccc cccc cccc}
\toprule
\multirow{2}{*}{Category} & \multirow{2}{*}{Model}
& \multicolumn{4}{c}{AAPL}
& \multicolumn{4}{c}{AMZN}
& \multicolumn{4}{c}{GOOG} \\
\cmidrule(lr){3-6} \cmidrule(lr){7-10} \cmidrule(lr){11-14}
&
& CR\%$\uparrow$ & ARR\%$\uparrow$ & SR$\uparrow$ & MDD\%$\downarrow$
& CR\%$\uparrow$ & ARR\%$\uparrow$ & SR$\uparrow$ & MDD\%$\downarrow$
& CR\%$\uparrow$ & ARR\%$\uparrow$ & SR$\uparrow$ & MDD\%$\downarrow$ \\
\midrule

Market & B\&H
& 13.97 & 27.95 & 0.61 & 22.99
& \third{14.26} & \third{28.51} & 0.63 & 14.64
& 53.19 & 106.38 & 1.98 & 7.98 \\
\midrule

\multirow{3}{*}{Rule-based} & MACD
& -11.24 & -22.48 & -0.73 & 17.46
& -16.72 & -33.43 & -0.94 & 19.25
& 15.40 & 30.80 & 0.94 & 12.06 \\
& ZMR
& 7.89 & 15.79 & \third{0.96} & \best{4.06}
& -3.54 & -7.08 & -0.97 & \best{4.16}
& 3.99 & 7.99 & 1.26 & \best{1.03} \\
& SMA
& 8.85 & 17.70 & 0.86 & \second{5.42}
& -4.47 & -8.93 & -0.49 & \third{7.36}
& 37.71 & 75.42 & \second{2.15} & 8.94 \\
\midrule

\multirow{2}{*}{ML/DL} & LSTM
& -3.73 & -7.46 & -0.02 & 22.99
& 1.13 & 2.26 & 0.16 & 14.64
& 30.44 & 60.88 & 1.52 & \third{7.74} \\
& Transformer
& 16.45 & 32.90 & 0.72 & 13.77
& 6.30 & 12.60 & 0.38 & 14.64
& \third{53.92} & \third{107.85} & \third{2.12} & 8.32 \\
\midrule

\multirow{2}{*}{RL} & DQN
& 16.04 & 32.07 & \second{1.06} & 12.18
& 9.57 & 19.14 & \third{0.72} & 9.84
& -5.77 & -11.54 & -0.47 & 13.77 \\
& PPO
& 13.60 & 27.20 & 0.68 & 9.48
& \second{14.52} & \second{29.03} & 0.70 & 12.93
& 21.87 & 43.73 & 1.14 & 7.98 \\
\midrule

\multirow{4}{*}{General LLMs} & Qwen3-8B
& 3.67 & 7.34 & 0.27 & 18.95
& 2.56 & 5.12 & 0.23 & 19.52
& 24.85 & 49.70 & 1.43 & 10.97 \\
& DeepSeek-R1
& 2.01 & 4.02 & 0.20 & 13.42
& 8.40 & 16.80 & 0.51 & 19.41
& 34.20 & 68.40 & 1.67 & 9.90 \\
& Llama4-Scout
& 12.12 & 24.23 & 0.58 & 18.95
& -10.23 & -20.46 & -0.39 & 22.50
& 39.84 & 79.69 & 1.87 & 9.96 \\
& GPT5-mini
& -9.44 & -18.88 & -0.42 & 25.83
& -2.64 & -5.27 & -0.06 & 18.01
& 11.92 & 23.85 & 0.74 & 9.59 \\
\midrule

\multirow{4}{*}{Financial LLMs} & FinGPT
& -3.92 & -7.85 & -0.06 & 26.25
& 13.89 & 27.78 & \second{0.86} & 8.53
& 32.99 & 65.99 & 1.40 & 9.26 \\
& FinAgent
& \third{16.88} & \third{33.77} & 0.90 & 13.16
& 12.38 & 24.76 & 0.57 & 14.63
& \second{54.08} & \second{108.16} & 2.00 & 7.98 \\
& TradingAgents
& -4.29 & -8.58 & -0.09 & 26.25
& 11.76 & 23.72 & 0.66 & 15.31
& 33.43 & 66.85 & 1.57 & 9.76 \\
& DeepFund
& \second{16.91} & \second{33.83} & 0.74 & 15.86
& 7.20 & 14.41 & 0.43 & 16.70
& 27.06 & 54.13 & 1.40 & 12.86 \\
\midrule

\textbf{Ours} & \textbf{\system}
& \best{25.04} & \best{50.08} & \best{1.22} & \third{7.83}
& \best{20.44} & \best{40.87} & \best{1.18} & 5.57
& \best{60.24} & \best{120.48} & \best{2.55} & \second{7.01} \\

\midrule
\multicolumn{2}{c}{Improvement(\%)}
& 48.08 & 48.03 & 15.09 & --
& 40.77 & 40.79 & 37.21 & --
& 11.39 & 11.39 & 18.60 & -- \\
\bottomrule
\end{tabular}
}

\vspace{2mm} 

\resizebox{\textwidth}{!}{
\begin{tabular}{ll cccc cccc cccc c}
\toprule
\multirow{2}{*}{Category} & \multirow{2}{*}{Model}
& \multicolumn{4}{c}{MSFT}
& \multicolumn{4}{c}{TSLA}
& \multicolumn{4}{c}{BTCUSD}
& \multirow{2}{*}{Avg Rank} \\
\cmidrule(lr){3-6} \cmidrule(lr){7-10} \cmidrule(lr){11-14}
&
& CR\%$\uparrow$ & ARR\%$\uparrow$ & SR$\uparrow$ & MDD\%$\downarrow$
& CR\%$\uparrow$ & ARR\%$\uparrow$ & SR$\uparrow$ & MDD\%$\downarrow$
& CR\%$\uparrow$ & ARR\%$\uparrow$ & SR$\uparrow$ & MDD\%$\downarrow$
& (ARR) \\
\midrule

Market & B\&H
& \third{35.26} & \third{70.52} & 1.76 & 8.03
& \second{65.64} & \second{131.28} & 1.32 & 21.54
& \third{31.78} & \third{43.76} & 1.27 & 11.70
& 3.50 \\
\midrule

\multirow{3}{*}{Rule-based} & MACD
& 6.80 & 13.61 & 0.67 & 6.79
& 25.74 & 51.48 & 0.94 & 17.29
& 21.73 & 29.92 & \third{1.40} & 8.29
& 13.67 \\
& ZMR
& 1.88 & 3.76 & 0.50 & \best{2.10}
& 10.42 & 20.84 & 0.60 & \third{14.26}
& 10.50 & 14.46 & 1.10 & \best{3.54}
& 14.33 \\
& SMA
& 1.75 & 3.50 & 0.67 & \best{2.10}
& 12.98 & 25.95 & 0.70 & 18.68
& -6.00 & -8.26 & -0.51 & 13.91
& 13.17 \\
\midrule

\multirow{2}{*}{ML/DL} & LSTM
& 29.13 & 58.26 & 1.55 & 8.03
& 9.57 & 19.13 & 0.42 & 21.53
& 18.94 & 26.09 & 1.00 & 9.07
& 11.17 \\
& Transformer
& 22.82 & 45.64 & 1.35 & 8.03
& 38.94 & 77.88 & 1.09 & 16.99
& 19.14 & 26.35 & 1.18 & \third{8.26}
& 7.00 \\
\midrule

\multirow{2}{*}{RL} & DQN
& 6.95 & 13.91 & 0.81 & 9.70
& 20.01 & 40.02 & 0.82 & 19.70
& 7.30 & 10.05 & 0.57 & \second{6.97}
& 11.83 \\
& PPO
& 19.27 & 38.54 & 1.51 & 7.55
& 17.34 & 34.68 & 0.80 & 20.10
& 17.60 & 24.24 & 0.89 & 11.59
& 9.33 \\
\midrule

\multirow{4}{*}{General LLMs} & Qwen3-8B
& 10.60 & 21.19 & 0.80 & 9.28
& 43.74 & 87.48 & 1.20 & 23.66
& 23.91 & 32.93 & 1.10 & 12.03
& 10.00 \\
& DeepSeek-R1
& 10.40 & 20.81 & 0.97 & 7.52
& 39.42 & 78.83 & 1.15 & 24.13
& 12.91 & 17.77 & 0.71 & 10.58
& 10.00 \\
& Llama4-Scout
& 15.95 & 31.90 & 1.19 & 9.73
& 51.55 & 103.09 & 1.34 & 21.24
& 25.72 & 35.42 & 1.15 & 13.98
& 8.33 \\
& GPT5-mini
& 19.60 & 39.21 & 1.63 & 7.55
& 52.58 & 105.17 & \third{1.55} & 19.44
& 11.68 & 16.08 & 0.71 & 12.39
& 11.50 \\
\midrule

\multirow{4}{*}{Financial LLMs} & FinGPT
& 13.26 & 26.52 & 0.98 & 10.03
& 47.17 & 94.35 & 1.20 & 21.54
& 25.75 & 35.46 & 1.20 & 11.28
& 8.17 \\
& FinAgent
& 34.93 & 69.87 & \best{2.19} & 8.02
& \third{58.44} & \third{116.87} & 1.24 & 24.43
& \second{36.36} & \second{50.07} & \second{1.47} & 14.82
& 3.17 \\
& TradingAgents
& 24.15 & 48.29 & 1.59 & \second{5.31}
& 25.90 & 51.80 & 0.93 & 22.64
& 28.29 & 38.96 & 1.37 & 8.74
& 8.17 \\
& DeepFund
& \second{40.30} & \second{80.60} & \third{2.07} & \third{5.94}
& 25.59 & 51.18 & \second{1.72} & \best{6.52}
& 5.84 & 8.05 & 0.38 & 16.11
& 8.67 \\
\midrule

\textbf{Ours} & \textbf{\system}
& \best{42.07} & \best{84.14} & \second{2.08} & 8.03
& \best{74.21} & \best{148.41} & \best{1.94} & \second{12.60}
& \best{38.90} & \best{53.57} & \best{1.52} & 9.79
& \textbf{1.00} \\

\midrule
\multicolumn{2}{c}{Improvement(\%)}
& 4.39 & 4.39 & -- & --
& 13.06 & 13.05 & 12.79 & --
& 6.99 & 6.99 & 3.40 & --
& -- \\

\bottomrule
\end{tabular}
}
\vspace{-2mm}
\end{table}

\label{app:full_results}
\begin{table}[H]
\centering
\caption{Performance comparison across four metrics on six assets. Top block reports results for AAPL, AMZN, and GOOG; Bottom block reports results for MSFT, TSLA, and BTCUSD. Best, second, and third results are marked by \best{bold-underlined}, \second{bold}, and \third{underlined} text, respectively. CR: Cumulative Return; ARR: Annualized Return; SR: Sharpe Ratio; MDD: Maximum Drawdown.}
\label{tab:main_results_split}

\resizebox{\textwidth}{!}{
\begin{tabular}{ll cccc cccc cccc}
\toprule
\multirow{2}{*}{Category} & \multirow{2}{*}{Model}
& \multicolumn{4}{c}{AAPL}
& \multicolumn{4}{c}{AMZN}
& \multicolumn{4}{c}{GOOG} \\
\cmidrule(lr){3-6} \cmidrule(lr){7-10} \cmidrule(lr){11-14}
&
& CR\%$\uparrow$ & ARR\%$\uparrow$ & SR$\uparrow$ & MDD\%$\downarrow$
& CR\%$\uparrow$ & ARR\%$\uparrow$ & SR$\uparrow$ & MDD\%$\downarrow$
& CR\%$\uparrow$ & ARR\%$\uparrow$ & SR$\uparrow$ & MDD\%$\downarrow$ \\
\midrule

Market & B\&H
& 13.97 & 27.95 & 0.61 & 22.99
& 14.26 & 28.51 & 0.63 & 14.64
& 53.19 & 106.38 & 1.98 & 7.98 \\
\midrule

\multirow{3}{*}{Rule-based} & MACD
& -11.24 & -22.48 & -0.73 & 17.46
& -16.72 & -33.43 & -0.94 & 19.25
& 15.40 & 30.80 & 0.94 & 12.06 \\
& ZMR
& 7.89 & 15.79 & 0.96 & \best{4.06}
& -3.54 & -7.08 & -0.97 & \best{4.16}
& 3.99 & 7.99 & 1.26 & \best{1.03} \\
& SMA
& 8.85 & 17.70 & 0.86 & \second{5.42}
& -4.47 & -8.93 & -0.49 & \third{7.36}
& 37.71 & 75.42 & \third{2.15} & 8.94 \\
\midrule

\multirow{2}{*}{ML/DL} & LSTM
& -3.73 & -7.46 & -0.02 & 22.99
& 1.13 & 2.26 & 0.16 & 14.64
& 30.44 & 60.88 & 1.52 & \third{7.74} \\
& Transformer
& 16.45 & 32.90 & 0.72 & 13.77
& 6.30 & 12.60 & 0.38 & 14.64
& \third{53.92} & \third{107.85} & 2.12 & 8.32 \\
\midrule

\multirow{2}{*}{RL} & DQN
& 16.04 & 32.07 & \third{1.06} & 12.18
& 9.57 & 19.14 & 0.72 & 9.84
& -5.77 & -11.54 & -0.47 & 13.77 \\
& PPO
& 13.60 & 27.20 & 0.68 & 9.48
& \third{14.52} & \third{29.03} & 0.70 & 12.93
& 21.87 & 43.73 & 1.14 & 7.98 \\
\midrule

\multirow{4}{*}{General LLMs} & Qwen3-8B
& 3.67 & 7.34 & 0.27 & 18.95
& 2.56 & 5.12 & 0.23 & 19.52
& 24.85 & 49.70 & 1.43 & 10.97 \\
& DeepSeek-R1
& 2.01 & 4.02 & 0.20 & 13.42
& 8.40 & 16.80 & 0.51 & 19.41
& 34.20 & 68.40 & 1.67 & 9.90 \\
& Llama4-Scout
& 12.12 & 24.23 & 0.58 & 18.95
& -10.23 & -20.46 & -0.39 & 22.50
& 39.84 & 79.69 & 1.87 & 9.96 \\
& GPT5-mini
& -9.44 & -18.88 & -0.42 & 25.83
& -2.64 & -5.27 & -0.06 & 18.01
& 11.92 & 23.85 & 0.74 & 9.59 \\
\midrule

\multirow{4}{*}{Financial LLMs} & FinGPT
& -3.92 & -7.85 & -0.06 & 26.25
& 13.89 & 27.78 & \third{0.86} & 8.53
& 32.99 & 65.99 & 1.40 & 9.26 \\
& FinAgent
& 16.88 & 33.77 & 0.90 & 13.16
& 12.38 & 24.76 & 0.57 & 14.63
& \second{54.08} & \second{108.16} & 2.00 & 7.98 \\
& TradingAgents
& -4.29 & -8.58 & -0.09 & 26.25
& 11.76 & 23.72 & 0.66 & 15.31
& 33.43 & 66.85 & 1.57 & 9.76 \\
& DeepFund
& \third{16.91} & \third{33.83} & 0.74 & 15.86
& 7.20 & 14.41 & 0.43 & 16.70
& 27.06 & 54.13 & 1.40 & 12.86 \\
\midrule


& \textbf{\system}
& \second{25.04} & \second{50.08} & \second{1.22} & 7.83
& \second{20.44} & \second{40.87} & \second{1.18} & \second{5.57}
& \best{60.24} & \best{120.48} & \second{2.55} & \second{7.01} \\

& PMRL\_Finance
& \best{51.28} & \best{102.56} & \best{2.77} & \third{5.72}
& \best{27.77} & \best{55.54} & \best{1.73} & 13.10
& \best{64.07} & \best{127.80} & \second{2.39} & 7.98 \\

\midrule
\multicolumn{2}{c}{Improvement(\%)}
& 48.08 & 48.03 & 15.09 & --
& 40.77 & 40.79 & 37.21 & --
& 11.39 & 11.39 & 18.60 & -- \\
\bottomrule
\end{tabular}
}
\end{table}

\begin{table}[t]
\centering
\caption{Classification performance comparison on AAPL, AMZN, and GOOG.}
\label{tab:acc_mcc_results}
\resizebox{0.75\textwidth}{!}{
\begin{tabular}{l cc cc cc}
\toprule
\multirow{2}{*}{Method}
& \multicolumn{2}{c}{AAPL}
& \multicolumn{2}{c}{AMZN}
& \multicolumn{2}{c}{GOOG} \\
\cmidrule(lr){2-3} \cmidrule(lr){4-5} \cmidrule(lr){6-7}
& ACC\%$\uparrow$ & MCC$\uparrow$
& ACC\%$\uparrow$ & MCC$\uparrow$
& ACC\%$\uparrow$ & MCC$\uparrow$ \\
\midrule
\system
& 50.00 & 0.0051
& 55.56 & 0.1055
& 54.76 & 0.0889 \\
PMRL\_Finance
& \textbf{64.29} & \textbf{0.2843}
& \textbf{56.35} & \textbf{0.1266}
& \textbf{59.52} & \textbf{0.1682} \\
\bottomrule
\end{tabular}
}
\end{table}

\begin{table}[t]
\centering
\caption{Ablation study on AAPL.}
\label{tab:ablation_aapl}
\resizebox{0.60\textwidth}{!}{
\begin{tabular}{lcc}
\toprule
\multirow{2}{*}{Method} 
& \multicolumn{2}{c}{AAPL} \\
\cmidrule(lr){2-3}
& ACC\%$\uparrow$ & MCC$\uparrow$ \\
\midrule
\system
& 50.00 & 0.0051 \\
PMRL\_Finance w/o robust
& 58.73 & 0.1686 \\
PMRL\_Finance
& \textbf{64.29} & \textbf{0.2843} \\
\bottomrule
\end{tabular}
}
\end{table}

\subsection{Experimental Results of Broader Asset Pool and Extended Test Horizons}
\label{app:add_stocks}

To further examine the generalization capability of \system{}, we extend the evaluation beyond the main assets---Apple Inc. (\textbf{AAPL}, Technology, mega-cap), Amazon.com Inc. (\textbf{AMZN}, Consumer Cyclical, mega-cap), Alphabet Inc. (\textbf{GOOG}, Communication Services, mega-cap), Microsoft Corp. (\textbf{MSFT}, Technology, mega-cap), Tesla Inc. (\textbf{TSLA}, Consumer Cyclical, mega-cap), and \textbf{BTCUSD} (Cryptocurrency, digital asset)---to seven additional equities: Grab Holdings (\textbf{GRAB}, Technology, large-cap), Alibaba Group (\textbf{BABA}, Consumer Cyclical, mega-cap), The Coca-Cola Company (\textbf{KO}, Consumer Defensive, mega-cap), Johnson \& Johnson (\textbf{JNJ}, Healthcare, mega-cap), United Parcel Service (\textbf{UPS}, Industrials, large-cap), Upwork (\textbf{UPWK}, Communication Services, small-cap), and Sprouts Farmers Market (\textbf{SFM}, Consumer Defensive, mid-cap).

First, we include \textbf{Grab Holdings (GRAB, Singapore)} and \textbf{Alibaba Group (BABA, China)} to assess whether \system{} remains effective beyond the primary U.S.-centric asset set. As shown in Table~\ref{tab:new_assets_appendix}, \system{} achieves the highest Annualized Return on both GRAB and BABA, reaching 105.43\% and 122.27\%, respectively. It also attains the best Sharpe Ratio on BABA and the second-best Sharpe Ratio on GRAB, suggesting strong risk-adjusted performance under different market narratives. In contrast, several ML/DL and LLM-based baselines show unstable behavior, including negative returns on BABA or GRAB. These results suggest that the proposed multi-agent fusion mechanism remains effective when the asset-specific information environment differs from the primary benchmark.

\input{tables/grab_baba_stock}

Second, we evaluate whether the performance advantage persists under a longer and more recent test horizon. Table~\ref{tab:main_results_reduced} reports results from 2025-04-01 to 2026-03-20 while keeping the original training and validation sets unchanged. Across AAPL, GOOG, TSLA, and BTCUSD, \system{} achieves the best CR, ARR, and SR on all four assets. Specifically, \system{} obtains 30.15\% ARR and 1.02 SR on AAPL, 108.75\% ARR and 2.88 SR on GOOG, and 78.49\% ARR and 1.47 SR on TSLA. The advantage is also evident on BTCUSD, where several baselines suffer negative returns, whereas \system{} achieves positive performance with 18.06\% CR, 12.58\% ARR, 0.84 SR, and the lowest MDD of 12.90\%. As shown in the improvement row, \system{} improves ARR by 42.35\%, 8.50\%, 16.54\%, and 24.43\% over the strongest alternatives on AAPL, GOOG, TSLA, and BTCUSD, respectively. Although \system{} does not achieve the lowest MDD on AAPL, it still provides the strongest return and Sharpe Ratio, indicating a favorable risk-return trade-off over the extended horizon.

\input{tables/longer_horizons}

Third, we further test \system{} on a broader stock pool beyond the main experimental assets. As reported in Table~\ref{tab:extended_results}, \system{} achieves the best CR, ARR, and SR on all newly added assets, including KO, JNJ, UPS, UPWK, and SFM. The advantage is particularly clear on UPWK and SFM, where the market benchmark and several baselines produce negative returns, while \system{} remains profitable. These results suggest that \system{} generalizes beyond mega-cap technology-oriented assets and remains effective across assets with different sector characteristics, volatility profiles, and market-cap ranges.

\input{tables/broader_asset_pool}

Overall, the additional experiments indicate that the performance of \system{} is not confined to the primary asset set or the original evaluation window. Instead, the model maintains strong profitability and competitive risk-adjusted behavior across broader assets and longer test horizons.

\section{Details of Random-seed Robustness and Statistical Significance}
\label{app:random_seed}

To further assess the robustness of \system{} beyond a single random seed, we conduct additional experiments with three different seeds on AAPL and BTCUSD. We compare \system{} with the Transformer baseline and the concatenation-based fusion baseline, which respectively represent a strong sequence modeling baseline and a simplified fusion variant. Tables~\ref{tab:seed_variance_raw} and~\ref{tab:seed_variance_mean_std} report the raw results and the corresponding mean--standard deviation statistics.

The raw seed-level results show that \system{} consistently achieves higher ARR and SR than both baselines across all tested seeds on both assets. On AAPL, \system{} obtains ARR values of 50.08\%, 47.60\%, and 55.59\%, all clearly above Transformer and Concat. Fusion. Its SR also remains highly stable around 1.20--1.22, while the two baselines show larger fluctuations. In particular, Concat. Fusion varies substantially across seeds, ranging from -3.32\% to 28.04\% ARR on AAPL, indicating that simple concatenation is more sensitive to initialization and training randomness.

The mean--standard deviation results further support this observation. On AAPL, \system{} achieves the highest average ARR of 51.09\% with a standard deviation of 4.09, compared with 31.20$\pm$7.68 for Transformer and 14.64$\pm$16.17 for Concat. Fusion. The difference is more pronounced for SR: \system{} obtains 1.2096$\pm$0.0102, showing very small seed-induced variation. In contrast, Transformer and Concat. Fusion have larger SR deviations, especially Concat. Fusion with 0.1465$\pm$0.4257. For MDD, \system{} also achieves the lowest average drawdown on AAPL, 8.04$\pm$0.37, suggesting that its stronger return does not come from substantially higher downside exposure.

On BTCUSD, which is more volatile than the equity asset, \system{} also maintains the best average ARR and SR. It achieves 50.33$\pm$3.80 ARR and 1.4984$\pm$0.0858 SR, outperforming Transformer and Concat. Fusion by clear margins. Although Transformer obtains a lower average MDD on BTCUSD, its ARR and SR are much lower than those of \system{}. Meanwhile, \system{} has an almost unchanged MDD across seeds, 9.7886$\pm$0.0029, indicating that its drawdown behavior is highly stable under different random initializations. These results suggest that \system{} provides a more reliable risk-return profile rather than relying on a favorable random seed.

Table~\ref{tab:significance_arr} reports a paired statistical significance analysis based on seed-wise ARR improvements. On AAPL, \system{} improves ARR over Transformer by 19.89 percentage points, with a 95\% confidence interval of [9.33, 30.44] and a paired t-test p-value of 0.0149. It also improves over Concat. Fusion by 36.45 percentage points, with a 95\% confidence interval of [5.05, 67.84] and a p-value of 0.0378. Both comparisons are statistically significant at the 0.05 level.

On BTCUSD, \system{} improves ARR over Concat. Fusion by 41.93 percentage points, with a 95\% confidence interval of [18.54, 65.33] and a p-value of 0.0164, again indicating a statistically significant improvement. The comparison against Transformer shows a positive mean improvement of 21.54 percentage points, but the 95\% confidence interval slightly crosses zero and the p-value is 0.0535. Therefore, this result should be interpreted as a strong positive trend rather than a statistically significant difference at the 0.05 level. Given the small number of seeds, we report these tests as supporting evidence rather than as a standalone proof of significance.

Overall, the random-seed experiments show that \system{} maintains strong performance across different initializations. It consistently achieves the highest average ARR and SR on both AAPL and BTCUSD, with lower or comparable variance relative to the baselines. The significance analysis further confirms that the improvements are statistically reliable in most comparisons, especially against the concatenation-based fusion baseline. These findings provide additional evidence that the proposed fusion and robustness mechanisms improve both performance and stability.
\input{tables/random_seed}

\section{Details of Case Study}
\label{app:case_study}
To qualitatively examine \system{}'s robustness to market noise and its cross-modal reasoning capability, we conduct a qualitative case study on Apple Inc. (AAPL) during a volatile period from April to October 2025 (Fig.~\ref{fig:case_analysis}). Compared with DeepFund, \system{} better navigates noisy narratives, structural risks, and high-confidence opportunities through agentic fusion.

For unsustainable news, on Sep~22, the market is driven by a high-noise positive narrative around iPhone 17 preorders while the news itself suggested the significance is exaggerated. DeepFund issued a BUY, whereas \system{} leveraged noise-robust consistency regularization to detect weak cross-modal confirmation and choose to HOLD, avoiding a potential bull trap. For bad news, on May~22, signals about expansion halting and partnership backlash implied a more persistent deterioration rather than a transient technical dip. DeepFund executed a contrarian BUY driven by oversold technical cues, while \system{}'s modality-aware fusion prioritized strategic risk evidence extracted by the News Agent and decisively issued a SELL signal to preserve capital under downside uncertainty. For good news, on Apr~10, valuation support and accumulation provided a coherent positive catalyst supported by both textual narratives and market dynamics. \system{} integrated these aligned cues and executed a confident BUY, whereas DeepFund misread the context and issued a SELL, missing the ensuing upside. 

\begin{figure}[htbp]
    \centering
    \includegraphics[width=0.9\textwidth]{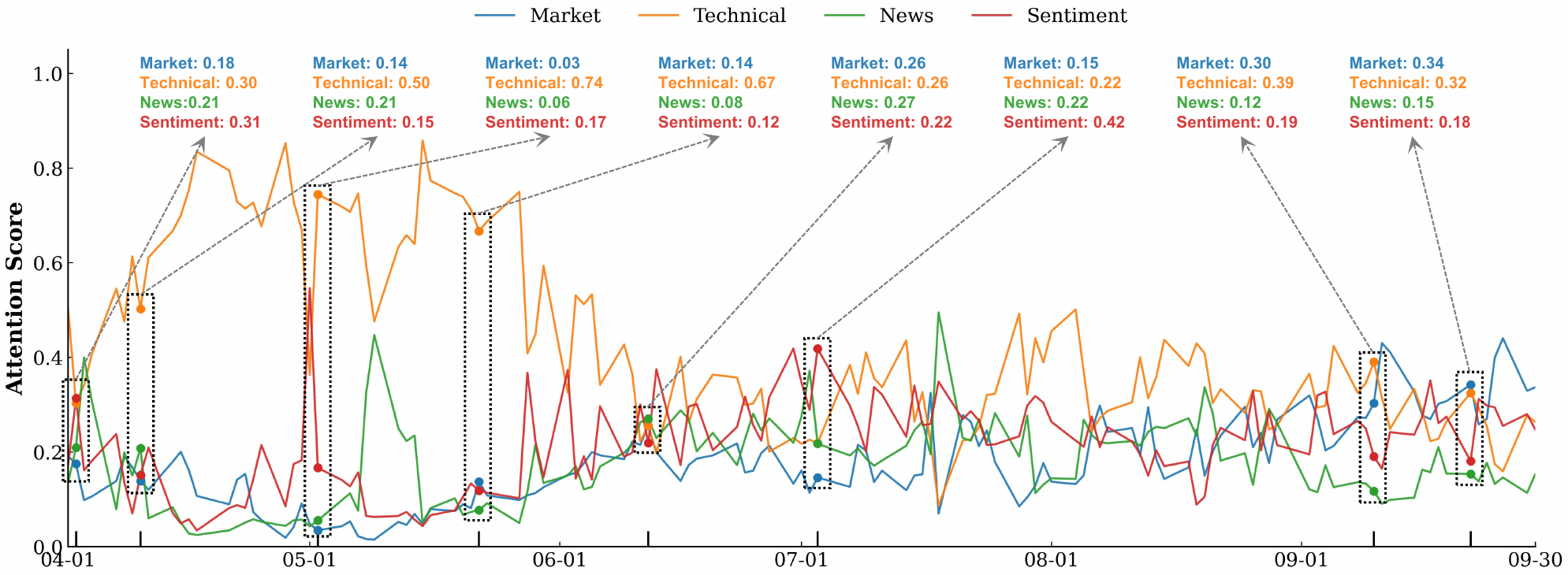}
    \caption{Attention scores of our proposed method.}
    \label{fig:case_analysis_attention}
\end{figure}

\begin{figure}[H]
    \centering
    \includegraphics[width=0.9\textwidth]{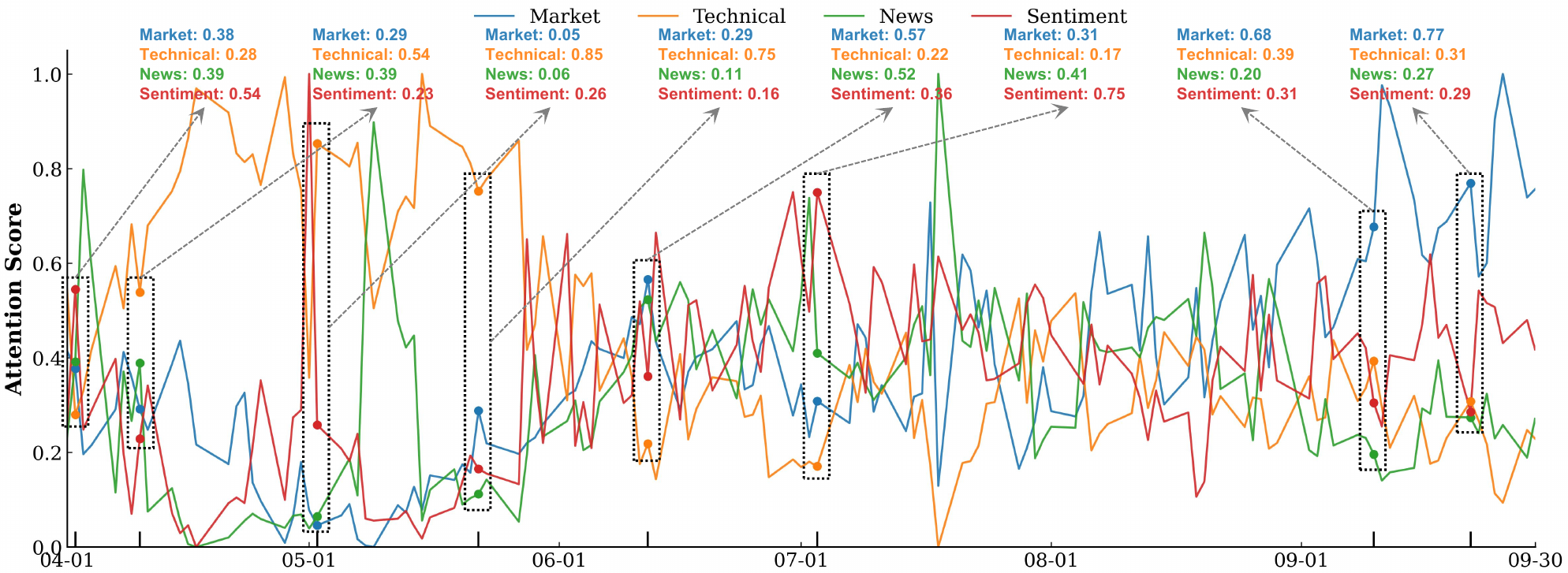}
    \caption{Attention scores of our proposed method (after standardized).}
    \label{fig:case_analysis_attention_standardized}
\end{figure}

To further interpret the case study, we visualize the modality-level attention scores before and after standardization. As shown in Fig.~\ref{fig:case_analysis_attention}, the raw attention scores reveal how \system{} dynamically allocates attention across market, technical, news, and sentiment modalities over time. After standardization, Fig.~\ref{fig:case_analysis_attention_standardized} makes the relative salience of each modality more comparable, highlighting when the model relies more on textual evidence or structured market signals. These results suggest that \system{} does not follow a fixed modality preference, but adaptively adjusts modality weights according to the market context.

Overall, this case study suggests that \system{} goes beyond single-modality reactions by integrating cross-modal evidence and robustness regularization to make more reliable trading decisions.

\section{Details of Stock Movement Predictions Enhancement}
\label{app:details_Stock_Movement}
\input{tables/smp_task}
Table~\ref{tab:acc_mcc_all} presents the ablation study investigating the impact of the Multi-agent Fusion (MaF) module. The results provide strong evidence for the importance of the fusion mechanism. When the MaF module is removed (``\system{} w/o MaF''), the system shows limited directional discrimination, as indicated by MCC scores hovering near zero (e.g., 0.0431 for AAPL and -0.0044 for AMZN). In contrast, integrating the MaF module leads to substantial performance gains across all six assets. Specifically, \textbf{Accuracy (ACC)} improves by approximately 4--6\% on major tech stocks (e.g., AAPL: $53.02\% \rightarrow 57.94\%$, MSFT: $51.70\% \rightarrow 57.14\%$). More importantly, the \textbf{Matthews Correlation Coefficient (MCC)} increases substantially (e.g., AAPL: $0.0431 \rightarrow 0.2122$, MSFT: $0.0159 \rightarrow 0.1840$), which suggests that the fusion module effectively synthesizes conflicting signals from different experts to generate robust, non-trivial predictive signals.

%% file: tex/6_related_work.tex
\section{Related Work}
\label{app:related_work}



{\bf Vanilla Trading Techniques. }The evolution of quantitative trading strategies has progressed from static heuritics~\cite{brock1992simple} to sophisticated data-driven decision making models~\cite{cavalcante2016computational}. Early methods were predominantly rule-based, relying on technical indicators such as Moving Average Convergence Divergence (MACD) and Relative Strength Index (RSI) derived from market microstructure theories~\cite{murphy1999technical,kakushadze2016101}. 
While computationally efficient and interpretable, these strategies suffer from structural rigidity, limiting their ability to adapt to non-stationary market dynamics or capture complex non-linear patterns. To address these limitations, research shifted towards Machine Learning (ML) and Deep Learning (DL) paradigms. 
Algorithms ranging from Support Vector Machines (SVMs)~\cite{patel2015predicting} to advanced sequence modeling architectures like~\cite{nelson2017stock} and Transformers~\cite{fischer2018deep,sezer2020financial,zhang2023crossformer}, became standard for capturing temporal dependencies in financial time series. Despite their superior pattern recognition capabilities, these prediction-focused discriminative models typically operate on isolated modalities or employ shallow fusion schemes, limiting their ability to captures fine-grained cross-modal dependencies and robustly integrate heterogeneous financial signals. 
Consequently, Deep Reinforcement Learning (DRL) emerged as more comprehensive framework, formulating as a Markov Decision Process (MDP)~\cite{sutton1998reinforcement} to optimize long-term cumulative rewards~\cite{jiang2017deep,liu2020finrl}. Although DRL methods excel at dynamic portfolio management~\cite{liu2020finrl}, they generally suffer from sample inefficiency and lack the semantic capacity to process unstructured information (e.g., news and policy reports), limiting their effectiveness in information-rich environments.




{\bf LLM-based Trading Techniques. }The advent of Large Language Models (LLMs) has significantly advanced financial analysis by introducing semantic reasoning capabilities~\cite{wu2023bloomberggpt}, yet challenges in multimodal integration persist. Early studies~\cite{chen2023chatgpt} leveraged general-purpose LLMs (e.g., GPT-4) for zero-shot sentiment analysis and market prediction, employing prompt engineering techniques such as Chain-of-Thought (CoT) to extract trading signals from textual data~\cite{lopez2023can,wei2022chain,hansen2024can}. 
However, without domain-specific fine-tuning, these general LLMs often hallucinate on quantitative tasks and struggle to align textual sentiment with precise market movements. To bridge this domain gap, researchers developed Financial LLMs such as FinGPT~\cite{liu2023fingpt} and BloombergGPT~\cite{wu2023bloomberggpt}, which are fine-tuned on extensive financial corpora to enhance domain alignment. Building upon these foundations, the field has recently advanced towards agentic trading, where LLMs function as autonomous agents equipped with memory, reflection, and tool-use capabilities~\cite{yu2025finmem,zhang2024multimodal,xiao2024tradingagents,li2025time}. Despite enabling dynamic interaction with market environments, most existing agents remain predominantly text-centric. They tend to process numerical and textual modalities in isolation
, a limitation that the proposed \system{} system is explicitly to address.

%% file: tables/prompt_template.tex
\begin{table}[H]
\centering
\scriptsize
\caption{Prompt template and filled example for the News Analysis Agent (Part II).}
\label{tab:news_agent_prompt_example_part2}
\renewcommand{\arraystretch}{1.08}
\setlength{\tabcolsep}{3pt}
\begin{tabularx}{\linewidth}{@{}L{0.22\linewidth}Y@{}}
\toprule
\textbf{Field} & \textbf{Content} \\
\midrule
Illustrative output
& \texttt{Analysis: The recent news indicates strong positive momentum for Apple (AAPL) following its Q4 earnings report, which showed record-breaking results and strong guidance for the upcoming quarter. Analysts have raised their price targets and forecast continued growth. Additionally, the stock has been hitting new all-time highs and setting records, indicating strong investor confidence. Technical indicators such as RSI and MACD also suggest upward momentum.} \newline
  \texttt{Signal: UP} \\

Purpose of example
& This example illustrates the 5-day input/output interface used by the News Analysis Agent in the paper: the model jointly uses summarized news and technical indicators, produces a concise rationale, and outputs a binary directional signal constrained to \texttt{UP}/\texttt{DOWN}. \\
\bottomrule
\end{tabularx}
\end{table}

\begin{table}[H]
\centering
\scriptsize
\caption{Prompt template and filled example for the News Analysis Agent (Part I).}
\label{tab:news_agent_prompt_example_part1}
\renewcommand{\arraystretch}{1.08}
\setlength{\tabcolsep}{3pt}
\begin{tabularx}{\linewidth}{@{}L{0.22\linewidth}Y@{}}
\toprule
\textbf{Field} & \textbf{Content} \\
\midrule
Task 
& Predict the next-day stock movement for a given asset using recent summarized news and recent price/technical indicators. \\

Asset 
& AAPL \\

Reference date
& 2025-11-03 \\

Example type
& Adapted from the logged AAPL news-analysis case on 2025-11-03, following the 5-day prompt template used in the paper. \\

System instruction
& \texttt{You are a financial analyst. Analyze the historical news and market data and provide a clear signal with concise reasoning.} \\

User instruction
& \texttt{Predict the next-day AAPL stock movement given the following information.} \\

News and Technical window 
& Past 5 trading days \\

Output format 
& \texttt{Analysis: <concise analysis>} \newline
  \texttt{Signal: UP/DOWN} \\

Filled input prompt
& \texttt{Predict the next-day AAPL stock movement given the following information:} \newline\newline
  \texttt{News for the past 5 days:} \newline
  \texttt{News on Day 1: Apple trades steady as BofA's Mohan boosts iPhone 17 unit forecasts, raises price target, and models stronger long-term revenue and margins.; Earnings Preview For Apple; Two Nvidia Shockers.} \newline
  \texttt{News on Day 2: Apple reports financial results for the fourth quarter on Thursday after the close. Here's a rundown of the iPhone maker's report.; Apple Inc. AAPL will release earnings results on 10/30. Analysts expect earnings of 1.77 per share and quarterly revenue of 102.17 billion.; U.S. stock futures mixed, Amazon, Alphabet, and Microsoft beat earnings expectations, while Apple and Meta post upbeat results.} \newline
  \texttt{News on Day 3: Apple stock hit new all-time highs Friday after reporting fourth-quarter results Thursday. The company is guiding for the first quarter to be even better and record-breaking.; Apple's blockbuster fourth-quarter results prompted bullish reactions across Wall Street, with Wedbush's Dan Ives calling it a pound-the-table moment, CNBC's Jim Cramer saying Apple bears owe us an explanation, and Gene Munster highlighting stronger iPhone demand and a solid holiday outlook.; Nasdaq Surges Over 1 Apple Earnings Top Views.} \newline
  \texttt{News on Day 4: Apple, Amazon, Microsoft, and Alphabet beat Q3 estimates with revenue and earnings growth. Trump and Xi conclude high-stakes meeting with trade, soybean, and resource agreements, potentially impacting U.S.-China economic relations and global markets.; As 2026 approaches, Apple is gearing up for a critical year filled with significant product launches and potential executive changes, all while grappling with regulatory and tariff-related challenges.; Apple breaks records in Q4 and ad revenue, Samsung teases iPhone users, and Jamie Dimon echoes Steve Jobs on office meetings.} \newline
  \texttt{News on Day 5: CNN Money Fear and Greed index remained in Fear zone on Friday. U.S. stocks closed higher, Nasdaq up 4.7 in October. Amazon reported earnings.; The weekend saw major tech milestones as Nvidia hit a 5 trillion valuation, Apple set iPhone upgrade records, Netflix eyed Warner Bros. Discovery, Meta doubled down on smart glasses, and OpenAI prepared for a potential 1 trillion IPO.; 10 Information Technology Stocks With Whale Alerts In Today's Session.} \newline\newline
  \texttt{Stock market data for the past 5 trading days: The following is the information about the stock price and technical indicators:} \newline
  \texttt{close: Day 1: 268.74, Day 2: 269.44, Day 3: 271.14, Day 4: 270.11, Day 5: 268.79.} \newline
  \texttt{RSI: Day 1: 33.90, Day 2: 34.25, Day 3: 100.00, Day 4: 69.65, Day 5: 64.61.} \newline
  \texttt{MACD: Day 1: 3.79, Day 2: 3.97, Day 3: 4.12, Day 4: 3.90, Day 5: 3.42.} \newline
  \texttt{Signal: Day 1: 2.24, Day 2: 2.58, Day 3: 2.89, Day 4: 3.09, Day 5: 3.16.} \newline\newline
  \texttt{Output in the following format:} \newline
  \texttt{Analysis: <concise analysis>} \newline
  \texttt{Signal: UP/DOWN} \newline
  \texttt{Note: You MUST output either UP or DOWN. NEUTRAL is NOT allowed.} \\
\bottomrule
\end{tabularx}
\end{table}

\begin{table}[H]
\centering
\scriptsize
\caption{Prompt template and one filled example for the Sentiment Analysis Agent.}
\label{tab:sentiment_agent_prompt_example}
\renewcommand{\arraystretch}{1.08}
\setlength{\tabcolsep}{3pt}
\begin{tabularx}{\linewidth}{@{}L{0.22\linewidth}Y@{}}
\toprule
\textbf{Field} & \textbf{Content} \\
\midrule
Task
& Classify the overall sentiment of the past 5-day news summaries for a stock as \texttt{POSITIVE} or \texttt{NEGATIVE}. \\

Asset
& AAPL \\

Reference date
& 2025-11-03 \\

Example type
& Adapted from the logged AAPL sentiment case on 2025-11-03, following the 5-day prompt template used in the paper. \\

System instruction
& \texttt{You are a financial sentiment classifier. You will receive a summary of news items for the last 5 days about a stock. Classify the overall sentiment as POSITIVE or NEGATIVE for the stock. Only return one word: POSITIVE or NEGATIVE.} \\

User instruction
& \texttt{Symbol: AAPL} \newline
  \texttt{News for the past 5 days:} \\

News window
& Past 5 days \\

Output constraint
& Only one word is allowed: \texttt{POSITIVE} or \texttt{NEGATIVE}. \\

Filled input prompt
& \texttt{<|im\_start|>system} \newline
  \texttt{You are a financial sentiment classifier. You will receive a summary of news items for the last 5 days about a stock. Classify the overall sentiment as POSITIVE or NEGATIVE for the stock. Only return one word: POSITIVE or NEGATIVE.} \newline
  \texttt{<|im\_end|>} \newline
  \texttt{<|im\_start|>user} \newline
  \texttt{Symbol: AAPL} \newline
  \texttt{News for the past 5 days:} \newline
  \texttt{News on Day 1: Apple reports Q4 results on Thursday with a key first look at iPhone 17 demand. Here's what experts are saying ahead of the report.; Analysts expect steady gains for Amazon, Apple, and Microsoft in Q3, driven by cloud, advertising, and retail performance. Price targets range from 280 to 625.; Big Nuclear Deal} \newline
  \texttt{News on Day 2: Apple trades steady as BofA's Mohan boosts iPhone 17 unit forecasts, raises price target, and models stronger long-term revenue and margins.; Earnings Preview For Apple; Two Nvidia Shockers} \newline
  \texttt{News on Day 3: Apple reports financial results for the fourth quarter on Thursday after the close. Here's a rundown of the iPhone maker's report.; Apple Inc. AAPL will release earnings results on 10/30. Analysts expect earnings of 1.77 per share and quarterly revenue of 102.17 billion.; U.S. stock futures mixed, Amazon, Alphabet, and Microsoft beat earnings expectations, while Apple and Meta post upbeat results.} \newline
  \texttt{News on Day 4: Apple stock hit new all-time highs Friday after reporting fourth-quarter results Thursday. The company is guiding for the first quarter to be even better and record-breaking.; Apple's blockbuster fourth-quarter results prompted bullish reactions across Wall Street, with Wedbush's Dan Ives calling it a pound-the-table moment, CNBC's Jim Cramer saying Apple bears owe us an explanation, and Gene Munster highlighting stronger iPhone demand and a solid holiday outlook.; Nasdaq Surges Over 1 Apple Earnings Top Views} \newline
  \texttt{News on Day 5: Apple, Amazon, Microsoft, and Alphabet beat Q3 estimates with revenue and earnings growth. Trump and Xi conclude high-stakes meeting with trade, soybean, and resource agreements, potentially impacting U.S.-China economic relations and global markets.; As 2026 approaches, Apple is gearing up for a critical year filled with significant product launches and potential executive changes, all while grappling with regulatory and tariff-related challenges.; Apple breaks records in Q4 and ad revenue, Samsung teases iPhone users, and Jamie Dimon echoes Steve Jobs on office meetings.} \newline
  \texttt{<|im\_end|>} \newline
  \texttt{<|im\_start|>assistant} \\

Reasoning
& \texttt{Alright, let's analyze the news for the past five days regarding Apple (AAPL).} \newline
  \texttt{Starting with Day 1, Apple is expected to report strong Q4 results, especially regarding iPhone 17 demand. Experts are optimistic about their performance.} \newline
  \texttt{On Day 2, there's positive news as BofA analyst Mohan boosts his iPhone 17 forecast and increases the price target, suggesting strength in the stock.} \newline
  \texttt{Day 3 mentions that Apple is set to release earnings, with analysts expecting \$1.77 per share and \$102.17 billion in revenue. That's already a positive indicator.} \newline
  \texttt{By Day 4, Apple hits new all-time highs after reporting strong Q4 results. Analysts like Dan Ives and Jim Cramer are bullish, and the company guides for an even better first quarter.} \newline
  \texttt{Looking at Day 5, Apple, along with other big tech companies, beats Q3 estimates. This shows consistent growth and positive momentum.} \newline
  \texttt{Overall, most news points toward positive developments for Apple---strong earnings, increased forecasts, stock highs, and positive market reactions. There doesn't seem to be any major negative news affecting the stock in this period.} \\

Final label
& \texttt{POSITIVE} \\

Purpose of example
& This example illustrates the input/output interface of the Sentiment Analysis Agent under the 5-day setting used in the paper: the model receives a news summary block together with a fixed role instruction, produces a free-form reasoning trace, and finally outputs a single-word polarity label constrained to \texttt{POSITIVE}/\texttt{NEGATIVE}. \\
\bottomrule
\end{tabularx}
\end{table}

%% file: tables/grab_baba_stock.tex
\begin{table}[H]
\centering
\caption{Performance comparison on GRAB and BABA.}
\label{tab:new_assets_appendix}
\footnotesize
\setlength{\tabcolsep}{3pt}
\renewcommand{\arraystretch}{1.0}

\begin{tabular}{@{}llcccccc@{}}
\toprule
\multirow{2}{*}{Category} & \multirow{2}{*}{Model}
& \multicolumn{3}{c}{GRAB}
& \multicolumn{3}{c}{BABA} \\
\cmidrule(lr){3-5} \cmidrule(lr){6-8}
&
& ARR\%$\uparrow$ & SR$\uparrow$ & MDD\%$\downarrow$
& ARR\%$\uparrow$ & SR$\uparrow$ & MDD\%$\downarrow$ \\
\midrule

Market & B\&H
& 64.62 & 0.96 & 24.51
& 73.93 & 1.07 & 25.12 \\
\midrule

\multirow{3}{*}{Rule-based} & MACD
& 35.80 & 0.86 & 15.22
& 46.30 & 1.05 & 22.02 \\
& ZMR
& 68.61 & \best{1.79} & \best{6.46}
& 16.06 & 0.57 & \second{11.26} \\
& SMA
& 28.10 & 1.21 & \second{12.25}
& 86.81 & 1.71 & \best{7.03} \\
\midrule

\multirow{2}{*}{ML/DL} & LSTM
& 58.90 & 0.96 & 24.51
& -17.84 & -0.24 & 26.85 \\
& Transformer
& 3.38 & 0.20 & 24.51
& -19.23 & -0.35 & 26.86 \\
\midrule

\multirow{2}{*}{RL} & DQN
& 26.78 & 0.69 & 18.31
& 16.94 & 0.44 & 20.42 \\
& PPO
& -24.87 & -0.54 & 15.87
& 65.88 & 1.37 & \third{13.39} \\
\midrule

\multirow{4}{*}{General LLMs} & Qwen3-8B
& 60.53 & 1.01 & 19.09
& 96.23 & 1.38 & 19.90 \\
& DeepSeek-R1
& -1.63 & 0.13 & 24.51
& \second{125.81} & \third{1.73} & 17.75 \\
& Llama4-Scout
& -24.72 & -0.32 & 29.65
& 16.34 & 0.43 & 25.90 \\
& GPT5-mini
& -20.99 & -0.26 & 18.31
& 48.71 & 0.99 & 16.29 \\
\midrule

\multirow{4}{*}{Financial LLMs} & FinGPT
& 70.43 & 1.20 & \third{13.27}
& 64.45 & 1.02 & 25.12 \\
& FinAgent
& \second{82.27} & \third{1.25} & 14.48
& \third{95.62} & \second{1.80} & 14.50 \\
& TradingAgents
& -19.42 & -0.33 & 29.65
& 27.20 & 74.31 & 15.39 \\
& DeepFund
& \third{74.31} & 1.16 & 18.15
& 84.90 & 1.43 & 19.22 \\
\midrule

\textbf{Ours} & \textbf{\system}
& \best{105.43} & \second{1.51} & 13.79
& \best{122.27} & \best{2.06} & \second{11.26} \\

\bottomrule
\end{tabular}
\vspace{-2mm}
\end{table}

%% file: tables/longer_horizons.tex
\begin{table}[H]
\centering
\caption{Performance comparison across four assets from different sectors, including AAPL (Technology, ultra-mega-cap), GOOG (Communication Services, ultra-mega-cap), TSLA (Consumer Cyclical, ultra-mega-cap), and BTCUSD (Cryptocurrency). The evaluation period is 2025-04-01 to 2026-03-20, while the original training and validation datasets remain unchanged. \textbf{Bold} indicates the best result in each column, and \underline{underlined} values indicate the second-best result.}
\label{tab:main_results_reduced}
\resizebox{\textwidth}{!}{
\begin{tabular}{llcccccccccccccccc}
\toprule
\multirow{2}{*}{Category} & \multirow{2}{*}{Model}
& \multicolumn{4}{c}{AAPL}
& \multicolumn{4}{c}{GOOG}
& \multicolumn{4}{c}{TSLA}
& \multicolumn{4}{c}{BTCUSD} \\
\cmidrule(lr){3-6} \cmidrule(lr){7-10}
\cmidrule(lr){11-14} \cmidrule(lr){15-18}
&
& CR(\%)$\uparrow$ & ARR(\%)$\uparrow$ & SR$\uparrow$ & MDD(\%)$\downarrow$
& CR(\%)$\uparrow$ & ARR(\%)$\uparrow$ & SR$\uparrow$ & MDD(\%)$\downarrow$
& CR(\%)$\uparrow$ & ARR(\%)$\uparrow$ & SR$\uparrow$ & MDD(\%)$\downarrow$
& CR(\%)$\uparrow$ & ARR(\%)$\uparrow$ & SR$\uparrow$ & MDD(\%)$\downarrow$ \\
\midrule

Market & B\&H
& 11.55 & 12.00 & 0.36 & 22.99
& 92.43 & 97.15 & 2.31 & 13.51
& 41.65 & 43.49 & 0.65 & 22.37
& -16.78 & -12.32 & -0.38 & 47.55 \\
\midrule

Rule-based
& SMA
& -2.13 & -2.21 & -0.15 & \underline{11.88}
& 68.84 & 72.15 & \underline{2.49} & \underline{8.94}
& 13.68 & 14.23 & 0.45 & 18.68
& -15.79 & -11.55 & -0.76 & 27.66 \\
\midrule

ML/DL
& Transformer
& 6.97 & 7.24 & 0.25 & 22.99
& 56.57 & 59.19 & 1.70 & 12.48
& \underline{64.30} & \underline{67.35} & 1.12 & \underline{17.64}
& -14.00 & -10.21 & -0.37 & 46.89 \\
\midrule

RL
& PPO
& 17.73 & 18.45 & \underline{0.67} & \textbf{9.48}
& 41.18 & 43.00 & 1.50 & 9.18
& -1.72 & -1.78 & -0.06 & 23.50
& -15.20 & -11.11 & -0.51 & 39.42 \\
\midrule

General LLMs
& Llama4-Scout-17B
& 3.62 & 3.76 & 0.13 & 18.95
& 77.50 & 81.31 & 2.32 & 9.96
& 10.86 & 11.28 & 0.24 & 29.20
& -3.00 & -2.15 & -0.10 & 30.99 \\
\midrule

\multirow{3}{*}{Financial LLMs}
& FinAgent
& 16.84 & 17.51 & 0.65 & 13.88
& \underline{95.33} & \underline{100.23} & 2.33 & 12.71
& 30.40 & 31.84 & 0.51 & 24.43
& -13.78 & -10.04 & -0.43 & 39.50 \\
& TradingAgents
& -12.04 & -12.46 & -0.53 & 26.25
& 56.50 & 59.12 & 2.00 & 9.76
& 49.15 & 51.38 & \underline{1.15} & 22.64
& \underline{16.33} & \underline{10.11} & \underline{0.76} & \underline{23.38} \\
& DeepFund
& \underline{20.35} & \underline{21.18} & 0.66 & 15.86
& 58.20 & 60.91 & 1.95 & 12.86
& 39.44 & 41.17 & 0.72 & 34.67
& -33.06 & -25.16 & -1.14 & 47.14 \\
\midrule

\textbf{Ours} & \textbf{\system}
& \textbf{28.93} & \textbf{30.15} & \textbf{1.02} & 20.05
& \textbf{103.33} & \textbf{108.75} & \textbf{2.88} & \textbf{8.85}
& \textbf{74.84} & \textbf{78.49} & \textbf{1.47} & \textbf{15.32}
& \textbf{18.06} & \textbf{12.58} & \textbf{0.84} & \textbf{12.90} \\
\midrule

\multicolumn{2}{c}{Improvement(\%)}
& 42.16 & 42.35 & 52.24 & --
& 8.39 & 8.50 & 15.66 & 1.01
& 16.39 & 16.54 & 27.83 & 13.15
& 10.59 & 24.43 & 10.53 & 44.82 \\

\bottomrule
\end{tabular}
}
\vspace{-2mm}
\end{table}

%% file: tables/broader_asset_pool.tex
\begin{table}[H]
\centering
\caption{Performance comparison on the five newly added assets KO (Consumer Defensive, Mega-cap), JNJ (Healthcare, Mega-cap), UPS (Industrials, Large-cap), UPWK (Communication Services, Small-cap), and SFM (Consumer Defensive, Mid-cap). The evaluation
period is 2025-04-01 to 2026-03-20, while the original training and validation datasets remain unchanged. \textbf{Bold} indicates the best result in each column, and \underline{underlined} values indicate the second-best result.}
\label{tab:extended_results}
\resizebox{\textwidth}{!}{
\begin{tabular}{llcccccccccccc}
\toprule
\multirow{2}{*}{Category} & \multirow{2}{*}{Model}
& \multicolumn{4}{c}{KO}
& \multicolumn{4}{c}{JNJ}
& \multicolumn{4}{c}{UPS} \\
\cmidrule(lr){3-6} \cmidrule(lr){7-10} \cmidrule(lr){11-14}
&
& CR(\%)$\uparrow$ & ARR(\%)$\uparrow$ & SR$\uparrow$ & MDD(\%)$\downarrow$
& CR(\%)$\uparrow$ & ARR(\%)$\uparrow$ & SR$\uparrow$ & MDD(\%)$\downarrow$
& CR(\%)$\uparrow$ & ARR(\%)$\uparrow$ & SR$\uparrow$ & MDD(\%)$\downarrow$ \\
\midrule

Market & B\&H
& 7.02 & 7.26 & 0.42 & 9.82
& \underline{57.86} & \underline{59.76} & 2.68 & 8.42
& -6.27 & -6.47 & -0.22 & 22.39 \\
\midrule

Rule-based
& SMA
& 3.47 & 3.60 & 0.32 & 7.62
& 36.50 & 38.08 & 2.40 & \underline{5.17}
& 6.00 & 6.23 & 0.27 & 23.80 \\
\midrule

ML/DL
& Transformer
& \underline{15.50} & \underline{16.12} & \underline{1.02} & \underline{6.73}
& 41.29 & 43.11 & 2.37 & 8.42
& -3.45 & -3.58 & -0.13 & 17.97 \\
\midrule

RL
& PPO
& -11.93 & -12.34 & -0.96 & 17.92
& 12.62 & 13.12 & 1.07 & 6.24
& -3.04 & -3.15 & -0.18 & \textbf{11.84} \\
\midrule

General LLMs
& Llama4-Scout-17B
& -2.62 & -2.71 & -0.22 & 13.03
& 50.94 & 53.26 & 2.57 & 8.67
& -15.11 & -15.62 & -0.65 & 26.36 \\
\midrule

\multirow{3}{*}{Financial LLMs}
& FinAgent
& 7.31 & 7.59 & 0.45 & 10.40
& 55.98 & 58.57 & 2.60 & 8.42
& \underline{12.61} & \underline{13.27} & \underline{0.50} & \underline{17.30} \\
& TradingAgents
& -2.30 & -2.40 & -0.35 & \textbf{5.83}
& 25.61 & 26.67 & \textbf{3.25} & \textbf{2.64}
& -15.98 & -16.53 & -0.81 & 21.17 \\
& DeepFund
& -5.72 & -5.95 & -0.46 & 13.36
& 22.43 & 23.68 & 1.42 & 10.25
& -0.19 & -0.20 & -0.01 & 22.46 \\
\midrule

\textbf{Ours} & \textbf{\system}
& \textbf{24.95} & \textbf{25.98} & \textbf{1.54} & 6.92
& \textbf{63.31} & \textbf{66.31} & \underline{2.92} & 9.70
& \textbf{14.50} & \textbf{15.08} & \textbf{0.63} & 20.09 \\
\midrule

\multicolumn{2}{c}{Improvement(\%)}
& 60.91 & 61.16 & 51.99 & --
& 9.42 & 10.95 & -- & --
& 15.04 & 13.61 & 23.98 & -- \\

\bottomrule
\end{tabular}
}
\vspace{-2mm}
\end{table}

\begin{table}[H]
\vspace{-2mm}
\centering
\resizebox{\textwidth}{!}{
\begin{tabular}{llcccccccc}
\toprule
\multirow{2}{*}{Category} & \multirow{2}{*}{Model}
& \multicolumn{4}{c}{UPWK}
& \multicolumn{4}{c}{SFM} \\
\cmidrule(lr){3-6} \cmidrule(lr){7-10}
&
& CR(\%)$\uparrow$ & ARR(\%)$\uparrow$ & SR$\uparrow$ & MDD(\%)$\downarrow$
& CR(\%)$\uparrow$ & ARR(\%)$\uparrow$ & SR$\uparrow$ & MDD(\%)$\downarrow$ \\
\midrule

Market & B\&H
& -13.29 & -13.75 & -0.25 & 48.67
& -45.56 & -46.77 & -1.35 & 63.48 \\
\midrule

Rule-based
& SMA
& \underline{10.65} & \underline{11.07} & \underline{0.49} & \underline{26.48}
& -17.12 & -17.70 & -1.10 & \underline{31.09} \\
\midrule

ML/DL
& Transformer
& -9.93 & -10.28 & -0.23 & 45.50
& -18.21 & -18.82 & -0.73 & 42.13 \\
\midrule

RL
& PPO
& -13.81 & -14.28 & -0.39 & 34.56
& -12.42 & -12.85 & -0.62 & 35.00 \\
\midrule

General LLMs
& Llama4-Scout-17B
& -18.61 & -19.23 & -0.58 & 34.56
& -34.55 & -35.57 & -1.02 & 54.93 \\
\midrule

\multirow{3}{*}{Financial LLMs}
& FinAgent
& 4.69 & 4.87 & 0.36 & 45.31
& -24.66 & -25.45 & -0.87 & 50.45 \\
& TradingAgents
& -14.51 & -15.00 & -0.37 & 39.27
& -54.33 & -55.63 & -2.05 & 61.67 \\
& DeepFund
& -14.88 & -15.33 & -0.32 & 46.32
& \underline{1.74} & \underline{1.80} & \underline{0.20} & 40.05 \\
\midrule

\textbf{Ours} & \textbf{\system}
& \textbf{22.04} & \textbf{22.94} & \textbf{0.83} & \textbf{14.68}
& \textbf{8.02} & \textbf{8.33} & \textbf{0.61} & \textbf{19.73} \\
\midrule

\multicolumn{2}{c}{Improvement(\%)}
& 106.95 & 107.23 & 69.39 & 44.56
& 360.92 & 362.78 & 205.00 & 36.54 \\

\bottomrule
\end{tabular}
}
\vspace{-2mm}
\end{table}

%% file: tables/random_seed.tex

\begin{table}[H]
\centering
\scriptsize
\caption{Random-seed variance analysis on AAPL and BTCUSD. ARR and MDD are reported in percentage (\%), and SR is reported as a unitless ratio.}
\label{tab:seed_variance_raw}
\setlength{\tabcolsep}{4pt}
\renewcommand{\arraystretch}{1.08}
\begin{tabular}{llcccccc}
\toprule
\multirow{2}{*}{Method} & \multirow{2}{*}{Seed} 
& \multicolumn{3}{c}{AAPL} 
& \multicolumn{3}{c}{BTCUSD} \\
\cmidrule(lr){3-5} \cmidrule(lr){6-8}
& 
& ARR(\%) $\uparrow$ & SR $\uparrow$ & MDD(\%) $\downarrow$
& ARR(\%) $\uparrow$ & SR $\uparrow$ & MDD(\%) $\downarrow$ \\
\midrule

\multirow{3}{*}{\system}
& 42 & 50.0757 & 1.2201 & 7.8252 & 53.5700 & 1.5200 & 9.7869 \\
& 43 & 47.5979 & 1.2088 & 7.8247 & 46.1502 & 1.4039 & 9.7869 \\
& 44 & 55.5921 & 1.1998 & 8.4610 & 51.2787 & 1.5714 & 9.7919 \\
\midrule

\multirow{3}{*}{Transformer}
& 42 & 32.9000 & 0.7200 & 13.7700 & 26.3500 & 1.1800 & 8.2600 \\
& 43 & 22.8142 & 0.5398 & 11.3903 & 34.9831 & 1.0527 & 9.7869 \\
& 44 & 37.8899 & 0.7937 & 11.9059 & 25.0519 & 0.9280 & 7.9304 \\
\midrule

\multirow{3}{*}{Concat. Fusion}
& 42 & 19.2080 & 0.3257 & 22.9865 & 7.1844  & 0.3963 & 10.7631 \\
& 43 & -3.3172 & -0.3395 & 3.3304  & 15.0368 & 0.5926 & 9.1551 \\
& 44 & 28.0350 & 0.4532  & 22.9865 & 2.9817  & 0.2276 & 10.0582 \\
\bottomrule
\end{tabular}
\end{table}

\begin{table}[H]
\centering
\scriptsize
\caption{Results over 3 random seeds on AAPL and BTCUSD. We report mean $\pm$ standard deviation. ARR and MDD are reported in percentage (\%).}
\label{tab:seed_variance_mean_std}
\setlength{\tabcolsep}{4pt}
\renewcommand{\arraystretch}{1.08}
\begin{tabular}{lcccccc}
\toprule
\multirow{2}{*}{Method} 
& \multicolumn{3}{c}{AAPL} 
& \multicolumn{3}{c}{BTCUSD} \\
\cmidrule(lr){2-4} \cmidrule(lr){5-7}
& ARR(\%) $\uparrow$ & SR $\uparrow$ & MDD(\%) $\downarrow$
& ARR(\%) $\uparrow$ & SR $\uparrow$ & MDD(\%) $\downarrow$ \\
\midrule
Concat. Fusion 
& 14.6419 $\pm$ 16.1672 
& 0.1465 $\pm$ 0.4257 
& 16.4345 $\pm$ 11.3485
& 8.4010 $\pm$ 6.1189 
& 0.4055 $\pm$ 0.1827 
& 9.9921 $\pm$ 0.8060 \\
Transformer 
& 31.2014 $\pm$ 7.6801 
& 0.6845 $\pm$ 0.1306 
& 12.3554 $\pm$ 1.2519
& 28.7950 $\pm$ 5.3982 
& 1.0536 $\pm$ 0.1260 
& 8.6591 $\pm$ 0.9905 \\
\system
& \textbf{51.0886 $\pm$ 4.0922} 
& \textbf{1.2096 $\pm$ 0.0102} 
& \textbf{8.0370 $\pm$ 0.3672}
& \textbf{50.3330 $\pm$ 3.7992} 
& \textbf{1.4984 $\pm$ 0.0858} 
& 9.7886 $\pm$ 0.0029 \\
\bottomrule
\end{tabular}
\end{table}

\begin{table}[H]
\centering
\scriptsize
\caption{Statistical significance analysis over 3 random seeds. We report the mean ARR improvement of \system{} over each baseline, together with the 95\% confidence interval (CI) and paired t-test p-value. ARR improvements are reported in percentage points.}
\label{tab:significance_arr}
\setlength{\tabcolsep}{4.5pt}
\renewcommand{\arraystretch}{1.08}
\begin{tabular}{llccc}
\toprule
Asset & Comparison & ARR(\%) Improvement $\uparrow$ & 95\% CI & p-value \\
\midrule
AAPL   & \system{} vs Transformer      & +19.8872 & [9.3330, 30.4414]  & 0.0149 \\
AAPL   & \system{} vs Concat. Fusion   & +36.4466 & [5.0498, 67.8435]  & 0.0378 \\
BTCUSD & \system{} vs Transformer      & +21.5380 & [-0.8072, 43.8832] & 0.0535 \\
BTCUSD & \system{} vs Concat. Fusion   & +41.9320 & [18.5369, 65.3271] & 0.0164 \\
\bottomrule
\end{tabular}
\end{table}

%% file: tables/smp_task.tex
\begin{table}[htbp]
\centering
\caption{Performance comparison of \system{} with and without the MaF module across different stocks.}
\label{tab:acc_mcc_all}
\footnotesize
\setlength{\tabcolsep}{4pt}
\renewcommand{\arraystretch}{0.95}

\begin{tabular}{lcccccc}
\toprule
\multirow{2}{*}{\textbf{Method}} 
& \multicolumn{2}{c}{\textbf{AAPL}} 
& \multicolumn{2}{c}{\textbf{AMZN}} 
& \multicolumn{2}{c}{\textbf{GOOG}} \\ 
\cmidrule(lr){2-3}
\cmidrule(lr){4-5}
\cmidrule(lr){6-7} 
& ACC(\%) & MCC 
& ACC(\%) & MCC 
& ACC(\%) & MCC \\ 
\midrule 
\system{} w/o MaF 
& 53.02 & 0.0431 
& 50.30 & -0.0044 
& 52.71 & 0.0533 \\ 

\system{} 
& 57.94 & 0.2122 
& 55.56 & 0.1055 
& 54.76 & 0.0889 \\ 

\midrule 
\multirow{2}{*}{\textbf{Method}} 
& \multicolumn{2}{c}{\textbf{MSFT}} 
& \multicolumn{2}{c}{\textbf{TSLA}} 
& \multicolumn{2}{c}{\textbf{BTCUSD}} \\ 
\cmidrule(lr){2-3}
\cmidrule(lr){4-5}
\cmidrule(lr){6-7} 
& ACC(\%) & MCC 
& ACC(\%) & MCC 
& ACC(\%) & MCC \\ 
\midrule 
\system{} w/o MaF 
& 51.70 & 0.0159 
& 50.60 & 0.0191 
& 51.67 & 0.0261 \\ 

\system{} 
& 57.14 & 0.1840 
& 56.35 & 0.0778 
& 53.55 & 0.0599 \\ 
\bottomrule 
\end{tabular}
\vspace{-2mm}
\end{table}

%% file: main/main.bbl
\begin{thebibliography}{57}
\providecommand{\natexlab}[1]{#1}
\providecommand{\url}[1]{\texttt{#1}}
\expandafter\ifx\csname urlstyle\endcsname\relax
  \providecommand{\doi}[1]{doi: #1}\else
  \providecommand{\doi}{doi: \begingroup \urlstyle{rm}\Url}\fi

\bibitem[BehnamGhader et~al.(2024)BehnamGhader, Adlakha, Mosbach, Bahdanau, Chapados, and Reddy]{behnamghader2024llm2vec}
Parishad BehnamGhader, Vaibhav Adlakha, Marius Mosbach, Dzmitry Bahdanau, Nicolas Chapados, and Siva Reddy.
\newblock Llm2vec: Large language models are secretly powerful text encoders.
\newblock \emph{arXiv preprint arXiv:2404.05961}, 2024.

\bibitem[Brock et~al.(1992)Brock, Lakonishok, and LeBaron]{brock1992simple}
William Brock, Josef Lakonishok, and Blake LeBaron.
\newblock Simple technical trading rules and the stochastic properties of stock returns.
\newblock \emph{The Journal of finance}, 47\penalty0 (5):\penalty0 1731--1764, 1992.

\bibitem[Cavalcante et~al.(2016)Cavalcante, Brasileiro, Souza, Nobrega, and Oliveira]{cavalcante2016computational}
Rodolfo~C Cavalcante, Rodrigo~C Brasileiro, Victor~LF Souza, Jarley~P Nobrega, and Adriano~LI Oliveira.
\newblock Computational intelligence and financial markets: A survey and future directions.
\newblock \emph{Expert Systems with Applications}, 55:\penalty0 194--211, 2016.

\bibitem[Chen et~al.(2023)Chen, Zheng, Lu, Yuan, and Zhu]{chen2023chatgpt}
Zihan Chen, Lei~Nico Zheng, Cheng Lu, Jialu Yuan, and Di~Zhu.
\newblock Chatgpt informed graph neural network for stock movement prediction.
\newblock \emph{arXiv preprint arXiv:2306.03763}, 2023.

\bibitem[Cont(2001)]{cont2001empirical}
Rama Cont.
\newblock Empirical properties of asset returns: stylized facts and statistical issues.
\newblock \emph{Quantitative finance}, 1\penalty0 (2):\penalty0 223, 2001.

\bibitem[Ding et~al.(2015)Ding, Zhang, Liu, and Duan]{ding2015deep}
Xiao Ding, Yue Zhang, Ting Liu, and Junwen Duan.
\newblock Deep learning for event-driven stock prediction.
\newblock In \emph{Ijcai}, volume~15, pp.\  2327--2333, 2015.

\bibitem[El-Baz et~al.(2013)El-Baz, Al~Awadhi, and Lasfer]{el2013sma}
Hazim El-Baz, Ibrahim Al~Awadhi, and Assia Lasfer.
\newblock Sma and macd combinations for stock investment decisions in frontier markets: evidence from dubai financial market.
\newblock \emph{International Journal of Financial Engineering and Risk Management}, 1\penalty0 (2):\penalty0 113--128, 2013.

\bibitem[Fama(1970)]{fama1970efficient}
Eugene~F Fama.
\newblock Efficient capital markets: A review of theory and empirical work.
\newblock \emph{The journal of Finance}, 25\penalty0 (2):\penalty0 383--417, 1970.

\bibitem[Feng et~al.(2019)Feng, He, Wang, Luo, Liu, and Chua]{feng2019temporal}
Fuli Feng, Xiangnan He, Xiang Wang, Cheng Luo, Yiqun Liu, and Tat-Seng Chua.
\newblock Temporal relational ranking for stock prediction.
\newblock \emph{ACM Transactions on Information Systems (TOIS)}, 37\penalty0 (2):\penalty0 1--30, 2019.

\bibitem[Feng et~al.(2021)Feng, Li, Luo, Ng, and Chua]{feng2021hybrid}
Fuli Feng, Moxin Li, Cheng Luo, Ritchie Ng, and Tat-Seng Chua.
\newblock Hybrid learning to rank for financial event ranking.
\newblock In \emph{Proceedings of the 44th International ACM SIGIR Conference on Research and Development in Information Retrieval}, pp.\  233--243, 2021.

\bibitem[Fischer \& Krauss(2018)Fischer and Krauss]{fischer2018deep}
Thomas Fischer and Christopher Krauss.
\newblock Deep learning with long short-term memory networks for financial market predictions.
\newblock \emph{European journal of operational research}, 270\penalty0 (2):\penalty0 654--669, 2018.

\bibitem[Grattafiori et~al.(2024)Grattafiori, Dubey, Jauhri, Pandey, Kadian, Al-Dahle, Letman, Mathur, Schelten, Vaughan, et~al.]{grattafiori2024llama}
Aaron Grattafiori, Abhimanyu Dubey, Abhinav Jauhri, Abhinav Pandey, Abhishek Kadian, Ahmad Al-Dahle, Aiesha Letman, Akhil Mathur, Alan Schelten, Alex Vaughan, et~al.
\newblock The llama 3 herd of models.
\newblock \emph{arXiv preprint arXiv:2407.21783}, 2024.

\bibitem[Gruver et~al.(2023)Gruver, Finzi, Qiu, and Wilson]{gruver2023large}
Nate Gruver, Marc Finzi, Shikai Qiu, and Andrew~G Wilson.
\newblock Large language models are zero-shot time series forecasters.
\newblock \emph{Advances in Neural Information Processing Systems}, 36:\penalty0 19622--19635, 2023.

\bibitem[Guo et~al.(2025)Guo, Yang, Zhang, Song, Zhang, Xu, Zhu, Ma, Wang, Bi, et~al.]{guo2025deepseek}
Daya Guo, Dejian Yang, Haowei Zhang, Junxiao Song, Ruoyu Zhang, Runxin Xu, Qihao Zhu, Shirong Ma, Peiyi Wang, Xiao Bi, et~al.
\newblock Deepseek-r1: Incentivizing reasoning capability in llms via reinforcement learning.
\newblock \emph{arXiv preprint arXiv:2501.12948}, 2025.

\bibitem[Hansen \& Kazinnik(2024)Hansen and Kazinnik]{hansen2024can}
Anne~Lundgaard Hansen and Sophia Kazinnik.
\newblock Can chatgpt decipher fedspeak?
\newblock \emph{Available at SSRN 4399406}, 2024.

\bibitem[Huang \& Wang(2025)Huang and Wang]{huang2025explainable}
Donghao Huang and Zhaoxia Wang.
\newblock Explainable sentiment analysis with deepseek-r1: Performance, efficiency, and few-shot learning.
\newblock \emph{IEEE Intelligent Systems}, 2025.

\bibitem[Huang et~al.(2023)Huang, Dong, Wang, Hao, Singhal, Ma, Lv, Cui, Mohammed, Patra, et~al.]{huang2023language}
Shaohan Huang, Li~Dong, Wenhui Wang, Yaru Hao, Saksham Singhal, Shuming Ma, Tengchao Lv, Lei Cui, Owais~Khan Mohammed, Barun Patra, et~al.
\newblock Language is not all you need: Aligning perception with language models.
\newblock \emph{Advances in Neural Information Processing Systems}, 36:\penalty0 72096--72109, 2023.

\bibitem[Hurst et~al.(2024)Hurst, Lerer, Goucher, Perelman, Ramesh, Clark, Ostrow, Welihinda, Hayes, Radford, et~al.]{hurst2024gpt}
Aaron Hurst, Adam Lerer, Adam~P Goucher, Adam Perelman, Aditya Ramesh, Aidan Clark, AJ~Ostrow, Akila Welihinda, Alan Hayes, Alec Radford, et~al.
\newblock Gpt-4o system card.
\newblock \emph{arXiv preprint arXiv:2410.21276}, 2024.

\bibitem[Jiang et~al.(2017)Jiang, Xu, and Liang]{jiang2017deep}
Zhengyao Jiang, Dixing Xu, and Jinjun Liang.
\newblock A deep reinforcement learning framework for the financial portfolio management problem.
\newblock \emph{arXiv preprint arXiv:1706.10059}, 2017.

\bibitem[Kakushadze(2016)]{kakushadze2016101}
Zura Kakushadze.
\newblock 101 formulaic alphas.
\newblock \emph{Wilmott}, 2016\penalty0 (84):\penalty0 72--81, 2016.

\bibitem[Koa et~al.(2023)Koa, Ma, Ng, and Chua]{koa2023diffusion}
Kelvin~JL Koa, Yunshan Ma, Ritchie Ng, and Tat-Seng Chua.
\newblock Diffusion variational autoencoder for tackling stochasticity in multi-step regression stock price prediction.
\newblock In \emph{Proceedings of the 32nd ACM International Conference on Information and Knowledge Management}, pp.\  1087--1096, 2023.

\bibitem[Li et~al.(2025)Li, Shi, Wang, Duan, Ruan, Huang, Long, Huang, Tang, and Luo]{li2025time}
Changlun Li, Yao Shi, Chen Wang, Qiqi Duan, Runke Ruan, Weijie Huang, Haonan Long, Lijun Huang, Nan Tang, and Yuyu Luo.
\newblock Time travel is cheating: Going live with deepfund for real-time fund investment benchmarking.
\newblock \emph{arXiv preprint arXiv:2505.11065}, 2025.

\bibitem[Li et~al.(2021)Li, Bao, Harimoto, Chen, Xu, and Su]{li2021modeling}
Wei Li, Ruihan Bao, Keiko Harimoto, Deli Chen, Jingjing Xu, and Qi~Su.
\newblock Modeling the stock relation with graph network for overnight stock movement prediction.
\newblock In \emph{Proceedings of the twenty-ninth international conference on international joint conferences on artificial intelligence}, pp.\  4541--4547, 2021.

\bibitem[Liang et~al.(2022)Liang, Lyu, Fan, Tsaw, Liu, Mo, Yogatama, Morency, and Salakhutdinov]{liang2022high}
Paul~Pu Liang, Yiwei Lyu, Xiang Fan, Jeffrey Tsaw, Yudong Liu, Shentong Mo, Dani Yogatama, Louis-Philippe Morency, and Ruslan Salakhutdinov.
\newblock High-modality multimodal transformer: Quantifying modality \& interaction heterogeneity for high-modality representation learning.
\newblock \emph{arXiv preprint arXiv:2203.01311}, 2022.

\bibitem[Liu et~al.(2020)Liu, Yang, Chen, Zhang, Yang, Xiao, and Wang]{liu2020finrl}
Xiao-Yang Liu, Hongyang Yang, Qian Chen, Runjia Zhang, Liuqing Yang, Bowen Xiao, and Christina~Dan Wang.
\newblock Finrl: A deep reinforcement learning library for automated stock trading in quantitative finance.
\newblock \emph{arXiv preprint arXiv:2011.09607}, 2020.

\bibitem[Liu et~al.(2023)Liu, Wang, Yang, and Zha]{liu2023fingpt}
Xiao-Yang Liu, Guoxuan Wang, Hongyang Yang, and Daochen Zha.
\newblock Fingpt: Democratizing internet-scale data for financial large language models.
\newblock \emph{arXiv preprint arXiv:2307.10485}, 2023.

\bibitem[Lopez-Lira \& Tang(2023)Lopez-Lira and Tang]{lopez2023can}
Alejandro Lopez-Lira and Yuehua Tang.
\newblock Can chatgpt forecast stock price movements? return predictability and large language models.
\newblock \emph{arXiv preprint arXiv:2304.07619}, 2023.

\bibitem[Mnih et~al.(2013)Mnih, Kavukcuoglu, Silver, Graves, Antonoglou, Wierstra, and Riedmiller]{mnih2013playing}
Volodymyr Mnih, Koray Kavukcuoglu, David Silver, Alex Graves, Ioannis Antonoglou, Daan Wierstra, and Martin Riedmiller.
\newblock Playing atari with deep reinforcement learning.
\newblock \emph{arXiv preprint arXiv:1312.5602}, 2013.

\bibitem[Mu \& Lin(2025)Mu and Lin]{mu2025comprehensive}
Siyuan Mu and Sen Lin.
\newblock A comprehensive survey of mixture-of-experts: Algorithms, theory, and applications.
\newblock \emph{arXiv preprint arXiv:2503.07137}, 2025.

\bibitem[Murphy(1999)]{murphy1999technical}
John~J Murphy.
\newblock \emph{Technical analysis of the financial markets: A comprehensive guide to trading methods and applications}.
\newblock Penguin, 1999.

\bibitem[Nelson et~al.(2017)Nelson, Pereira, and De~Oliveira]{nelson2017stock}
David~MQ Nelson, Adriano~CM Pereira, and Renato~A De~Oliveira.
\newblock Stock market's price movement prediction with lstm neural networks.
\newblock In \emph{2017 International joint conference on neural networks (IJCNN)}, pp.\  1419--1426. Ieee, 2017.

\bibitem[Nti et~al.(2020)Nti, Adekoya, and Weyori]{nti2020systematic}
Isaac~Kofi Nti, Adebayo~Felix Adekoya, and Benjamin~Asubam Weyori.
\newblock A systematic review of fundamental and technical analysis of stock market predictions.
\newblock \emph{Artificial Intelligence Review}, 53\penalty0 (4):\penalty0 3007--3057, 2020.

\bibitem[Patel et~al.(2015)Patel, Shah, Thakkar, and Kotecha]{patel2015predicting}
Jigar Patel, Sahil Shah, Priyank Thakkar, and Ketan Kotecha.
\newblock Predicting stock and stock price index movement using trend deterministic data preparation and machine learning techniques.
\newblock \emph{Expert systems with applications}, 42\penalty0 (1):\penalty0 259--268, 2015.

\bibitem[Rehman et~al.(2025)Rehman, Sanyal, and Chattopadhyay]{rehman2025green}
Tohida Rehman, Debarshi~Kumar Sanyal, and Samiran Chattopadhyay.
\newblock How green are neural language models? analyzing energy consumption in text summarization fine-tuning.
\newblock \emph{arXiv preprint arXiv:2501.15398}, 2025.

\bibitem[Schulman et~al.(2017)Schulman, Wolski, Dhariwal, Radford, and Klimov]{schulman2017proximal}
John Schulman, Filip Wolski, Prafulla Dhariwal, Alec Radford, and Oleg Klimov.
\newblock Proximal policy optimization algorithms.
\newblock \emph{arXiv preprint arXiv:1707.06347}, 2017.

\bibitem[Sezer et~al.(2020)Sezer, Gudelek, and Ozbayoglu]{sezer2020financial}
Omer~Berat Sezer, Mehmet~Ugur Gudelek, and Ahmet~Murat Ozbayoglu.
\newblock Financial time series forecasting with deep learning: A systematic literature review: 2005--2019.
\newblock \emph{Applied soft computing}, 90:\penalty0 106181, 2020.

\bibitem[Shi et~al.(2023)Shi, Chen, Misra, Scales, Dohan, Chi, Sch{\"a}rli, and Zhou]{shi2023large}
Freda Shi, Xinyun Chen, Kanishka Misra, Nathan Scales, David Dohan, Ed~H Chi, Nathanael Sch{\"a}rli, and Denny Zhou.
\newblock Large language models can be easily distracted by irrelevant context.
\newblock In \emph{International Conference on Machine Learning}, pp.\  31210--31227. PMLR, 2023.

\bibitem[Sun et~al.(2023)Sun, Qin, Zhang, Xia, Zong, Ying, Xie, Zhao, Wang, and An]{sun2023trademaster}
Shuo Sun, Molei Qin, Wentao Zhang, Haochong Xia, Chuqiao Zong, Jie Ying, Yonggang Xie, Lingxuan Zhao, Xinrun Wang, and Bo~An.
\newblock Trademaster: A holistic quantitative trading platform empowered by reinforcement learning.
\newblock \emph{Advances in Neural Information Processing Systems}, 36:\penalty0 59047--59061, 2023.

\bibitem[Sutton et~al.(1998)Sutton, Barto, et~al.]{sutton1998reinforcement}
Richard~S Sutton, Andrew~G Barto, et~al.
\newblock \emph{Reinforcement learning: An introduction}, volume~1.
\newblock MIT press Cambridge, 1998.

\bibitem[Team et~al.(2024)]{team2024qwen2}
Qwen Team et~al.
\newblock Qwen2 technical report.
\newblock \emph{arXiv preprint arXiv:2407.10671}, 2\penalty0 (3), 2024.

\bibitem[Tetlock(2007)]{tetlock2007giving}
Paul~C Tetlock.
\newblock Giving content to investor sentiment: The role of media in the stock market.
\newblock \emph{The Journal of finance}, 62\penalty0 (3):\penalty0 1139--1168, 2007.

\bibitem[Wei et~al.(2022)Wei, Wang, Schuurmans, Bosma, Xia, Chi, Le, Zhou, et~al.]{wei2022chain}
Jason Wei, Xuezhi Wang, Dale Schuurmans, Maarten Bosma, Fei Xia, Ed~Chi, Quoc~V Le, Denny Zhou, et~al.
\newblock Chain-of-thought prompting elicits reasoning in large language models.
\newblock \emph{Advances in neural information processing systems}, 35:\penalty0 24824--24837, 2022.

\bibitem[Wu et~al.(2023)Wu, Irsoy, Lu, Dabravolski, Dredze, Gehrmann, Kambadur, Rosenberg, and Mann]{wu2023bloomberggpt}
Shijie Wu, Ozan Irsoy, Steven Lu, Vadim Dabravolski, Mark Dredze, Sebastian Gehrmann, Prabhanjan Kambadur, David Rosenberg, and Gideon Mann.
\newblock Bloomberggpt: A large language model for finance.
\newblock \emph{arXiv preprint arXiv:2303.17564}, 2023.

\bibitem[Xiao et~al.(2024)Xiao, Sun, Luo, and Wang]{xiao2024tradingagents}
Yijia Xiao, Edward Sun, Di~Luo, and Wei Wang.
\newblock Tradingagents: Multi-agents llm financial trading framework.
\newblock \emph{arXiv preprint arXiv:2412.20138}, 2024.

\bibitem[Xu \& Cohen(2018)Xu and Cohen]{xu2018stock}
Yumo Xu and Shay~B Cohen.
\newblock Stock movement prediction from tweets and historical prices.
\newblock In \emph{Proceedings of the 56th Annual Meeting of the Association for Computational Linguistics (Volume 1: Long Papers)}, pp.\  1970--1979, 2018.

\bibitem[Yang et~al.(2025)Yang, Li, Yang, Zhang, Hui, Zheng, Yu, Gao, Huang, Lv, et~al.]{yang2025qwen3}
An~Yang, Anfeng Li, Baosong Yang, Beichen Zhang, Binyuan Hui, Bo~Zheng, Bowen Yu, Chang Gao, Chengen Huang, Chenxu Lv, et~al.
\newblock Qwen3 technical report.
\newblock \emph{arXiv preprint arXiv:2505.09388}, 2025.

\bibitem[Yang et~al.(2020)Yang, Liu, Zhou, Bian, and Liu]{yang2020qlib}
Xiao Yang, Weiqing Liu, Dong Zhou, Jiang Bian, and Tie-Yan Liu.
\newblock Qlib: An ai-oriented quantitative investment platform.
\newblock \emph{arXiv preprint arXiv:2009.11189}, 2020.

\bibitem[Yu et~al.(2024)Yu, Yao, Li, Deng, Jiang, Cao, Chen, Suchow, Cui, Liu, et~al.]{yu2024fincon}
Yangyang Yu, Zhiyuan Yao, Haohang Li, Zhiyang Deng, Yuechen Jiang, Yupeng Cao, Zhi Chen, Jordan~W Suchow, Zhenyu Cui, Rong Liu, et~al.
\newblock Fincon: A synthesized llm multi-agent system with conceptual verbal reinforcement for enhanced financial decision making.
\newblock \emph{Advances in Neural Information Processing Systems}, 37:\penalty0 137010--137045, 2024.

\bibitem[Yu et~al.(2025)Yu, Li, Chen, Jiang, Li, Suchow, Zhang, and Khashanah]{yu2025finmem}
Yangyang Yu, Haohang Li, Zhi Chen, Yuechen Jiang, Yang Li, Jordan~W Suchow, Denghui Zhang, and Khaldoun Khashanah.
\newblock Finmem: A performance-enhanced llm trading agent with layered memory and character design.
\newblock \emph{IEEE Transactions on Big Data}, 2025.

\bibitem[Zerveas et~al.(2021)Zerveas, Jayaraman, Patel, Bhamidipaty, and Eickhoff]{zerveas2021transformer}
George Zerveas, Srideepika Jayaraman, Dhaval Patel, Anuradha Bhamidipaty, and Carsten Eickhoff.
\newblock A transformer-based framework for multivariate time series representation learning.
\newblock In \emph{Proceedings of the 27th ACM SIGKDD conference on knowledge discovery \& data mining}, pp.\  2114--2124, 2021.

\bibitem[Zhang et~al.(2024{\natexlab{a}})Zhang, Luan, Hu, Lee, Qiao, Chen, Su, and Chang]{zhang2024magiclens}
Kai Zhang, Yi~Luan, Hexiang Hu, Kenton Lee, Siyuan Qiao, Wenhu Chen, Yu~Su, and Ming-Wei Chang.
\newblock Magiclens: Self-supervised image retrieval with open-ended instructions.
\newblock \emph{arXiv preprint arXiv:2403.19651}, 2024{\natexlab{a}}.

\bibitem[Zhang et~al.(2024{\natexlab{b}})Zhang, Zhao, Xia, Sun, Sun, Qin, Li, Zhao, Zhao, Cai, et~al.]{zhang2024multimodal}
Wentao Zhang, Lingxuan Zhao, Haochong Xia, Shuo Sun, Jiaze Sun, Molei Qin, Xinyi Li, Yuqing Zhao, Yilei Zhao, Xinyu Cai, et~al.
\newblock A multimodal foundation agent for financial trading: Tool-augmented, diversified, and generalist.
\newblock In \emph{Proceedings of the 30th acm sigkdd conference on knowledge discovery and data mining}, pp.\  4314--4325, 2024{\natexlab{b}}.

\bibitem[Zhang \& Yan(2023)Zhang and Yan]{zhang2023crossformer}
Yunhao Zhang and Junchi Yan.
\newblock Crossformer: Transformer utilizing cross-dimension dependency for multivariate time series forecasting.
\newblock In \emph{The eleventh international conference on learning representations}, 2023.

\bibitem[Zhao et~al.(2023)Zhao, Kong, and Shen]{zhao2023doubleadapt}
Lifan Zhao, Shuming Kong, and Yanyan Shen.
\newblock Doubleadapt: A meta-learning approach to incremental learning for stock trend forecasting.
\newblock In \emph{Proceedings of the 29th ACM SIGKDD Conference on Knowledge Discovery and Data Mining}, pp.\  3492--3503, 2023.

\bibitem[Zhao et~al.(2021)Zhao, Wallace, Feng, Klein, and Singh]{zhao2021calibrate}
Zihao Zhao, Eric Wallace, Shi Feng, Dan Klein, and Sameer Singh.
\newblock Calibrate before use: Improving few-shot performance of language models.
\newblock In \emph{International conference on machine learning}, pp.\  12697--12706. PMLR, 2021.

\bibitem[Zhu et~al.(2025)Zhu, Ng, Liu, Liu, Zeng, Wang, Tan, Yao, Shao, Xu, et~al.]{zhu2025findeepresearch}
Fengbin Zhu, Xiang~Yao Ng, Ziyang Liu, Chang Liu, Xianwei Zeng, Chao Wang, Tianhui Tan, Xuan Yao, Pengyang Shao, Min Xu, et~al.
\newblock Findeepresearch: Evaluating deep research agents in rigorous financial analysis.
\newblock \emph{arXiv preprint arXiv:2510.13936}, 2025.

\bibitem[Zong et~al.(2024)Zong, Wang, Qin, Feng, Wang, and An]{zong2024macrohft}
Chuqiao Zong, Chaojie Wang, Molei Qin, Lei Feng, Xinrun Wang, and Bo~An.
\newblock Macrohft: Memory augmented context-aware reinforcement learning on high frequency trading.
\newblock In \emph{Proceedings of the 30th ACM SIGKDD Conference on Knowledge Discovery and Data Mining}, pp.\  4712--4721, 2024.

\end{thebibliography}
